\documentclass[%
 reprint,
 amsmath,amssymb,
 aps,
]{revtex4-1}

\usepackage{graphicx}
\usepackage{dcolumn}
\usepackage{bm}

\begin{document}

\preprint{APS/123-QED}

\title{Some Aspects of Tachyon Field Cosmology}

\author{Kourosh Nozari}
 \homepage{knozari@umz.ac.ir}
\author{Narges Rashidi}
\homepage{n.rashidi@umz.ac.ir}%
\affiliation{Department of Physics, Faculty of Basic Sciences,\\
University of Mazandaran,\\
P. O. Box 47416-95447, Babolsar, IRAN}

\date{\today}

\begin{abstract}
We study inflation, perturbations, non-gaussinity and late-time
cosmological dynamics of a tachyon field both minimally and
non-minimally coupled to gravity. By analyzing the parameters space
of the model, the viability of the model in confrontation with
recent observational data is considered. In a dynamical system
technique, we study the phase space dynamics of both minimally and
non-minimally coupled tachyon field. We find the fixed points (lines
in our setup) and explore their stability. Also, we perform a
statefinder diagnostic to both cases and show that the trajectories
of the state finder pairs reach a stable state which is
corresponding to a $\Lambda$CDM scenario.
\begin{description}
\item[PACS numbers]
98.80.Cq\,, 95.36.+x
\item[Key Words]
Inflation, Cosmological Perturbations, Non-Gaussinity, Accelerated
Expansion, Tachyon Field, Dynamical System, Statefinder Diagnostic
\end{description}
\end{abstract}

\maketitle


\section{Introduction}

Despite the great successes of the standard big bang cosmology in
confrontation with observational data, it suffers from some
shortcomings such as the flatness, horizon and relics problems. It
has been shown that an inflationary stage in the early time
evolution of the universe can address successfully at least some of
these problems. In a simple inflationary model, the universe is
dominated with a scalar field called inflaton whose potential energy
dominates over the kinetic term (the slow-roll conditions), followed
by a reheating period
~\cite{Gut81,Lin82,Alb82,Lin90,Lid00a,Lid97,Rio02,Lyt09}. Inflation
also provides a mechanism for production of density perturbations
required to seed the formation of structures in the universe. But
there are several problems with no concrete solutions in inflation
paradigm such as natural realization of inflation in a fundamental
theory, cosmological constant and dark energy problem, unexpected
low power spectrum at large scales and egregious running of the
spectral index ~\cite{Bra05}. Another unsolved problem in the spirit
of the inflationary scenario is that yet we don't know how to
integrate it with ideas of the particle physics. For example, we
would like to identify the inflaton, the scalar field that drives
inflation, with one of the known fields of particle physics. Also,
it is important that the inflaton potential to be emerged naturally
from underlying fundamental theory~\cite{Lid97}. In this respect,
tachyon fields associated with unstable D-branes could be
responsible for inflation in early time.

Focusing on late-time dynamics, recent cosmological observations
show that our universe is undergoing an accelerating phase of
expansion and transition to the accelerated phase has been occurred
in the recent cosmological past ~\cite{Rie98,Rie04,Spr06}. The
simplest way to describe the accelerating phase of universe
expansion is to adopt a cosmological constant in Einstein's field
equations. However, there are some problems with cosmological
constant that make it unfavorable, such as huge amount of
fine-tuning required for its magnitude and other theoretical
problems such as unknown origin and lake of dynamics
~\cite{Pad03,Car01}. To explain this remarkable behavior of the
universe, several theoretical approaches have been proposed, some of
which can be seen in Refs.
~\cite{Cop06,Pia02,Wei06,Ani05,Wan05,Noj06a,Zha06,Ala04,Sam09a,
Cal09,Sah04,Zha09,Cap03,Noj09,Sot10,Ame07,Tsu10}. Specially, dark
energy models based on scalar fields have attracted much attention
in recent years, see for instance~\cite{Cop06,Cal98,Lid99,Rat88}.\\
The scalar field we are going to explore its cosmological dynamics
is the tachyon field described by the Dirac-Born-Infeld (DBI)
action. Such a scalar field is associated with D-branes in string
theory ~\cite{Sen99,Sen02b,Sen02c}. This field can be responsible
for early time inflation in the history of the universe (see for
instance ~\cite{Sam02,Fei02}) and also can be considered as a dark
energy in the late time (see for instance
~\cite{Pad02,Gor04,Edm05}). As has been indicated in
Ref.~\cite{Gib02}, in the case that the tachyon condensate starts to
roll down slowly the potential, a universe dominated by this field
evolves smoothly from a phase of accelerated expansion to an era
dominated by a non-relativistic fluid. These features show that
tachyon fields may provide suitable candidates to realize initial
inflation and late time cosmic speed-up. This is the
reason why we consider a tachyon field for our purposes.\\
Since scalar fields can interact with other fields, such as the
gravitational sector of the theory, in the spirit of scalar-tensor
theories we can consider a non-minimal coupling (NMC) of the tachyon
field with intrinsic (Ricci) curvature. There are compelling reasons
to include an explicit non-minimal coupling in the action. For
instance, it is necessary for the renormalizability of the scalar
field theory in curved space. Also it arises at the quantum level
when quantum corrections to the scalar field theory are considered.
In most theories used to describe inflationary scenarios, a
non-vanishing value of the coupling constant cannot be avoided
~\cite{Far96,Far00}. Many authors have studied the models with
scalar field non-minimally coupled to gravity (see for instance
~\cite{Spo84,Fut89,Sal89,Fak90,Mak91,Hwa99,Tsu00b,
Pal11,Noz07b,Noz10,Noz12}). The non-minimally coupled tachyon field
also has been considered by some authors~\cite{Pia03,Pra04,Alb05}.\\
In this paper, we study cosmological dynamics of a tachyon field
both in the inflationary and late time epochs. Firstly, we consider
a tachyon field which is minimally coupled to the Ricci scalar. We
study its cosmological dynamics as a field responsible for cosmic
inflation. The perturbation will be considered in details and a
comparison with recent observational data will be done. Also the
issue of non-gaussinity of perturbations will be considered with
details. Then we study the role of a tachyon field as a dark energy
in the late time history of the universe. We study the cosmological
dynamics of the model in a dynamical system approach. By obtaining
the autonomous system of equations we consider if there is any late
time attractor in the model. Also, by studying the cosmological
diagnostic pair (dubbed statefinder) we compare the model, which
contains a tachyon as dark energy component, with $\Lambda$CDM
scenario. Secondly, we consider a tachyon field which is
non-minimally coupled to the Ricci scalar. As the minimal case, we
study the role of tachyon field as deriver of inflation and also a
possible candidate for dark energy responsible for late time
acceleration. Here also, the perturbations and their non-gaussinity
will be studied in details. On the other hand, in a dynamical system
technique and phase space trajectories we study the cosmological
dynamics in the presence of a non-minimally coupled tachyon field.
As we will show, in the non-minimal coupling case we find an
attractor in the phase space of the model which has capability to
explain the late time accelerating phase of the universe expansion.
By exploring the diagnostic pair we shall see that a model with a
non-minimally coupled tachyon field reaches a stable
$\Lambda$CDM-like state and remains there forever.

\section{Minimally coupled tachyon field}
\subsection{Inflation}

The 4-dimensional action for a tachyon field, which is minimally
coupled to gravity, can be written as follows
\begin{equation}
S=\int\sqrt{-g}\Bigg[\,\frac{1}{\kappa^{2}}R-V(\phi)\sqrt{1-\partial^{\mu}\phi\partial_{\mu}\phi}\,\Bigg]d^{4}x\,,
 \label{1}
\end{equation}
where $R$ is the 4-dimensional Ricci scalar, $\phi$ is the tachyon
field and $V(\phi)$ is its potential. Einstein's field equations
calculated from action \eqref{1} are given by
\begin{equation}
\label{2} G_{\mu\nu}=\kappa^{2}T_{\mu\nu}\,,
\end{equation}
where $T_{\mu\nu}$ is the energy momentum tensor given by
\begin{equation}
\label{3}
T_{\mu\nu}=-\frac{V{(\phi)}}{\sqrt{1-\partial_{\alpha}\phi\partial^{\alpha}\phi}}\,\partial_{\mu}\phi\partial_{\nu}\phi
+g_{\mu\nu}\Big(-V(\phi)\sqrt{1-\partial_{\alpha}\phi\partial^{\alpha}\phi}\Big)\,.
\end{equation}
The energy momentum tensor \eqref{3} leads to the following energy
density and pressure for the tachyon field
\begin{equation}
\label{4} \rho_{\phi}=\frac{V(\phi)}{\sqrt{1-\dot{\phi}^{2}}}\,,
\end{equation}
and
\begin{equation}
\label{5} p_{\phi}=-V(\phi)\sqrt{1-\dot{\phi}^{2}}\,.
\end{equation}
In order to formulate cosmological dynamics, we assume the following
spatially flat FRW line element
\begin{equation}
\label{6} ds^{2}=-n^{2}(t)dt^{2}+a^{2}(t)\gamma_{ij}dx^{i}dx^{j}\,,
\end{equation}
where $\gamma_{ij}$ is a maximally symmetric 3-dimensional metric
defined as $\gamma_{ij}=\delta_{ij}+k\frac{x_{i}x_{j}}{1-kr^{2}}$
where $k=-1,0,+1$ parameterizes the spatial curvature and
$r^{2}=x_{i}x^{i}$. By using this line element with $n^{2}(t)=1$ and
considering the $(0,0)$ component of Einstein's field equations, the
Friedmann equation of a model with minimally coupled tachyon field
and flat spatial section is obtained as follows
\begin{equation}
\label{7}
H^{2}=\frac{\kappa^{2}}{3}\,\frac{V(\phi)}{\sqrt{1-\dot{\phi}^{2}}}\,.
\end{equation}
Variation of the action \eqref{1} with respect to the scalar field
gives the following equation of motion
\begin{equation}
\label{8}
\frac{\ddot{\phi}}{1-\dot{\phi}^{2}}+3H\dot{\phi}+\frac{V'}{V}=0\,,
\end{equation}
where a prime refers to a derivative with respect to the tachyon
field and a dot marks derivative with respect to the cosmic time .
Also, the energy conservation equation of the model is given by
\begin{equation}
\label{9} \dot{\rho}_{\phi}+3H(\rho_{\phi}+p_{\phi})=0\,.
\end{equation}

During the inflationary era and in the slow-roll approximation,
where $\dot{\phi}^{2}\ll 1$ and $\ddot{\phi}\ll|3H\dot{\phi}|$,
energy density and equation of motion for scalar field take the
following forms respectively
\begin{equation}
\label{10} \rho_{\phi}=V(\phi)\,,
\end{equation}
and
\begin{equation}
\label{11} 3H\dot{\phi}+\frac{V'}{V}=0\,.
\end{equation}
So, the Friedmann equation is reduced to the following form
\begin{equation}
\label{12} H^{2}=\frac{\kappa^{2}}{3}V(\phi)\,.
\end{equation}
Now, we introduce the slow-roll parameters as follows
\begin{equation}
\label{13} \epsilon\equiv -\frac{\dot{H}}{H^{2}}\,\,,
\end{equation}
and
\begin{equation}
\label{14} \eta\equiv -\frac{1}{H}\frac{\ddot{H}}{\dot{H}}\,\,,
\end{equation}
which in our setup and within the slow-roll approximation these
parameters become
\begin{equation}
\label{15} \epsilon=\frac{1}{2\kappa^{2}}\frac{V'^{2}}{V^{3}}\,,
\end{equation}
and
\begin{equation}
\label{16}
\eta=\frac{1}{\kappa^{2}}\left(\frac{V''}{V^{2}}-\frac{1}{2}\frac{V'^{2}}{V^{3}}\right)\,.
\end{equation}
The condition under which the inflation epoch could take place is
$\{\epsilon,\eta\}<1$; as soon as one of these parameters reaches
unity, the inflation phase terminates.

The number of e-folds during inflation is given by
\begin{equation}
\label{17} N=\int_{t_{hc}}^{t_{f}} H dt\,,
\end{equation}
which, within the slow-roll approximation can be written as
\begin{equation}
\label{18} N\simeq-\int_{\phi_{hc}}^{\phi_{f}} 3H^{2}\frac{V}{V'}
d\phi\,,
\end{equation}
where $\phi_{hc}$ denotes the value of $\phi$ when the universe
scale observed today crosses the Hubble horizon during inflation and
$\phi_{f}$ is the value of $\phi$ when the universe exits the
inflationary phase. In a model with a minimally coupled tachyon
field, the number of e-folds in the slow-roll approximation takes
the following form
\begin{equation}
\label{19}
N\simeq-\int_{\phi_{hc}}^{\phi_{f}}\kappa^{2}\frac{V^{2}}{V'}d\phi\,.
\end{equation}

The key test of any inflation model is the spectrum of perturbations
produced due to quantum fluctuations of the fields about their
homogeneous background values. So, in the following section, we
consider the scalar perturbation of the metric in our setup.

\subsection{Perturbations}

In this section, we explore the linear perturbation theory in
inflation with tachyon field. There are many different ways of
characterizing cosmological perturbations, which depend on the
choice of gauge (coordinates). In longitudinal gauge, the scalar
metric perturbations of the FRW background are given by
~\cite{Bar80,Muk92,Ber95}
\begin{equation}
\label{20}
ds^{2}=-\big(1+2\Phi\big)dt^{2}+a^{2}(t)\big(1-2\Psi\big)\delta_{i\,j}\,dx^{i}dx^{j}\,,
\end{equation}
where $a(t)$ is the scale factor, $\Phi=\Phi(t,x)$ and
$\Psi=\Psi(t,x)$, the metric perturbations, are gauge-invariant
variables. The spatial dependence of all perturbed quantities are of
the form of plane waves $e^{ikx}$, where $k$ is the wave number. The
perturbation of the metric leads to the perturbation in the
energy-momentum tensor through Einstein's field equations. For the
perturbed metric \eqref{20}, the perturbed Einstein's field
equations can be obtained as follows
\begin{equation}
\label{21}
-3H(H\Phi+\dot{\Psi})-\frac{k^{2}}{a^{2}}=\frac{\kappa_{4}^{2}}{2}\delta
\rho_{\phi}\,,
\end{equation}
\begin{equation}
\label{22}
\ddot{\Psi}+3H(H\Phi+\dot{\Psi})+H\dot{\Phi}+2\dot{H}\Phi+\frac{1}{3a^{2}}k^{2}(\Phi-\Psi)=
\frac{\kappa_{4}^{2}}{2}\delta p_{\phi}\,,
\end{equation}
\begin{equation}
\label{23}
\dot{\Psi}+H\Phi=-\frac{\kappa^{2}V(\phi)}{\sqrt{1-\dot{\phi}^{2}}}\frac{\dot{\phi}\delta\phi}{2}.
\end{equation}
\begin{equation}
\label{24} \Psi-\Phi=0.
\end{equation}
As we see from last equation, in the minimally coupled tachyon field
model, the two metric perturbations are equal. In equations
\eqref{21} and \eqref{22}, $\delta\rho_{\phi}$ and $\delta
p_{\phi}$, the perturbed energy density and pressure are given by
\begin{equation}
\label{25}
\delta\rho_{\phi}=\frac{V'\,\delta\phi}{\sqrt{1-\dot{\phi}^{2}}}-V\,\frac{\dot{\phi}\,\delta\dot{\phi}
+\dot{\phi}^{2}\,\Phi}{\big(1-\dot{\phi}^{2}\big)^{\frac{3}{2}}}\,,
\end{equation}
and
\begin{equation}
\label{26} \delta
p_{\phi}=-V'\,\sqrt{1-\dot{\phi}^{2}}\,\delta\phi-V\,\frac{\dot{\phi}\,\delta\dot{\phi}
+\dot{\phi}^{2}\,\Phi}{\sqrt{1-\dot{\phi}^{2}}}\,,
\end{equation}
which are obtained by perturbing of equations \eqref{4} and
\eqref{5}. The variation of the scalar field's equation of motion
\eqref{8} leads to the following expression
\begin{eqnarray}
\delta\ddot{\phi}+3H\delta\dot{\phi}+2\dot{\phi}\,\ddot{\phi}\,\frac{\delta\dot{\phi}+\dot{\phi}^{2}\Phi}{1-\dot{\phi}^{2}}+
\frac{\sqrt{1-\dot{\phi}^{2}}}{V}\,\Bigg[\Big(\frac{ka^{2}}{a^{2}}\hspace{1cm}\nonumber\\
-3\dot{H}\Big)\Phi-\frac{2k^{2}}{a^{2}}\Psi-3\Big(\ddot{\Psi}+4H\dot{\Psi}+H\dot{\Phi}
\hspace{2cm}\nonumber\\
+\dot{H}\Phi+4H^{2}\Phi\Big)\Bigg]=\Bigg(6H\dot{\phi}^{3}-\frac{2V'}{V}(1-\dot{\phi}^{2})\Bigg)\Phi\hspace{1cm}\nonumber\\
+\dot{\phi}\Big(\dot{\Phi}+3\dot{\Psi}\Big)+(1-\dot{\phi}^{2})\delta\phi\Bigg(\frac{V''}{V}-\frac{V'^{2}}{V^{2}}\Bigg).
\label{27}\hspace{1cm}
\end{eqnarray}

One can decompose the scalar perturbations into the entropy
(isocurvature) perturbations which are projection orthogonal to the
trajectory, and adiabatic (curvature) perturbations which are
projection parallel to the trajectory. If inflation is driven by
more than one scalar field ~\cite{Lan00,Lan07,Bas99,Gor01} or it
interacts with other fields such as the scalar Ricci term
~\cite{Lop04,Kal05}, we deal with the isocurvature perturbations. If
there is only one scalar field during the inflationary period, we
deal with the adiabatic perturbations
~\cite{Bas99,Gor01,Lop04,Kal05,Maa00}. In this section, since the
tachyon field is the only field in the inflationary period, the
perturbations are adiabatic perturbations. Since we are dealing with
the linear perturbation (first order cosmological perturbations), we
can define a gauge-invariant primordial curvature perturbation
$\zeta$, on scales outside the horizon, as follows ~\cite{Bar83}
\begin{equation}
\label{28}
\zeta=\Psi-\frac{H}{\dot{\rho}_{\phi}}\delta\rho_{\phi}\,\,.
\end{equation}
On uniform density hypersurfaces where $\delta\rho_{\phi}=0$, the
above quantity reduces to the curvature perturbation, $\Psi$.
Equation \eqref{28} leads to the following equation for time
evolution of $\zeta$ \cite{Wan00}
\begin{equation}
\label{29} \dot{\zeta}=H\left(\frac{\delta
p_{nad}}{\rho_{\phi}+p_{\phi}}\right)\,.
\end{equation}
Equation \eqref{29} shows that the change in the curvature
perturbation on uniform-density hypersurfaces, on large scales, is
due to the non-adiabatic part of the pressure perturbation,
independent of the form of the gravitational field equations.
$\zeta$ is constant if the pressure perturbation is adiabatic on the
large scales.

The pressure perturbation (in any gauge) can be decomposed into
adiabatic and entropic (non-adiabatic) parts ~\cite{Wan00}
\begin{equation}
\label{30} \delta
p=c_{s}^{2}\delta\rho_{\phi}+\dot{p}_{\phi}\Gamma\,\,,
\end{equation}
where $c_{s}^{2}=\frac{\dot{p}_{\phi}}{\dot{\rho}_{\phi}}$ is the
sound effective velocity. The non-adiabatic part is $\delta
p_{nad}=\dot{p}_{\phi}\Gamma$\,, where $\Gamma$ marks the
displacement between hypersurfaces of uniform pressure and density.
From equations \eqref{30} and within the slow-roll conditions, we
can deduce
\begin{equation}
\label{31} \delta p_{nad}=0\,.
\end{equation}
That is, the non-adiabatic part of the pressure perturbation is
zero; the pressure perturbation is adiabatic. So, from equation
\eqref{29} we find
\begin{equation}
\label{32} \dot{\zeta}=0\,.
\end{equation}
It has been shown that the curvature perturbation on uniform density
hypersurfaces, in terms of the scalar field fluctuations on
spatially flat hypersurfaces, is given by ~\cite{Lyt09}
\begin{equation}
\label{33} \zeta=-\frac{H\delta\phi}{\dot{\phi}}\,.
\end{equation}
Also, the field fluctuations at Hubble crossing and within the
slow-roll limit are given by the following expression, which is
independent of the underlying gravity theory for a massless field in
de Sitter space
\begin{equation}
\label{34} \langle\delta\phi^{2}\rangle=\frac{H^{2}}{4\pi^{2}}\,.
\end{equation}
Since in the minimally coupled tachyon field setup, perturbations
are adiabatic, $\zeta$ can be related to the density perturbations
by the following equation ~\cite{Lid93}
\begin{equation}
\label{35} A_{s}^{2}=\frac{\langle\zeta^{2}\rangle}{V}
\end{equation}
So, from equations \eqref{33}-\eqref{35}, we find
\begin{equation}
\label{36}
A_{s}^{2}=\frac{\kappa^{6}}{12\pi^{2}}\frac{V^{4}}{V'^{2}}\,.
\end{equation}
The scale-dependence of the perturbations is described by the
spectral index as
\begin{equation}
\label{37} n_{s}-1=\frac{d \ln A_{s}^{2}}{d \ln k}\,.
\end{equation}
The interval in wave number is related to the number of e-folds by
the relation
\begin{equation}
\label{38} d \ln k(\varphi)=d N(\varphi)\,.
\end{equation}
So, from equations \eqref{36}-\eqref{38} we obtain (by regarding the
definition of the slow-roll parameters)
\begin{equation}
\label{39} n_{s}=1-6\epsilon+2\eta\,.
\end{equation}

The tensor perturbations amplitude of a given mode when leaving the
Hubble radius are given by
\begin{equation}
\label{40} A_{T}^{2}=\frac{4\kappa^{2}}{25\pi}H^{2}\Bigg|_{k=aH}\,.
\end{equation}
In our setup and within the slow-roll approximation, we find
\begin{equation}
\label{41} A_{T}^{2}=\frac{4\kappa^4}{75\pi}V
\end{equation}
The tensor spectral index is given by
\begin{equation}
\label{42} n_{T}=\frac{d \ln A_{T}^{2}}{d \ln k}\,,
\end{equation}
that in our model and in terms of the slow-roll parameters, the
tensor (gravitational wave) spectral index is given by the following
expression
\begin{equation}
\label{43} n_{T}=-2\epsilon\,.
\end{equation}

Another important parameter is the ratio between the amplitudes of
tensor and scalar perturbations (tensor-to-scalar ratio) which is
given by
\begin{equation}
\label{44}
r_{t-s}=\frac{A_{T}^{2}}{A_{s}^{2}}=\frac{32}{25\pi}\epsilon\,.
\end{equation}

In the following we perform numerical analysis in our setup, by
considering three types of potential.

\begin{figure*}
\flushleft\leftskip1em{
\includegraphics[width=.35\textwidth,origin=c,angle=0]{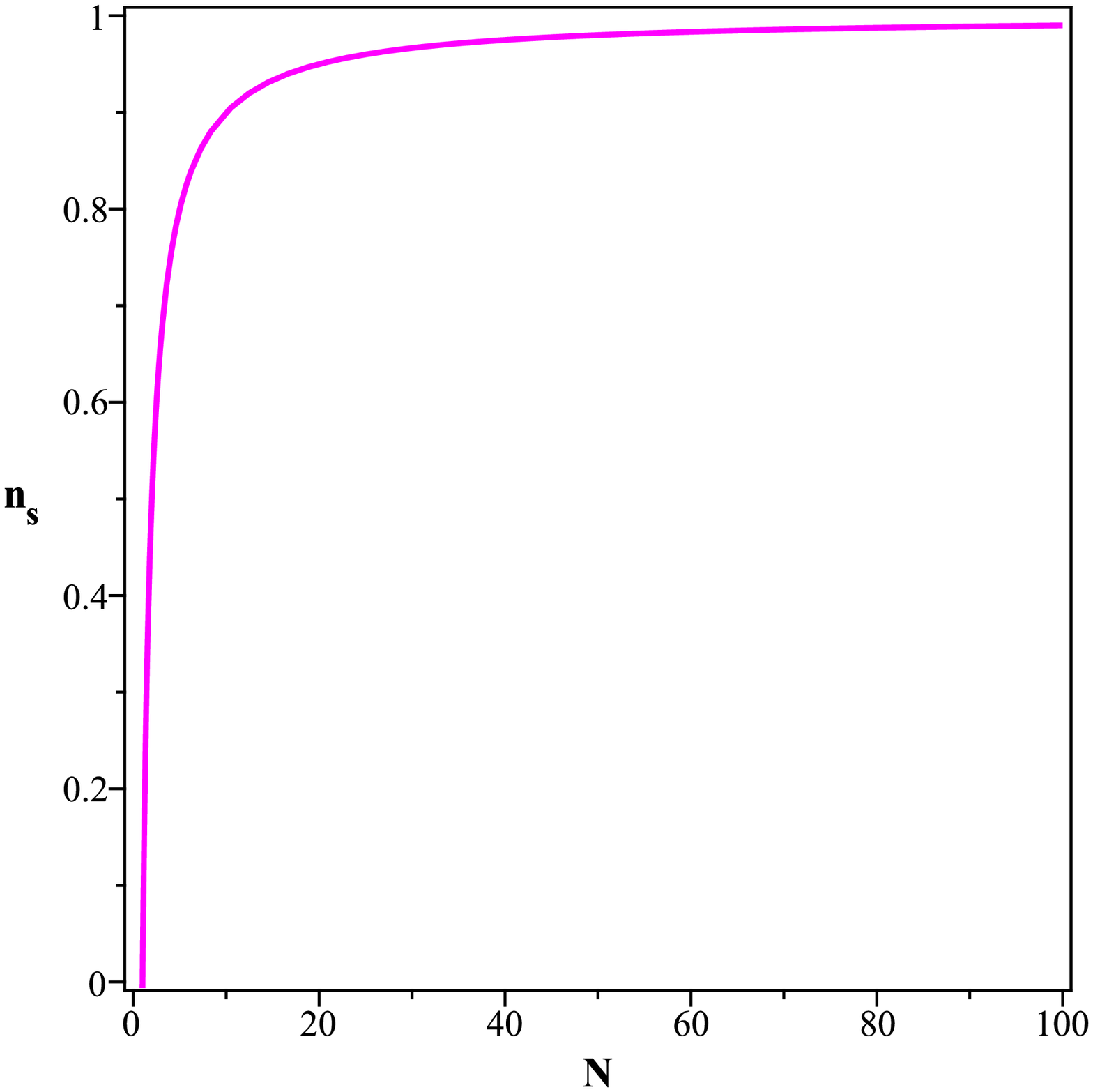}}
\hspace{3.6cm}{
\includegraphics[width=.35\textwidth,origin=c,angle=0]{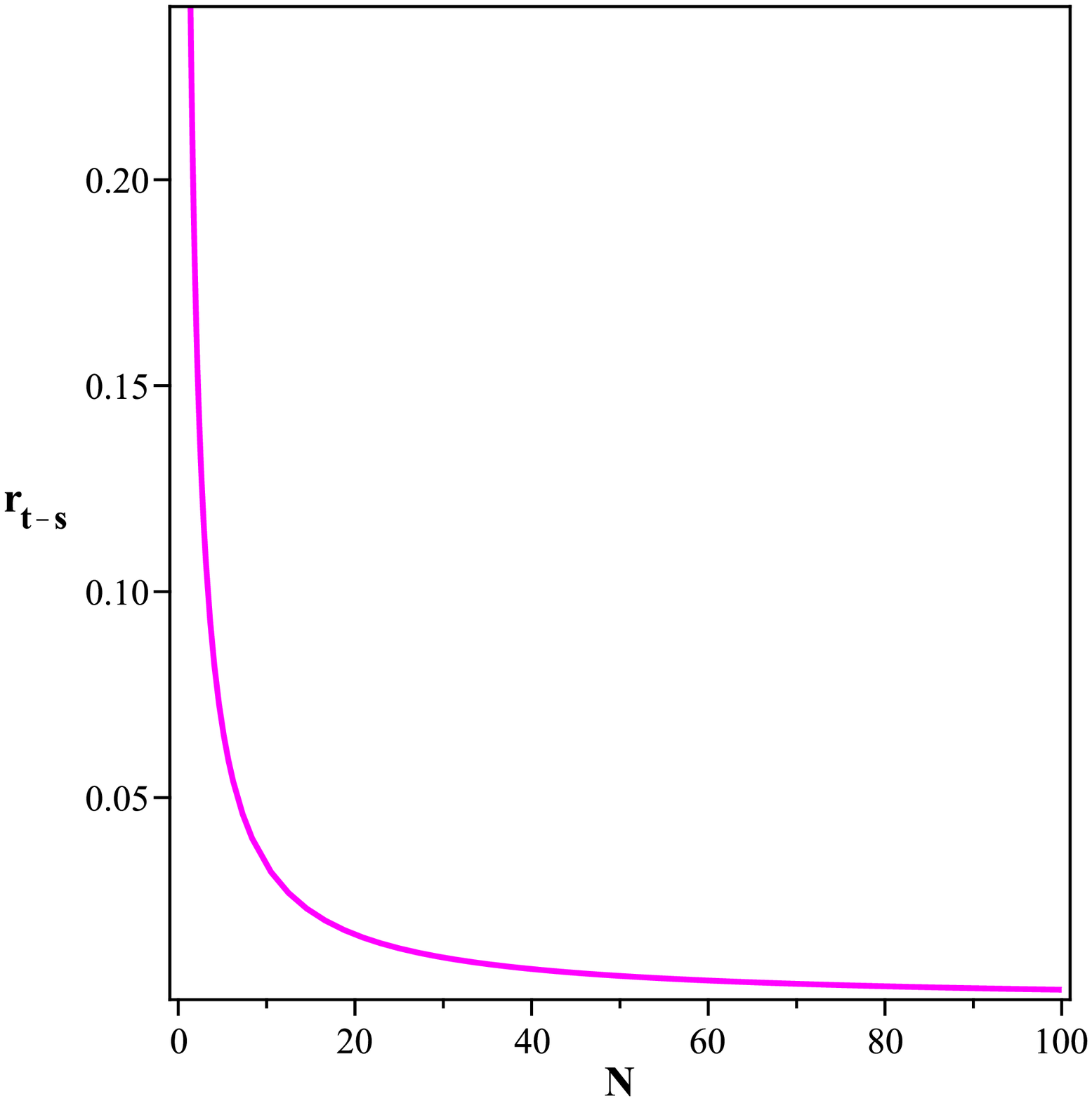}}
\caption{\label{fig1} Evolution of the scalar spectral index (left
panel) and tensor to scalar ratio (right panel) versus the number of
e-folds for the quadratic potential. For a minimally coupled tachyon
field, the spectral index is red tilted and nearly scale invariant.}
\end{figure*}
\begin{figure}
\flushleft\leftskip0em{
\includegraphics[width=.40\textwidth,origin=c,angle=0]{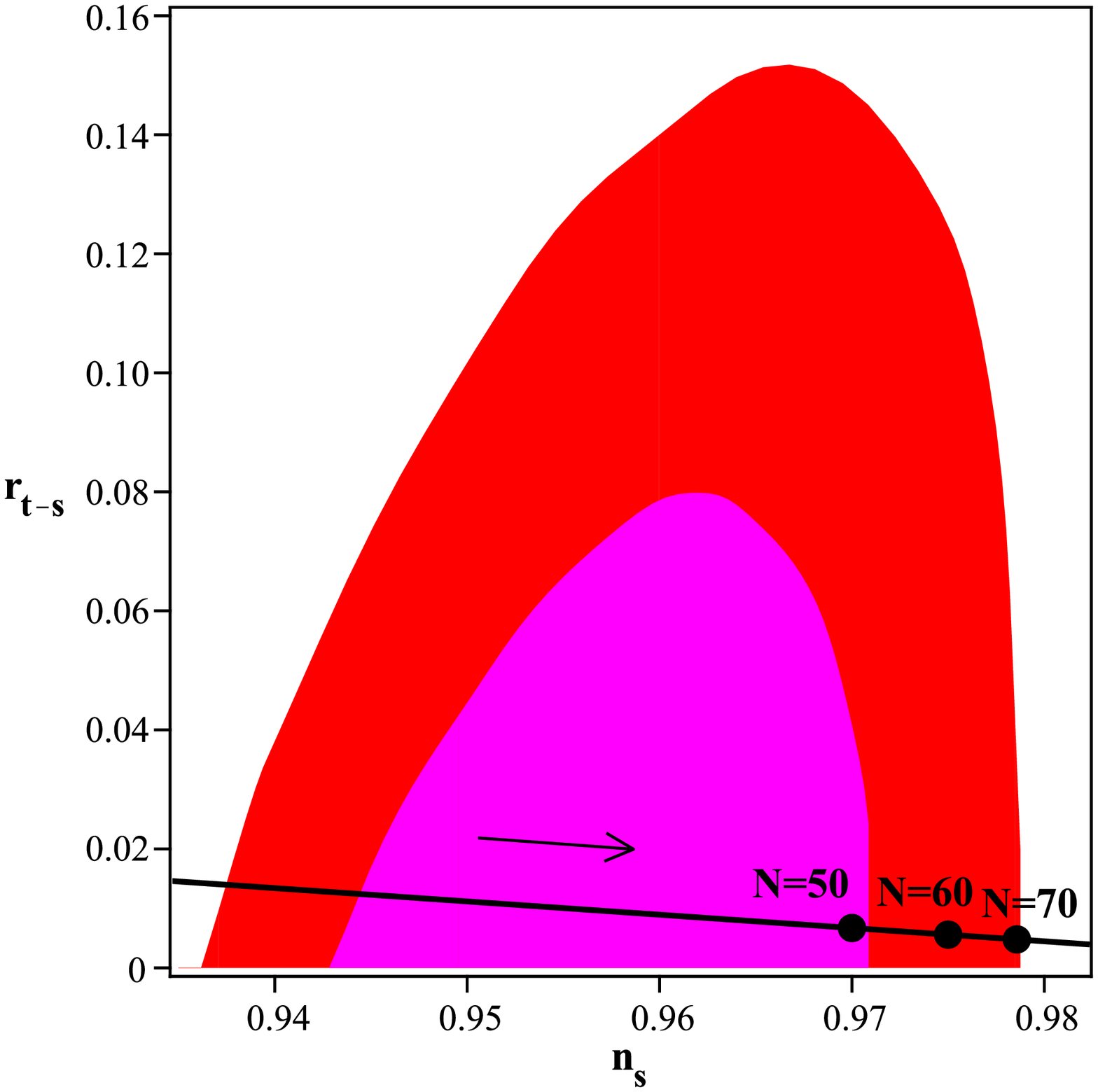}}
\caption{\label{fig2} Behavior of the tensor to scalar ratio with
respect to the scalar spectral index for various $N$ and for
quadratic potential in the background of WMAP9+eCMB+BAO+H$_{0}$
data. The two contours are corresponding to the 68$\%$ and 95$\%$
levels of confidence. $N$ increases in the direction of the arrow.}
\end{figure}
\begin{figure*}
\flushleft\leftskip1em{
\includegraphics[width=.35\textwidth,origin=c,angle=0]{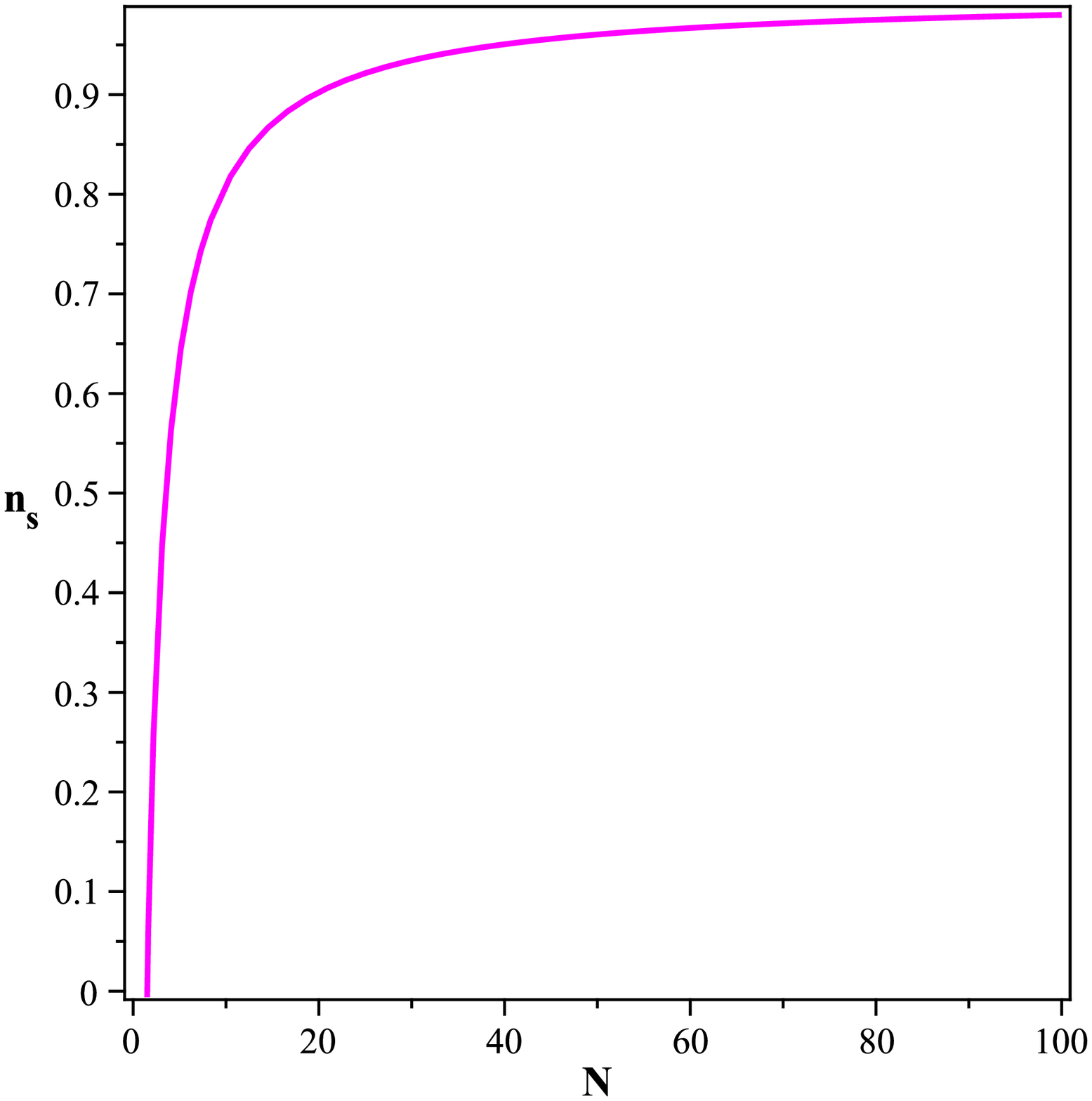}}
\hspace{3.6cm}{
\includegraphics[width=.35\textwidth,origin=c,angle=0]{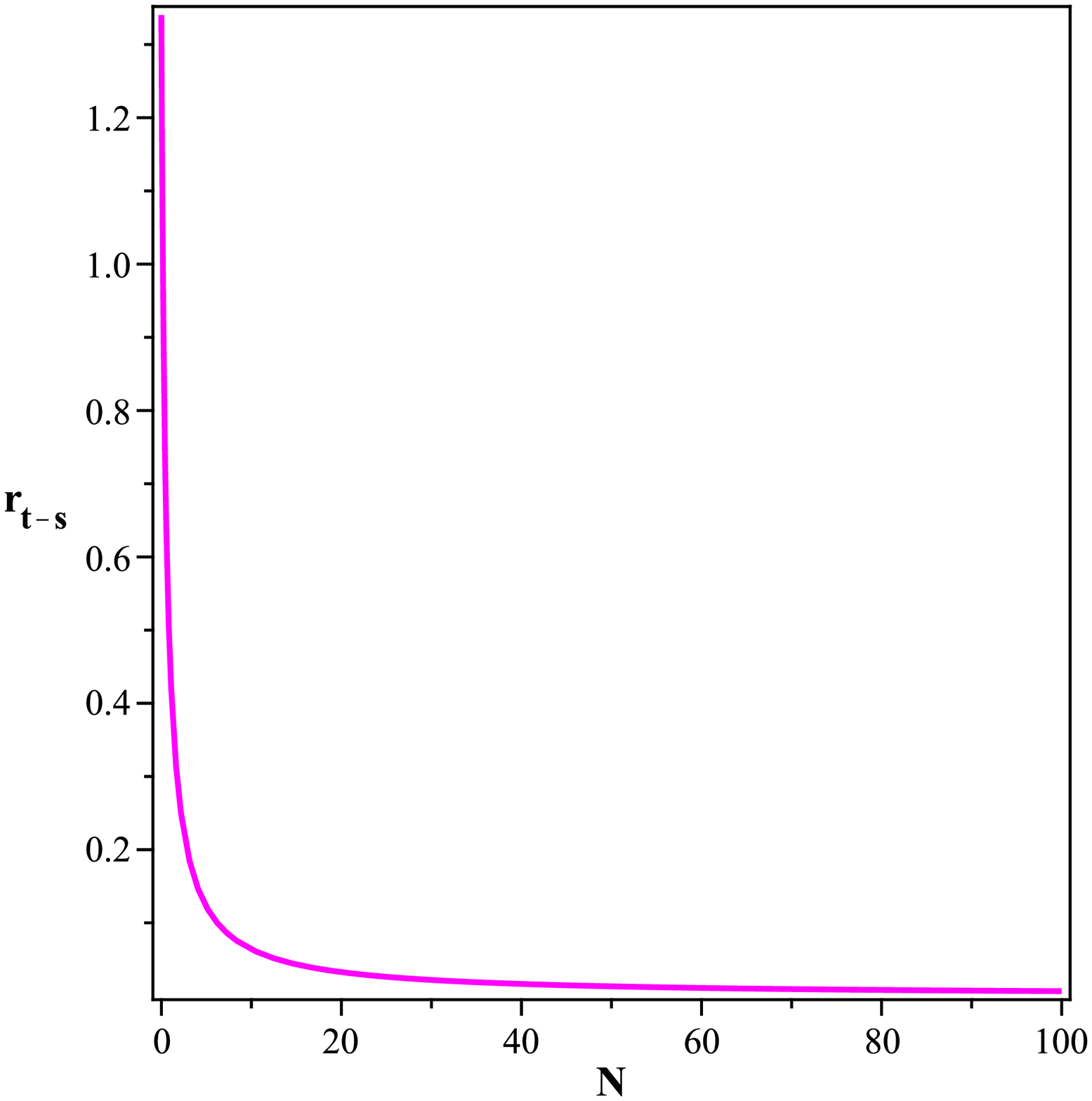}}
\caption{\label{fig3} Evolution of the scalar spectral index (left
panel) and tensor to scalar ratio (right panel) versus the number of
e-folds, for an exponential type potential.}
\end{figure*}
\begin{figure}
\flushleft\leftskip0em{
\includegraphics[width=.40\textwidth,origin=c,angle=0]{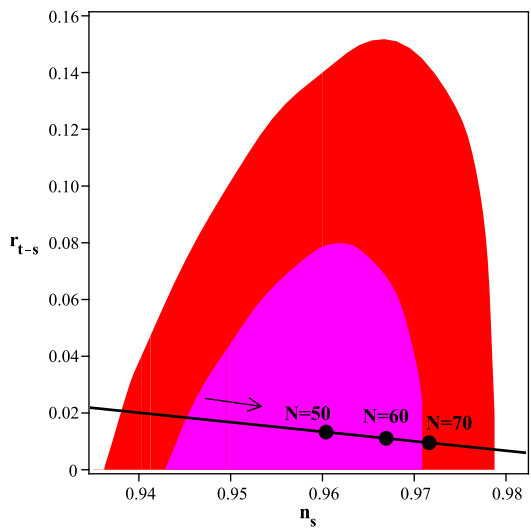}}
\caption{\label{fig4} Behavior of the tensor to scalar ratio with
respect to the scalar spectral index for various $N$ and for
exponential type potential in the background of
WMAP9+eCMB+BAO+H$_{0}$ data. The two contours are corresponding to
the 68$\%$ and 95$\%$ levels of confidence.}
\end{figure}

\subsubsection{$V=\frac{1}{2}\sigma\phi^{2}$}
In this subsection, we consider the quadratic tachyon potential.
With this choice, solving the integral of equation \eqref{19} gives
\begin{equation}
\label{45} N=-\frac{1}{16}\,{\kappa}^{2}\sigma
\left(\phi_{f}^{4}-\phi_{hc}^{4}\right)\,.
\end{equation}
By assuming $\phi_{hc}\gg\phi_{f}$, we find $\phi_{hc}$ from
equation \eqref{45}. Then we substitute this obtained parameter into
equation \eqref{39} and plot the evolution of the scalar spectral
index versus the number of e-folds parameter. One can see the result
in the left panel of figure~\ref{fig1}. By increasing the number of
e-folds parameter, the scalar spectral index increases and reaches
unity asymptotically. As figure shows, in the case of minimally
coupled tachyon field, the scalar spectral index is red-tilted and
nearly scale invariant (note that in all figures we have set
$\kappa=\sigma=1$). The evolution of the tensor to scalar ratio with
respect to the number of e-folds parameter is shown in the right
panel of the figure~\ref{fig1}. The value of this ratio decreases as
$N$ increases. To compare with observational data, we plot the
evolution of the tensor to scalar ratio versus the scalar spectral
index for various $N$ in the background of WMAP9+eCMB+BAO+H$_{0}$
data (figure~\ref{fig2})(see ref~\cite{Hin13}). Note that the
observational parameters are defined at $k_{0}=0.002$ Mpc$^{-1}$
where $k_{0}$ refers to the value of $k$ when universe scale crosses
the Hubble horizon during inflation. In the case of quadratic
potential, for $24\leq N\leq70$ the values of $r_{t-s}$ and $n_{s}$
are compatible with observational data. We see that for these
values, the scalar spectral index, at the time that physical scales
crossed the horizon, is red-tilted and nearly scale invariant. In
the plot we have highlighted three values of $N$ that are considered
usually in literature ($N=50$, $N=60$, $N=70$).

\subsubsection{$V=V_{0}\,e^{-\kappa\sigma\phi}$}
Another potential which is considered in this paper, is an
exponential type potential. With an exponential potential, solving
the integral of equation \eqref{19} gives
\begin{equation}
\label{45-2} N=\frac{V_{0}\left( e^{-\kappa\sigma\phi_{hc}}-e^{-
\kappa\sigma\phi_{f}}\right)}{\sigma^{2}}\,.
\end{equation}
If we set \eqref{15} equal to 1 (corresponding to the end of
inflation), we obtain $\phi_{f}$ and by substituting it in equation
\eqref{45-2} we find $\phi_{hc}$. Now, we substitute this obtained
parameter into equation \eqref{39} and plot the evolution of the
scalar spectral index versus the number of e-folds (left panel of
figure~\ref{fig3}). As figure shows, by increasing the number of
e-folds parameter, the scalar spectral index increases and also, it
is red-tilted. The right panel of figure~\ref{fig3} shows the
evolution of the tensor to scalar ratio with respect to the number
of e-folds parameter. The value of this ratio decreases as $N$
increases. Here also, to compare with observational data, we plot
the evolution of the tensor to scalar ratio versus the scalar
spectral index for various $N$ in the background of
WMAP9+eCMB+BAO+H$_{0}$ data in figure~\ref{fig4}. In the case of
exponential potential, for $31\leq N\leq90$ the values of $r_{t-s}$
and $n_{s}$ are compatible with observational data. Here also, we
have highlighted three values of $N$; $N=50$, $N=60$ and $N=70$. In
table~\ref{tab1} we have summarized the value of $n_{s}$ and
$r_{t-s}$ for the mentioned values of $N$ and both for quadratic and
exponential potentials.
\begin{table*}
\caption{\label{tab1} Comparing the value of inflationary parameters
in the horizon crossing, for quadratic and exponential potential.
Note that $r < 0.13$ ($95\%$ CL) and $n_{s}=0.9636 \pm 0.0084$ with
WMAP9+eCMB+BAO+H$_0$.}
\begin{ruledtabular}
\begin{tabular}{ccccc}
Potential& Inflationary parameter & $N=50$ & $N=60$ & $N=70$\\
\hline $V=\sigma\phi^{2}$ & $n_{s}$ & $0.97$ & $0.975$ & $0.97857$ \\
& $r$ & $0.0067020$ & $0.0055851$ & $0.0047872$ \\
\hline $V=V_{0}\,e^{-\kappa\sigma\phi}$ &$n_{s}$ & $0.96040$ & $0.96694$ & $0.97163$ \\
&$r$ & $0.013271$ & $0.011078$ & $0.0095065$ \\
\end{tabular}
\end{ruledtabular}
\end{table*}

\subsubsection{Intermediate inflation}
Intermediate inflation is an interesting scenario in which the scale
factor evolves slower than the standard de Sitter inflation
($a=\exp(Ht)$) and faster than the power law inflation ($a=t^{p}$
with $p>1$). The evolution of the scale factor in an intermediate
inflation is given by $a=a_{0}\exp(\vartheta t^{l})$ with $0<l<1$
and positive constant $\vartheta$. With this scale factor and by
using equations \eqref{12} and \eqref{13} we find the intermediate
potential as follows
\begin{equation}
\label{45-3} V=b\,\phi^{-\beta}\,,
\end{equation}
where
\begin{equation}
\label{45-4} \beta=\frac{4l-4}{l-2}\,,
\end{equation}
and
\begin{equation}
\label{45-5} b=\frac{3{\vartheta}^{2}{l}^{2}}{\kappa^{2}} \left(
\frac{8}{3}\frac{1-l}{\left( l-2
 \right) ^{2}\vartheta l}\right)^{{\frac {2l-2}{l-2}}}\,.
\end{equation}
\begin{figure*}
\flushleft\leftskip1em{
\includegraphics[width=.35\textwidth,origin=c,angle=0]{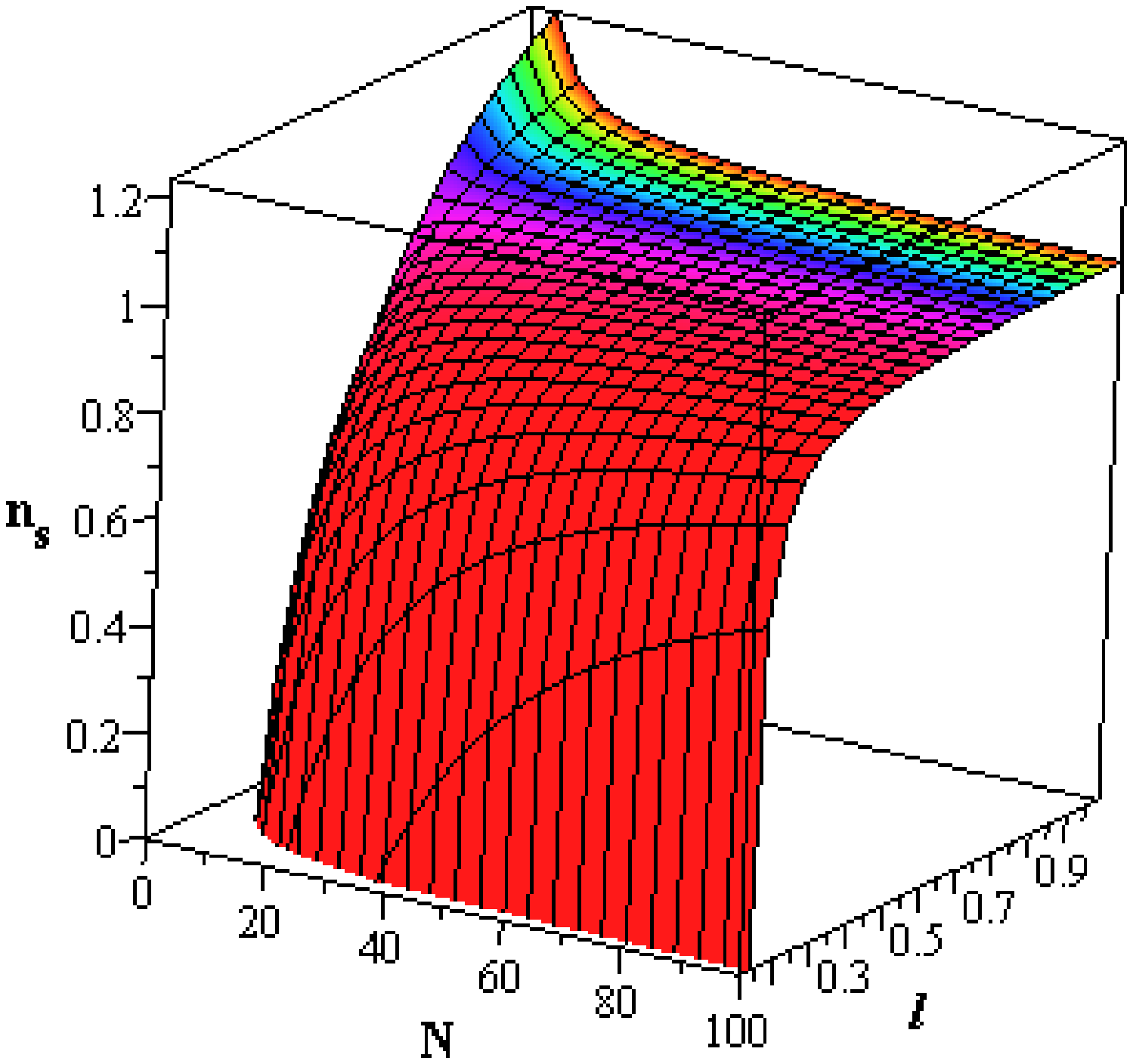}}
\hspace{3.6cm}
\includegraphics[width=.35\textwidth,origin=c,angle=0]{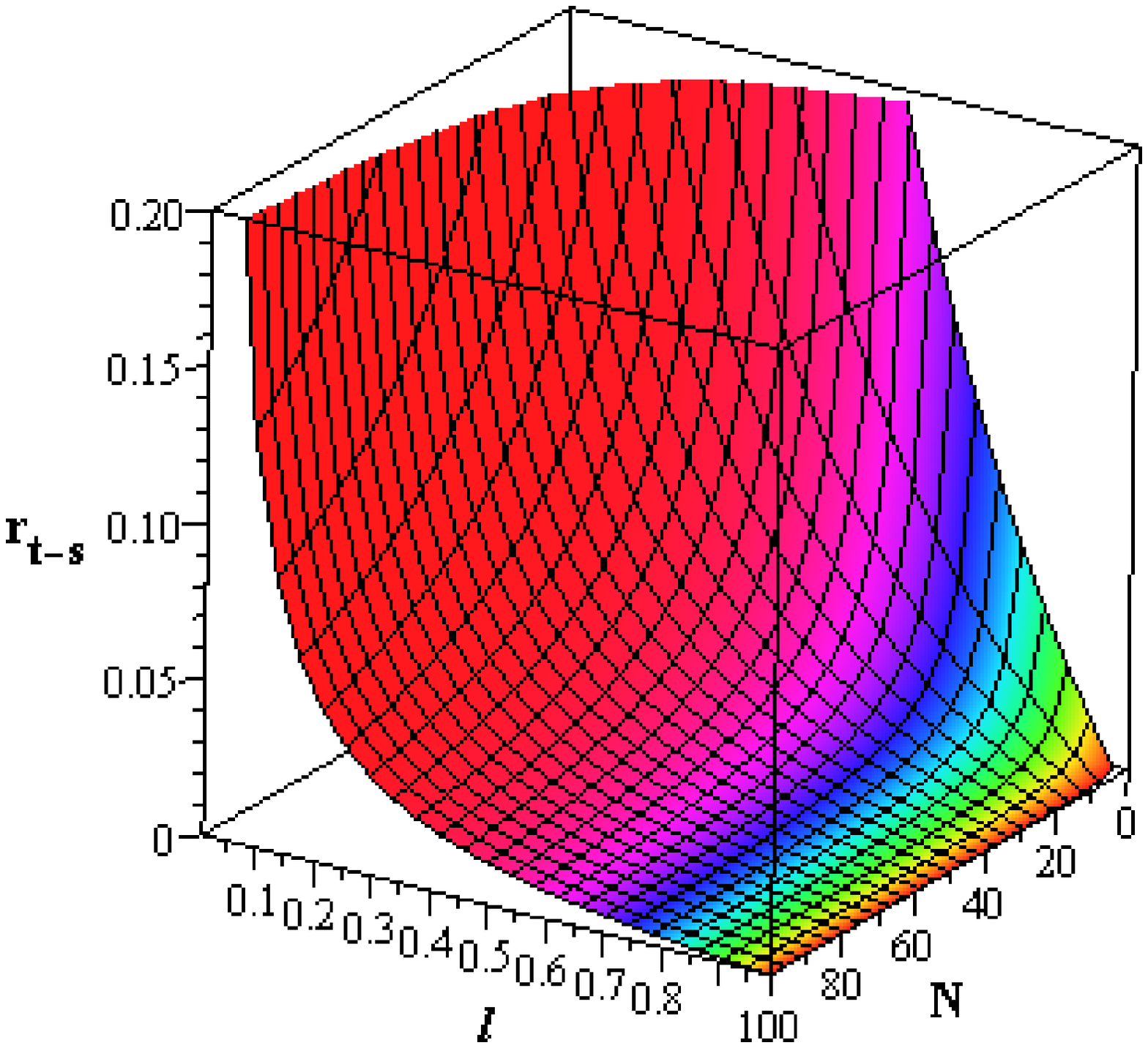}
\caption{\label{fig5} Evolution of the scalar spectral index (left
panel) and tensor to scalar ratio (right panel) versus the number of
e-folds and the intermediate parameter $l$, for the intermediate
potential.}
\end{figure*}
\begin{figure}
\flushleft\leftskip0em{
\includegraphics[width=.40\textwidth,origin=c,angle=0]{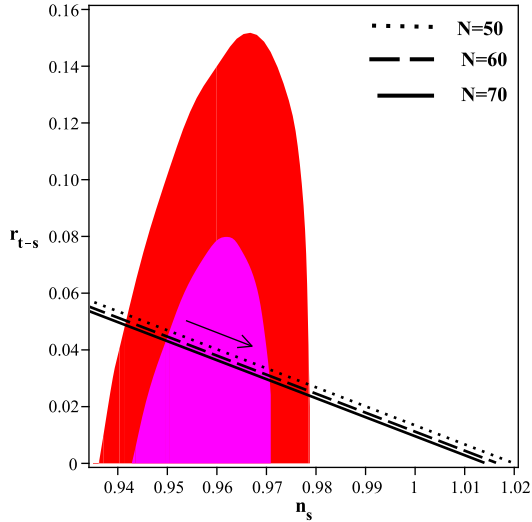}}
\caption{\label{fig6} Behavior of the tensor to scalar ratio with
respect to the scalar spectral index for various $l$ for
intermediate potential in the background of WMAP9+eCMB+BAO+H$_{0}$
data. The two contours are corresponding to the 68$\%$ and 95$\%$
levels of confidence. $l$ increases in the direction of the arrow.}
\end{figure}
With intermediate potential \eqref{45-3} we solve the integral of
equation \eqref{19} to obtain
\begin{equation}
\label{45-6} N=-\frac{\left(\phi_{hc}^{-\beta+2}-\phi_{f}^{-\beta+2}
\right){\kappa}^{2}b}{\left( \beta-2 \right) \beta}\,.
\end{equation}
By finding $\phi_{f}$ from equation \eqref{15} and substituting it
in equation \eqref{45-6} we obtain $\phi_{hc}$. Then, we substitute
this obtained parameter into equation \eqref{39} and plot the
evolution of the scalar spectral index versus the number of $N$ and
$l$ (left panel of figure~\ref{fig5}). As figure shows, depending on
the value of the number of e-folds parameter, the scalar spectral
index is red-tilted for some values of $l$. The right panel of
figure~\ref{fig5} shows the evolution of the tensor to scalar ratio
with respect to the number of e-folds and intermediate parameter,
$l$. The value of this ratio decreases as $N$ and $l$ increase.
Figure~\ref{fig6} shows the evolution of the tensor to scalar ratio
versus the scalar spectral index in the background of
WMAP9+eCMB+BAO+H$_{0}$ data for various $l$ and three values of $N$
($N=50$, $N=60$ and $N=70$). Note that a similar figure has been
plotted in ref~\cite{Cam09} for three specified values of $l$ but
various $N$. Here, we fix the value of $N$ and find the range of $l$
in which the model is compatible with observational data (see
table~\ref{tab2}).

\begin{table*}
 \caption{\label{tab2} The range of intermediate
parameter for which the values of inflationary parameters are
compatible with observational data from WMAP9+eCMB+BAO+H$_0$.}
\begin{ruledtabular}
\begin{tabular}{cccc}
 & $N=50$ & $N=60$ & $N=70$\\
\hline $l$ & $0.325<l<0.489$ & $0.295<l<0.463$ & $0.269<l<0.44$ \\
\end{tabular}
\end{ruledtabular}
\end{table*}

\subsection{Non-Guassianity}
Now we focus on the issue of non-gaussianity in the context of a
minimally coupled tachyon field model by exploring the bispectrum or
three-point correlation functions of the density perturbation
distribution in Fourier space. In this regard we follow the
reference~\cite{Gup02}. Equation of motion of the tachyon field, by
considering the gradient term, is given by the following expression
\begin{equation}
\label{45-7}
\dot{\phi}=\frac{1}{3H}\Big[a^{-2}\,\nabla^{2}\phi-\frac{V'}{V}+\Xi\Big]\,,
\end{equation}
where $\Xi$ is a stochastic term~\cite{Lyt09}. Now, to expand
equation~\eqref{45-7}, the background tachyon field is treated with
fluctuations, $\phi(x,t)=\phi_{0}(t)+\delta\phi(x,t)$, with
$\delta\phi=\delta\phi^{(1)}+\delta\phi^{(2)}$. So,
equation~\eqref{45-7}, in a time interval
$t_{n}-t_{n-1}=\frac{1}{H}$, gives the following expressions for the
first and second order fluctuations
\begin{equation}
\label{45-8}
\frac{d}{dt}\Big(\delta\phi^{(1)}(\textbf{k},t)\Big)=\frac{1}{3H}\left\{\left[-k^{2}-\frac{V''}{V}
+\frac{V'^{2}}{V^{2}}\right]\delta\phi^{(1)}+\Xi\right\}\,,
\end{equation}
\begin{eqnarray}
 \label{45-9}\frac{d}{dt}\Big(\delta\phi^{(2)}(\textbf{k},t)\Big)=\frac{1}{3H}\Bigg\{\left[-k^{2}-\frac{V''}{V}
+\frac{V'^{2}}{V^{2}}\right]\delta\phi^{(2)}\hspace{0.7cm}\nonumber\\
+\left[-\frac{V'''}{V}
+\frac{3V'V''}{V^{2}}-\frac{2V'^{3}}{V^{3}}\right]
\Big(\delta\phi^{(1)}\Big)^{2}\Bigg\}\,.\hspace{1.7cm}
\end{eqnarray}
Note that, in order to focus on the computation of those
non-Gaussian features which are directly produced by non-linearities
(self-interactions) of the scalar field itself, we have perturbed
the tachyon field's equation of motion in a particular gauge where
metric perturbations can be neglected compared with those of the
scalar field itself~\cite{Fal93}.

The solution of the equations~\eqref{45-8} and~\eqref{45-9} are
given by
\begin{eqnarray}
\delta\phi^{(1)}(\textbf{k},t)={\cal{F}}(k,t-t_{n-1})\int_{t_{n-1}}^{t}dt'\frac{\Xi}{3H}{\cal{F}}^{-1}(k,t-t_{n-1})
\hspace{0.1cm}\nonumber\\
+{\cal{F}}(k,t-t_{n-1})\delta\phi^{(1)}(\textbf{k}e^{-H(t_{n}-t_{n-1})},t_{n-1})\,,\hspace{2cm}
\label{45-10}
\end{eqnarray}
and
\begin{widetext}
\begin{eqnarray}
\delta\phi^{(2)}(\textbf{k},t)={\cal{F}}(k,t-t_{n-1})\int_{t_{n-1}}^{t}dt'{\cal{G}}(k,t')
\Bigg[\int\frac{d^{3}p}{(2\pi)^{3}}\delta\phi^{(1)}(\textbf{p},t')\delta\phi^{(1)}(\textbf{k}-\textbf{p},t')\Bigg]
{\cal{F}}^{-1}(k,t'-t_{n-1})\nonumber\\
+{\cal{F}}(k,t-t_{n-1})\delta\phi^{(2)}(\textbf{k}e^{-H(t_{n}-t_{n-1})},t_{n-1})\,,
\label{45-11}
\end{eqnarray}
\end{widetext}
where
\begin{equation}
\label{45-12}
{\cal{F}}=\exp\left[-\int_{t_{0}}^{t}\Bigg(\frac{k^{2}}{3H}+\frac{V''}{3HV}
-\frac{V'^{2}}{3HV^{2}}\Bigg)dt'\right]
\end{equation}
and
\begin{equation}
\label{45-13} {\cal{G}}=-\frac{2V'''}{3HV}
+\frac{2V'V''}{HV^{2}}-\frac{4V'^{3}}{3HV^{3}}\,.
\end{equation}
The second term on the right hand side of equations~\eqref{45-10}
and~\eqref{45-11} are memory terms which give the concept of
freeze-out. When $k\geq k_{F}$ this term damps away in a Hubble time
and when $k\leq k_{F}$ its effect is not negligible. In this regard,
the freeze-out momentum $k_{F}$ is expressed by the following
condition
\begin{equation}
\label{45-14} \frac{k_{F}^{2}}{3H^{2}}+\frac{V''}{3H^{2}V}
-\frac{V'^{2}}{3H^{2}V^{2}}=1
\end{equation}
So, the freeze-out momentum $k_{F}$ is given by
\begin{equation}
\label{45-15} k_{F}=\sqrt{\frac{3H^{2}V^{2}-V''V+V'^{2}}{V^{2}}}
\end{equation}
Now, we compute the three-point correlation function of the tachyon
fluctuations at large scale, at the time about 50 e-folds before the
end of inflation, by using equations~\eqref{45-10}
and~\eqref{45-11}. The result is as follows
\begin{eqnarray}
\langle\delta\phi(\textbf{k}_{1},t)\,\delta\phi(\textbf{k}_{2},t)\,\delta\phi(\textbf{k}_{3},t)\rangle=
{\cal{F}}(k_{3},t-t_{50}-\frac{1}{H})
\hspace{0.4cm}\nonumber\\
\int_{t_{50}-\frac{1}{H}}^{t_{50}}dt'
{\cal{F}}^{-1}(k_{3},t'-t_{50}-\frac{1}{H}){\cal{G}}(k_{3},t')\hspace{2cm}\nonumber\\
\Bigg[\int\frac{dp^{3}}{(2\pi)^{3}}\langle\delta\phi^{(1)}(\textbf{k}_{1},t_{1})\,
\delta\phi^{(1)}(\textbf{p},t')\rangle\hspace{2.5cm}\nonumber\\
\langle\delta\phi^{(1)}(\textbf{k}_{2},t_{2})\delta\phi^{(1)}(\textbf{k}_{3}-\textbf{p},t')\rangle\Bigg]
+{\cal{F}}(k_{3},t-t_{60}-\frac{1}{H})
\hspace{0.01cm}\nonumber\\
\langle\delta\phi^{(1)}(\textbf{k}_{1},t_{60})
\,\delta\phi^{(1)}(\textbf{k}_{2},t_{60})\,\delta\phi^{(1)}(\textbf{k}_{3}e^{-1},t_{60}-\frac{1}{H})\rangle
\hspace{0.01cm}\nonumber\\
+(\textbf{k}_{1}\leftrightarrow
\textbf{k}_{3})+(\textbf{k}_{2}\leftrightarrow
\textbf{k}_{3})\hspace{1cm} \label{45-16}
\end{eqnarray}
\begin{figure*}
\flushleft\leftskip0em{
\includegraphics[width=.31\textwidth,angle=0]{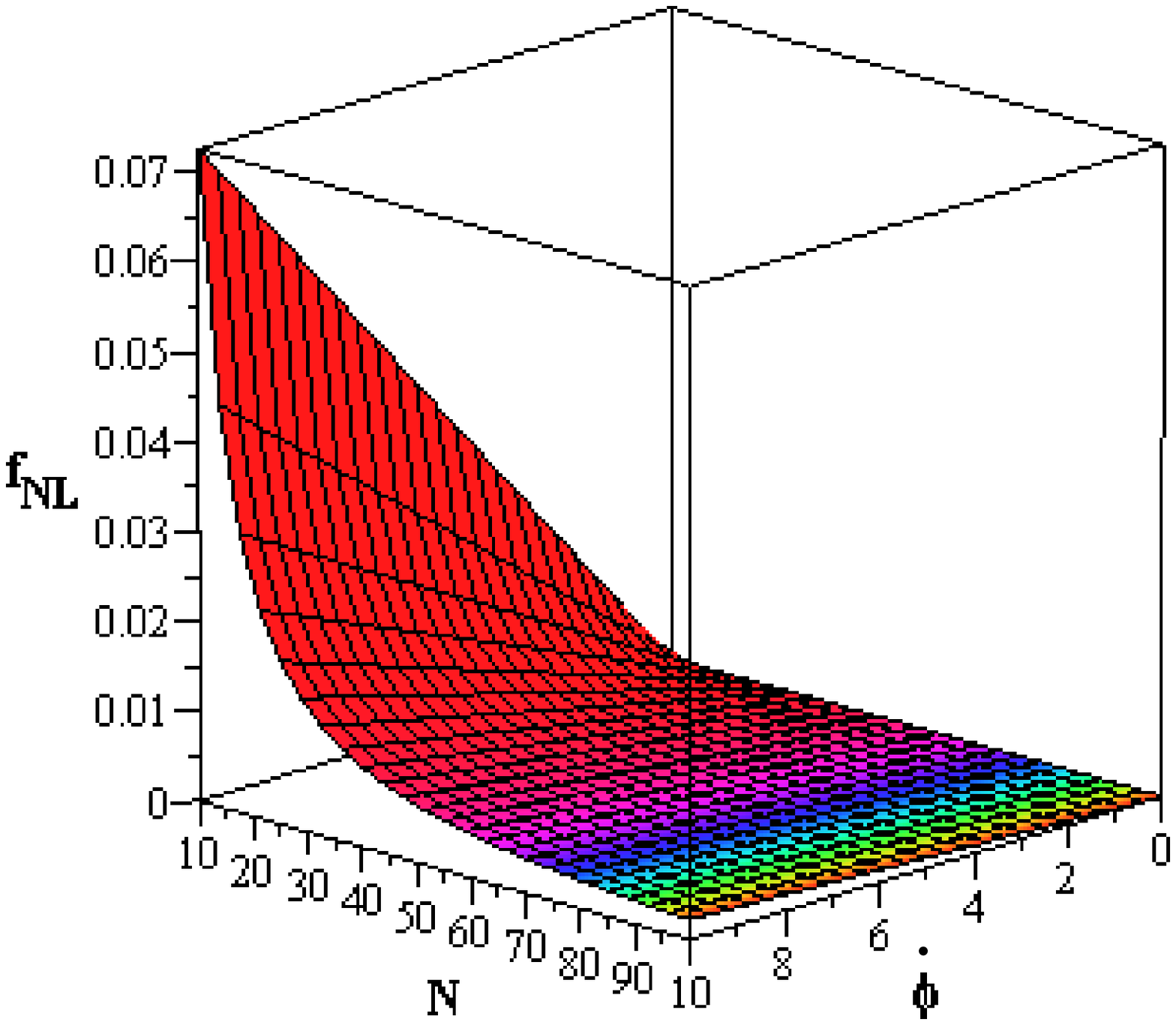}
\includegraphics[width=.36\textwidth,angle=0]{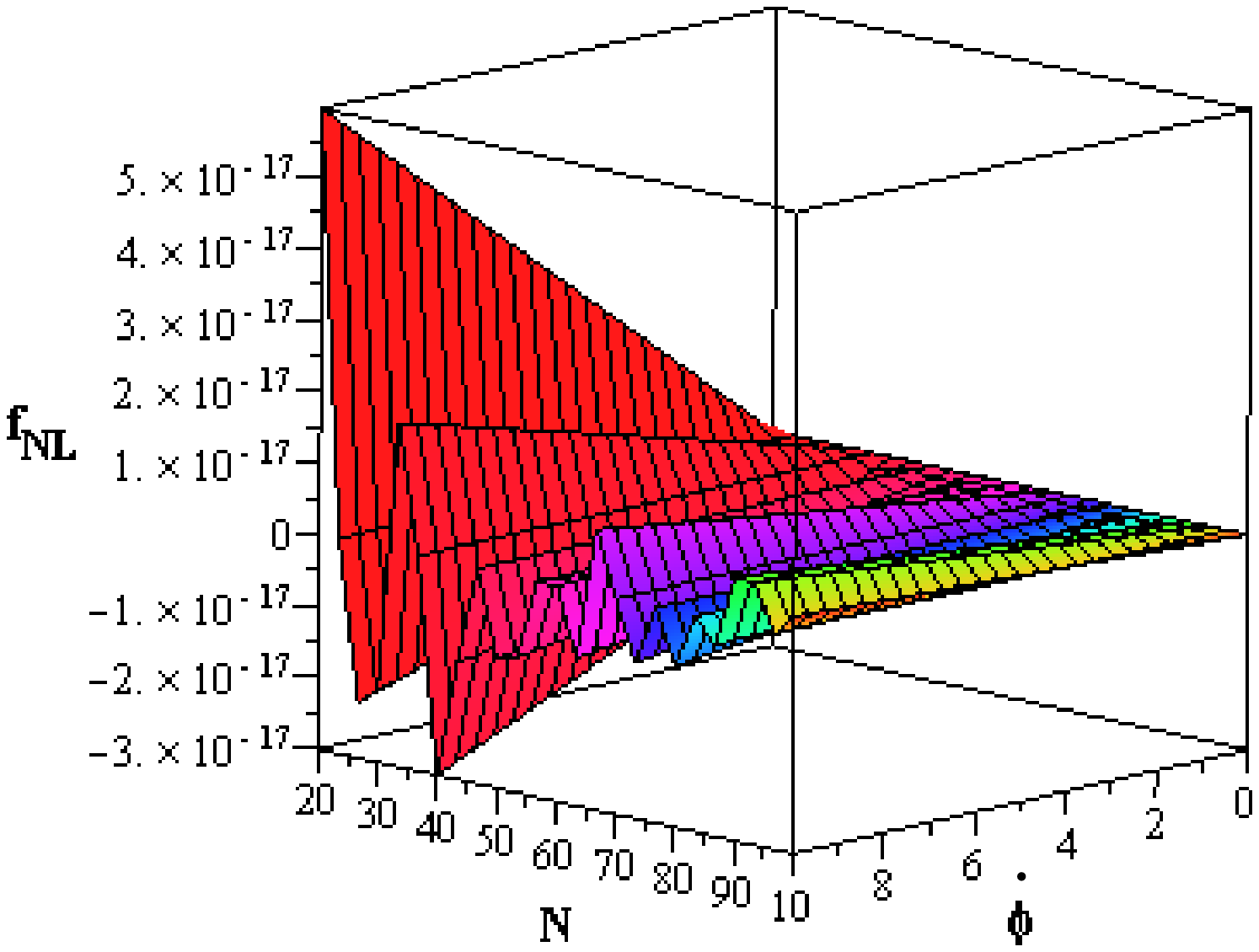}
\includegraphics[width=.31\textwidth,angle=0]{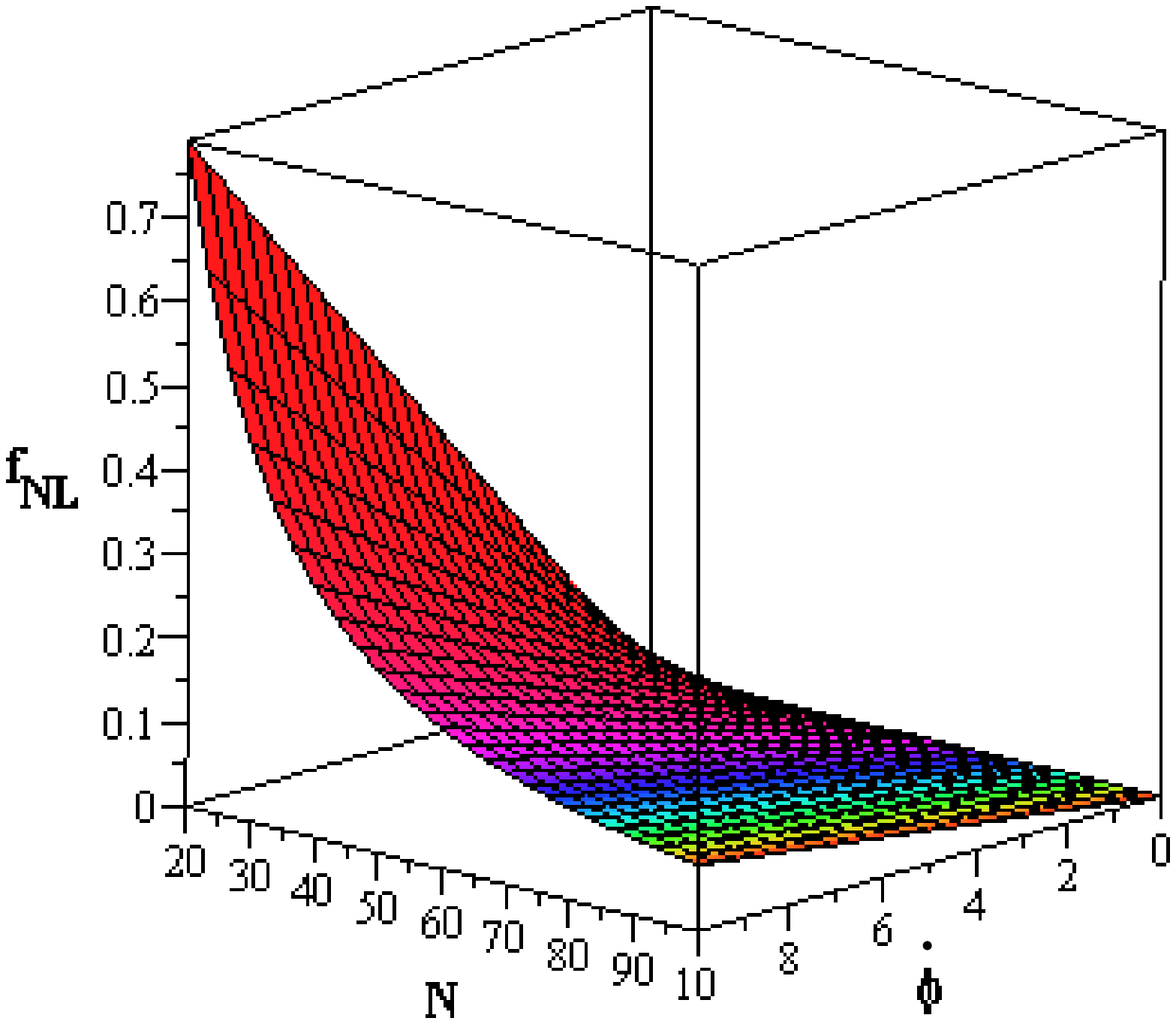}}
\caption{\label{fig2-3-1} Evolution of the non-gaussianity with
respect to $N$ and $\dot{\phi}$ for a quadratic potential (left
panel), an exponential potential (meddle panel) and an intermediate
potential (right panel) in minimal coupling setup.}
\end{figure*}
On the scale $k<k_{F}$, we can take ${\cal{F}}$ about unity and
${\cal{G}}$ as a constant and so we have
\begin{widetext}
\begin{eqnarray}
\langle\delta\phi(\textbf{k}_{1},t)\,\delta\phi(\textbf{k}_{2},t)\,\delta\phi(\textbf{k}_{3},t)\rangle\approx
{\cal{G}}(k_{F},t_{F})\frac{1}{H}\ln\Big(\frac{k_{F}}{H}\Big)\times
\hspace{7cm}\nonumber\\
\Bigg[\int\frac{d^{3}p}{(2\pi)^{3}}\langle\delta\phi^{(1)}(\textbf{k}_{1},t_{1})\,
\delta\phi^{(1)}(\textbf{p},t')\rangle
\langle\delta\phi^{(1)}(\textbf{k}_{2},t_{2})\delta\phi^{(1)}(\textbf{k}_{3}-\textbf{p},t')\rangle
+(\textbf{k}_{1}\leftrightarrow
\textbf{k}_{3})+(\textbf{k}_{2}\leftrightarrow \textbf{k}_{3})\Bigg]
\label{45-17}
\end{eqnarray}
\end{widetext}
The bispectrum for the single field inflation models in the
slow-roll limit is given by the following expression~\cite{Lyt09}
\begin{widetext}
\begin{equation}
\label{45-18}
\langle\Phi(\textbf{k}_{1})\,\Phi(\textbf{k}_{2})\,\Phi(\textbf{k}_{3})\rangle
=2f_{NL}(2\pi)^{3}
\delta^{3}(\textbf{k}_{1}+\textbf{k}_{2}+\textbf{k}_{3})\big[P_{\Phi}(\textbf{k}_{1})+P_{\Phi}(\textbf{k}_{2})+
permutations\big].
\end{equation}
\end{widetext}
On the other hand, the gravitational field potential is given
by~\cite{Rio02,Muk92}
\begin{equation}
\label{45-19}
\Phi(\textbf{k})=-\frac{3}{2}\frac{H}{\dot{\phi}}\delta\phi(\textbf{k})\,.
\end{equation}
So, from equations~\eqref{45-17}, ~\eqref{45-18} and ~\eqref{45-19}
we obtain
\begin{equation}
\label{45-20}
f_{NL}=-\frac{2}{3}\frac{\dot{\phi}}{H}\Bigg[\frac{1}{H}\ln\bigg(\frac{k_{F}}{H}\bigg)\bigg(\frac{2V'''}{3HV}
-\frac{2V'V''}{HV^{2}}+\frac{4V'^{3}}{3HV^{3}}\bigg)\Bigg]
\end{equation}

In figure~\ref{fig2-3-1} we have plotted the behavior of $f_{NL}$
versus $\dot{\phi}$ and $N$ for three types of potentials:
quadratic, exponential and intermediate potential. As figure shows,
both for quadratic and intermediate potentials, $f_{NL}$ is positive
and decreases as the number of e-folds parameter increases. Note
that for the intermediate potential we have set $l=0.4$. With an
exponential potential the non-gaussianity can be either positive or
negative, depending on the values of $N$ and $\dot{\phi}$. Also, for
this type of potential the value of non-gaussianity is so small.

\subsection{Cosmological dynamics and late time acceleration}

In this section, we study cosmological dynamics of the minimally
coupled tachyon field model by using the dynamical system analysis
and phase space trajectories of the model. The Friedmann equation in
the presence of the tachyon field and ordinary components (matter
and radiation) is given by
\begin{equation}
\label{46}
H^{2}=\frac{\kappa^{2}}{3}\bigg[\frac{V(\phi)}{\sqrt{1-\dot{\phi}^{2}}}+\rho\bigg]\,,
\end{equation}
where $\rho$ is the energy density of the ordinary components. In
order to analyze the phase space of the model, we firstly introduce
some new convenient and dimensionless variables which help us to
translate the equations of the cosmological dynamics in the language
of the autonomous dynamical system. In this regard, we define the
following dimensionless parameters in our setup
\begin{equation}
\label{47} x=\dot{\phi}\,\,,\quad\quad
y=\frac{\kappa\sqrt{V}}{\sqrt{3}H}\,\,,\quad\quad
d=\frac{\kappa\sqrt{\rho}}{\sqrt{3}H}\,,\quad\quad
\alpha=\frac{V'}{\kappa_{4}V^{\frac{3}{2}}},
\end{equation}
We can rewrite the Friedmann equation \eqref{46} in terms of these
dimensionless parameters and obtain the following constraint on the
model's parameter space
\begin{equation}
\label{48} 1=\frac{y^{2}}{\sqrt{1-x^{2}}}+d^{2}\,.
\end{equation}
We note that, the constraint equation \eqref{48} allows us to
express one of the dimensionless variables in terms of the others.

Now, we rewrite other important equations of the setup in terms of
the new variables. The acceleration equation in terms of new
variables takes the following form
\begin{equation}
\label{49}
\frac{\dot{H}}{H^{2}}=-\frac{3}{2}\frac{y^{2}x^{2}}{\sqrt{1-x^{2}}}-\frac{3}{2}d^{2}\big(1+\omega\big)\,.
\end{equation}
Also, the scalar field's equation of motion \eqref{8} is given by
the following expression
\begin{equation}
\label{50}
\frac{\ddot{\phi}}{H}=-3x\big(1-x^{2}\big)-\sqrt{3}\,y\alpha\big(1-x^{2}\big)\,.
\end{equation}
We can reformulate the effective equation of state parameter in
terms of the dimensionless parameters as
\begin{equation}
\label{51}
\omega_{eff}=-y^{2}\,\sqrt{1-x^{2}}+\omega\Bigg(1-\frac{y^{2}}{\sqrt{1-x^{2}}}\Bigg)\,.
\end{equation}
In the next step, we introduce a new time variable $\tau\equiv \ln
a$, and obtain the following autonomous system of equations in our
setup
\begin{equation}
\label{52}
\frac{dx}{d\tau}=-3x\big(1-x^{2}\big)-\sqrt{3}\,y\alpha\big(1-x^2\big)\,,
\end{equation}
and
\begin{equation}
\label{53} \frac{dy}{d\tau}=\frac{\sqrt{3}}{2}y^{2}\alpha
x-y\Bigg(1-\frac{3y^{2}x^{2}}{\sqrt{1-x^{2}}}\Bigg)\,.
\end{equation}
We choose the inverse power law potential
$V(\phi)=\frac{1}{\kappa_{4}^{2}\phi^{2}}$ (see~\cite{Cop06} for
reliability of this choice), which leads to a constant $\alpha$ and
so $\frac{d\alpha}{d\tau}=0$.

\begin{table*}
\caption{\label{tab3} Location, Eigenvalues and Dynamical Characters
of the Critical Points.}
\begin{ruledtabular}
\begin{tabular}{cccccc}
Point & ($x$,$y$) &eigenvalues & $\omega_{eff}$&stability\\
\hline $N$ & (0.84480,\,0.73159) &$-0.8594-3\omega$,\,$-1.9296$& $-0.2863887966$ &stable \\
& && $-0.000264873\omega$&\\
\hline $O$ & (0,0) &$-3$,\,$\frac{3}{2}+\frac{3}{2}\omega$& $\omega$& unstable\\
\hline $M$ & (-0.84480,\,-0.73159)
&$-0.8594-3\omega$,\,$-1.9296$& $-0.2863887966$& stable\\
& && $-0.000264873\omega$&\\
\hline $P$ & (1,0) &$6$\,,\,$\frac{3}{2}+\frac{3}{2}\omega$& $\omega$& unstable\\
\hline $Q$ & (-1,0) &$6$\,,\,$\frac{3}{2}+\frac{3}{2}\omega$& $\omega$& unstable\\
\end{tabular}
\end{ruledtabular}
\end{table*}

\begin{figure}
\flushleft\leftskip0em{
\includegraphics[width=.40\textwidth,origin=c,angle=0]{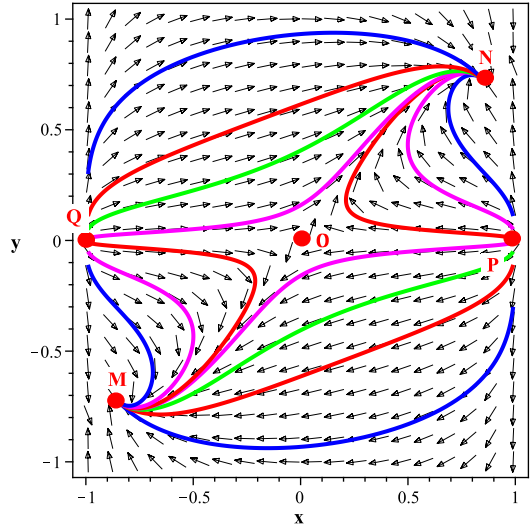}}
\caption{\label{fig7} The phase space trajectories of the minimally
coupled tachyon field model. The critical point $O$ is an
effectively ordinary components (matter or radiation) dominated
solution and is an unstable saddle point. Points $P$ and $Q$ are
unstable too. The points $N$ and $M$ are effectively tachyon field
dominated solutions and are stable critical points. There is no
accelerated phase of expansion.}
\end{figure}

Now, we analyze the cosmological evolution of this setup in the
dynamical system approach. To this end, we should find the fixed (or
critical) points in the phase space of the model. Fixed points are
defined as the points where the autonomous equations \eqref{52} and
\eqref{53} are vanishing. In the minimally coupled tachyon field
setup, there is 5 critical points; $M$, $N$, $O$, $P$ and $Q$. The
properties of these critical points are summarized in
table~\ref{tab3}. The points $P$ and $Q$ are effectively ordinary
components dominated solutions. Their eigenvalues are positive, so
these critical points are repeller nodes. These solutions can be
corresponding to the early time radiation or matter domination era
in the history of the universe. The point $O$ is effectively
ordinary components dominated. This point is an unstable saddle
point, since one of its eigenvalues is positive and the other one is
negative. If the universe during its evolution reaches this point,
it doesn't remain there and evolves to another state. Note that,
although $P$, $Q$ and $O$, all are effectively ordinary components
dominated phases, there is a difference between them. At the point
$O$, other components of the universe (the kinetic and potential
terms of the tachyon field) have no contribution. While, at the
points $P$ and $Q$ the kinetic term of the tachyon field has
significant contribution. The points $M$ and $N$ are stable critical
points, meaning that if the universe reaches these states, remains
there forever. However, in this case the universe expansion is not
accelerating (as the effective equation of state parameter shows).
The trajectories of the phase space of the minimally coupled tachyon
field model, for $\omega=0$, is shown in figure~\ref{fig7}. Note
that, the location of the fixed points are independent of $\omega$.
For both $\omega=0$ and $\omega=\frac{1}{3}$, the trajectories of
the phase space are the same but the eigenvalues and effective
equation of state parameters are different.
\begin{figure*}
\flushleft\leftskip0em{
\includegraphics[width=.35\textwidth,origin=c,angle=0]{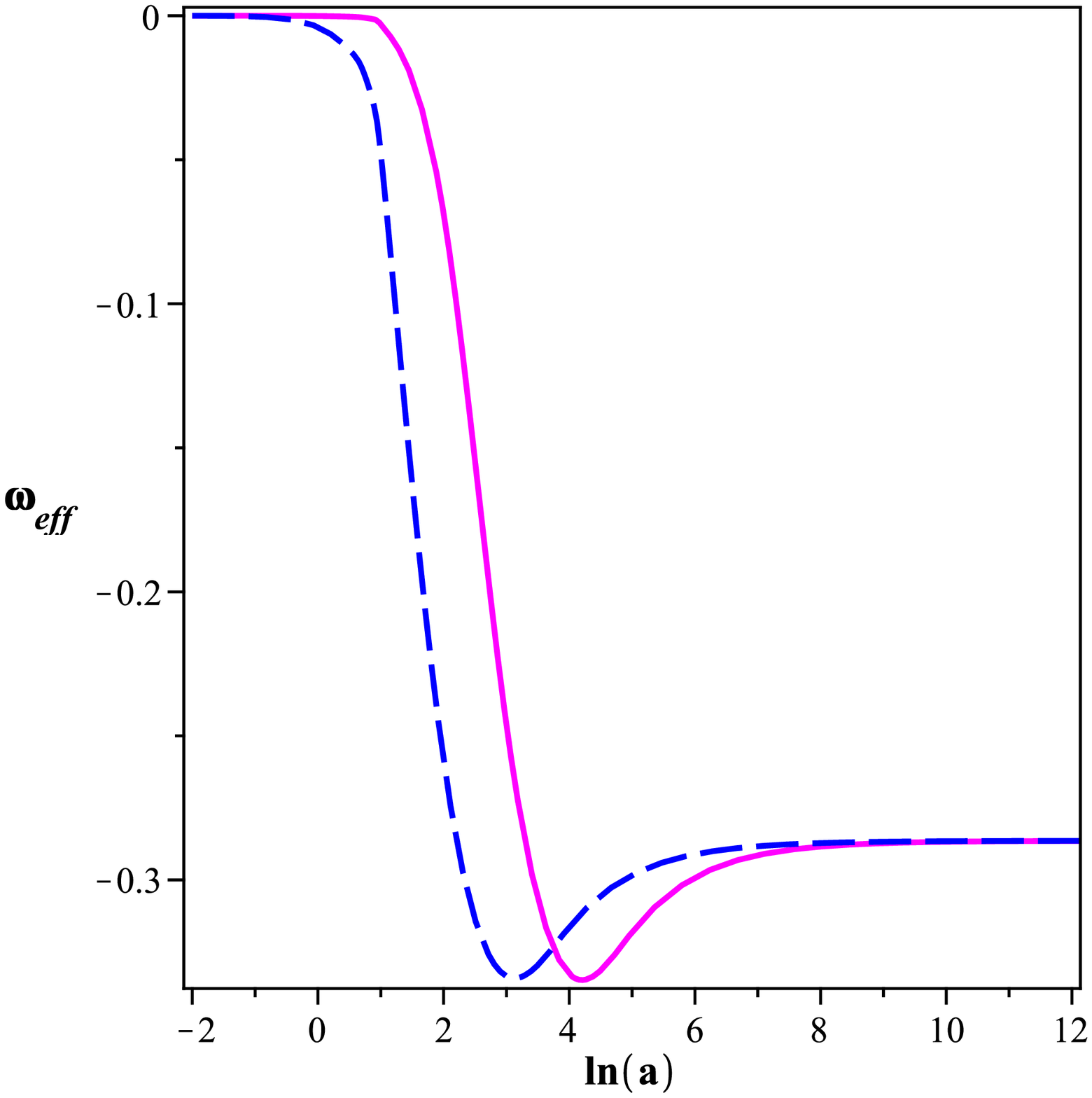}
\hspace{3.6cm}
\includegraphics[width=.35\textwidth,origin=c,angle=0]{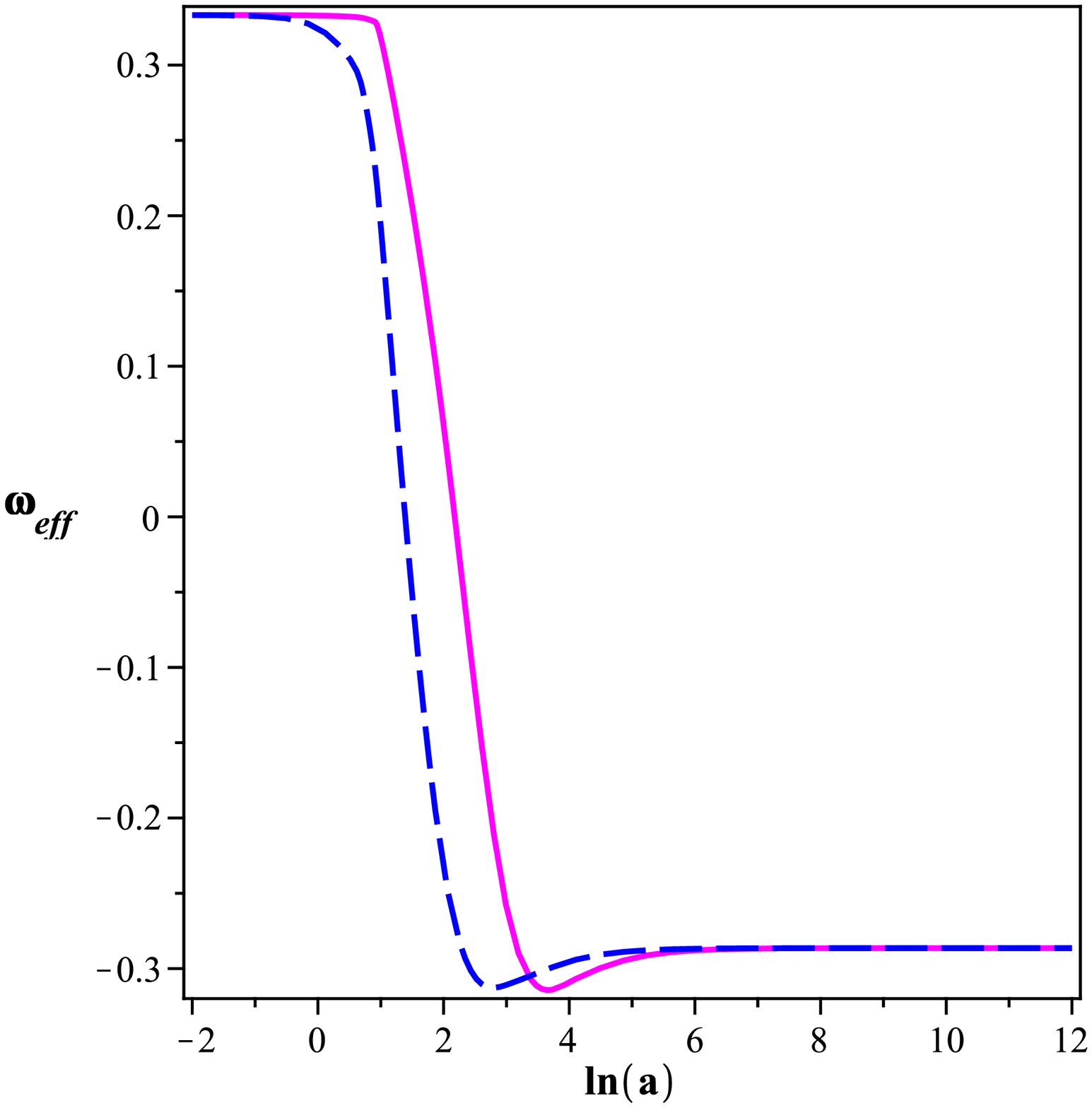}}
\caption{\label{fig8} Behavior of the effective equation of state
parameter versus the cosmic time. The left panel is corresponding to
$\omega=0$ (dust) and the right panel is corresponding to
$\omega=\frac{1}{3}$ (radiation). $\omega_{eff}$ evolves and tends
to the value $-0.286$. These figures have been plotted with
$x(0)=0.99$ and $y(0)=0.01$ (the magenta curve) and $x(0)=0.75$ and
$y(0)=0.05$ (the blue curve).}
\end{figure*}
\begin{figure*}
\flushleft\leftskip0em{
\includegraphics[width=.35\textwidth,origin=c,angle=0]{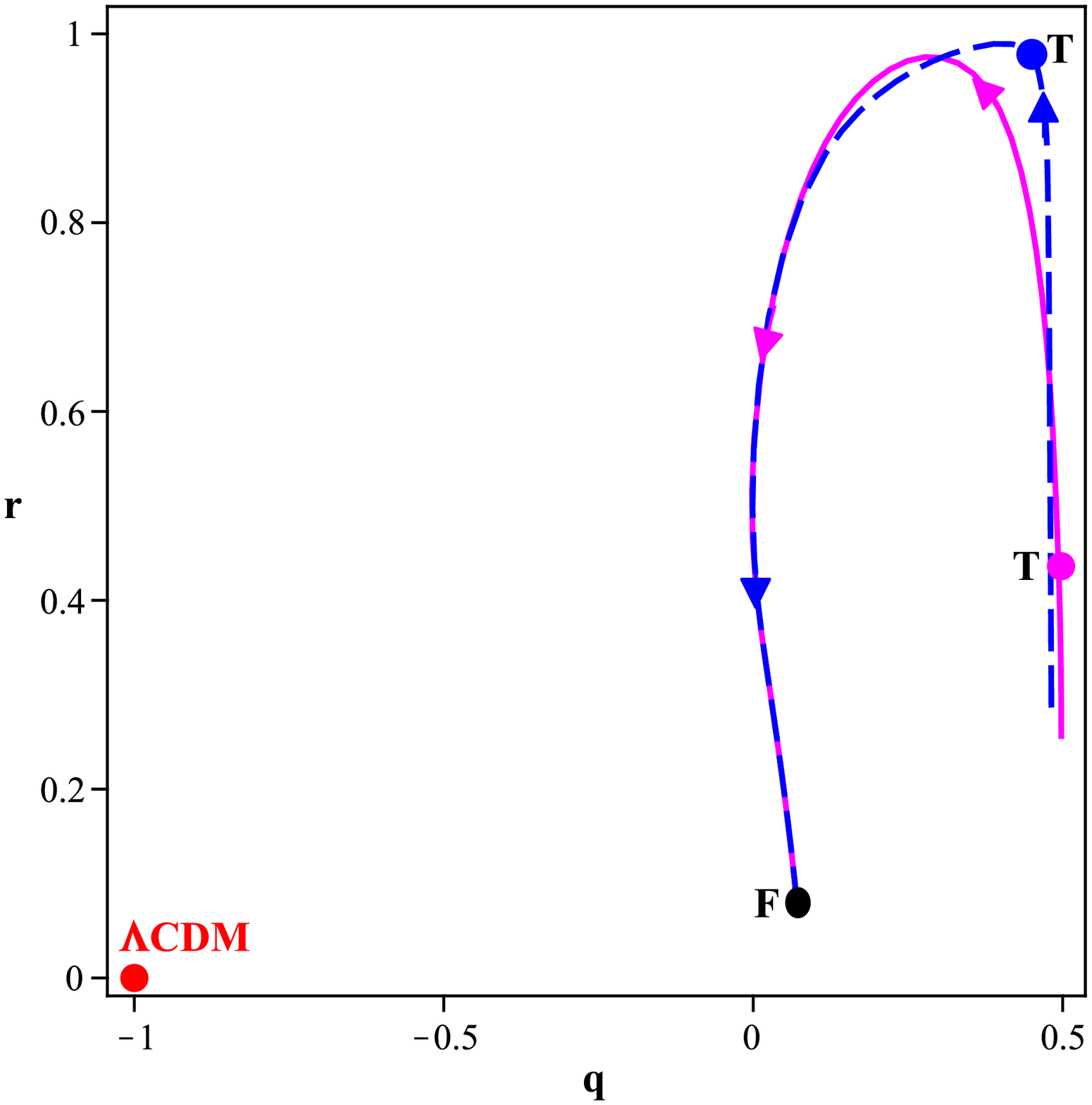}
\hspace{3.6cm}
\includegraphics[width=.35\textwidth,origin=c,angle=0]{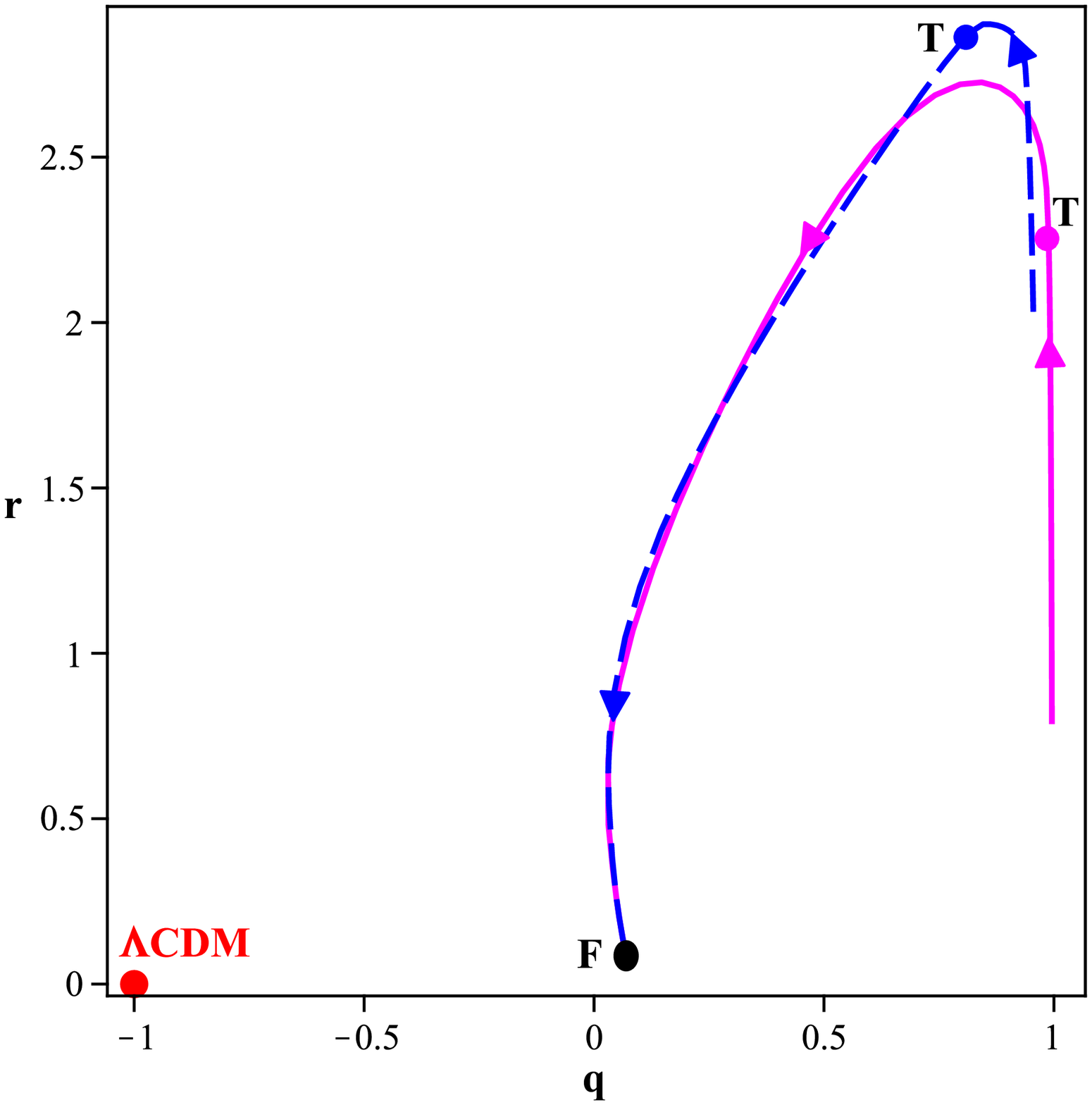}}
\caption{\label{fig9} Trajectories in $\{r,\,q\}$ phase plan with
$\omega=0$ (left panel, corresponding to the dust) and
$\omega=\frac{1}{3}$ (right panel, corresponding to the radiation).
The initial conditions are $x(0)=0.99$ and $y(0)=0.01$ for magenta
trajectory and $x(0)=0.75$ and $y(0)=0.05$ for blue trajectory. The
magenta and blue highlighted dots (specified by $T$) are current
values of $\{r,\,q\}$ in the model. The black dot (specified by $F$)
is the stable state of $\{r,\,q\}$ in the future. Also the red dot
(remarked by $\Lambda$CDM) is the value of statefinder $\{r,\,q\}$
in a $\Lambda$CDM scenario.}
\end{figure*}
\begin{figure*}
\flushleft\leftskip0em{
\includegraphics[width=.35\textwidth,origin=c,angle=0]{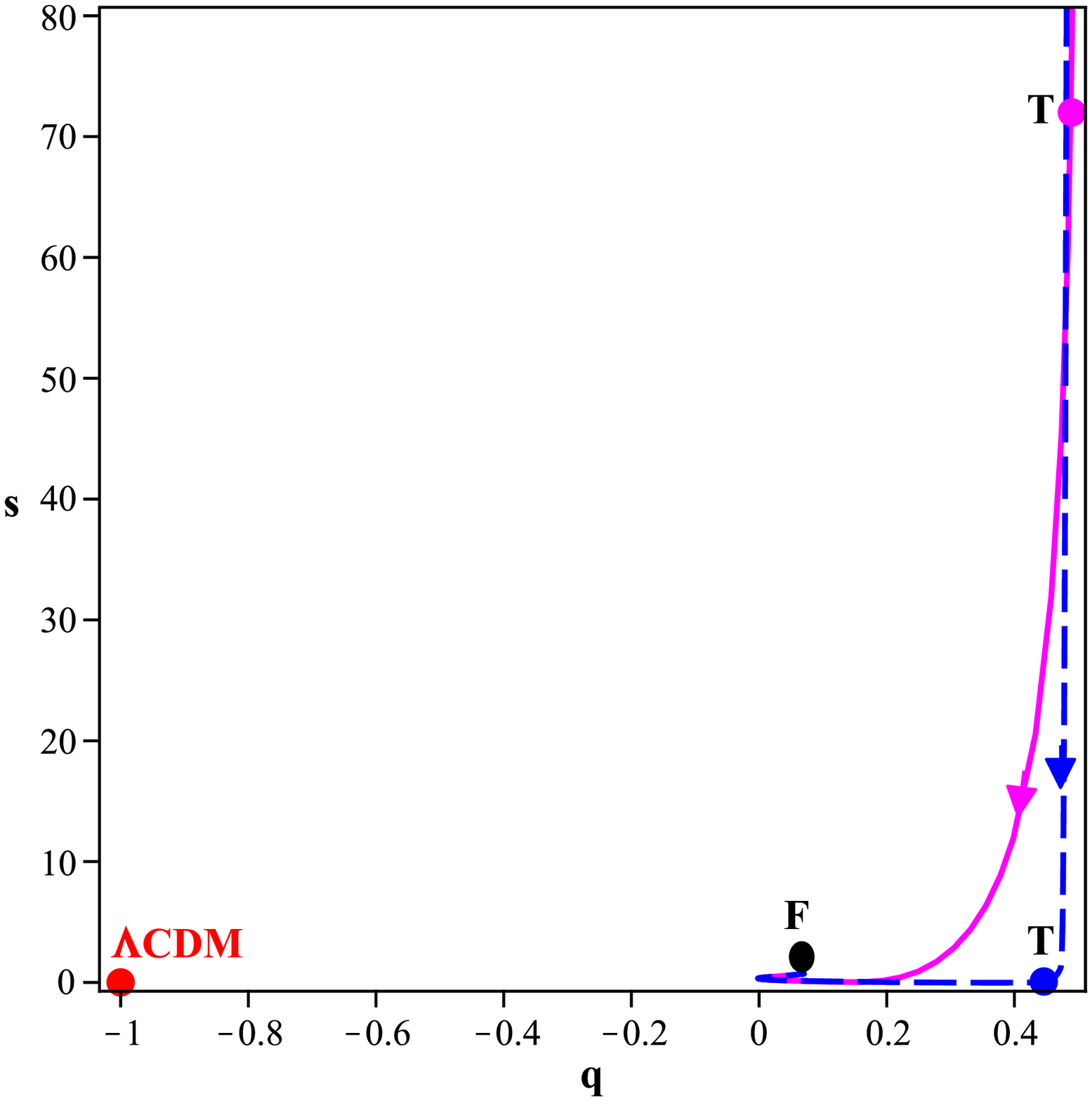}
\hspace{3.6cm}
\includegraphics[width=.35\textwidth,origin=c,angle=0]{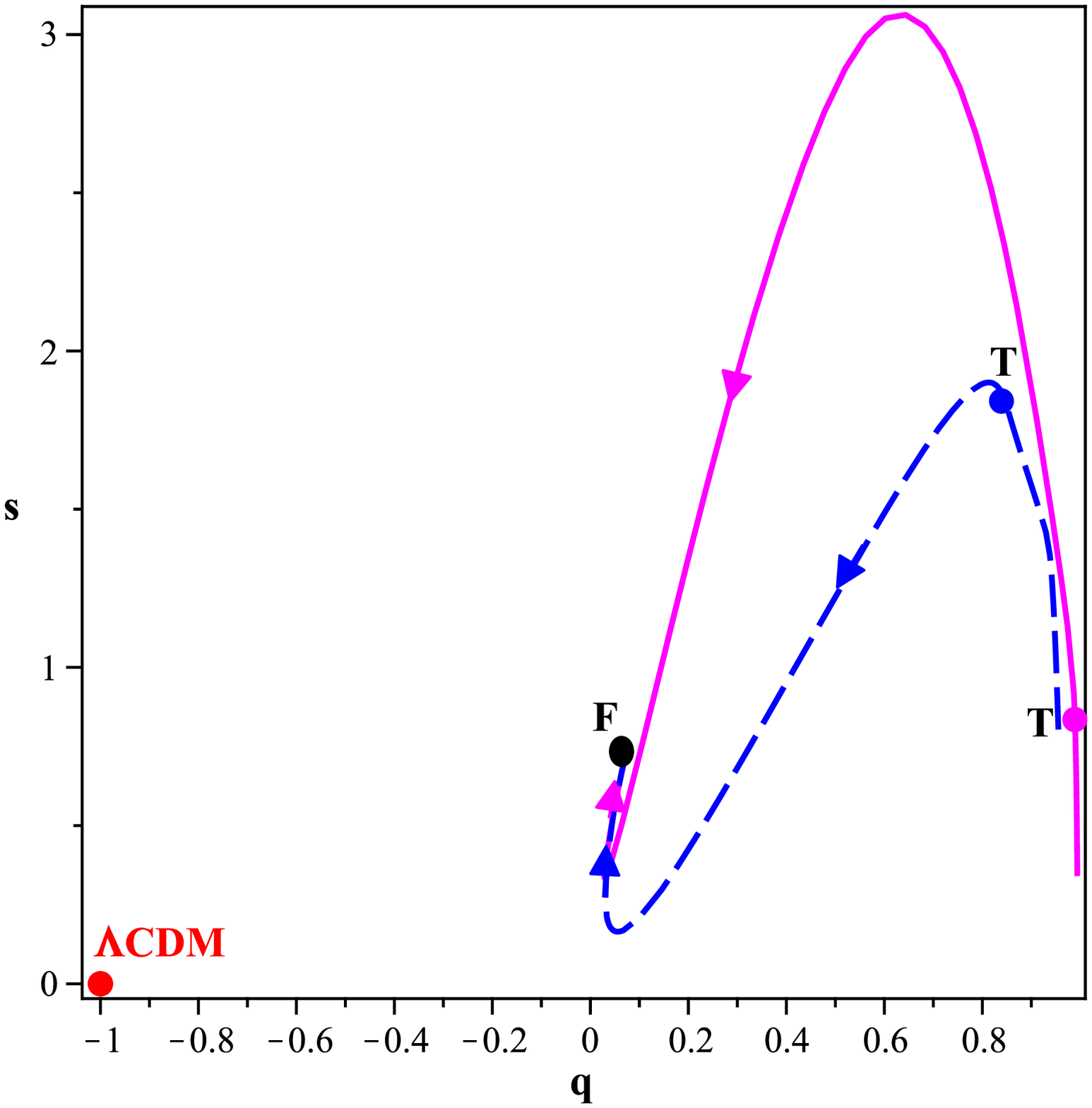}}
\caption{\label{fig10} Trajectories in $\{s,\,q\}$ phase plan with
$\omega=0$ (left panel) and $\omega=\frac{1}{3}$ (right panel). The
initial conditions are $x(0)=0.99$ and $y(0)=0.01$ for magenta
trajectory and $x(0)=0.75$ and $y(0)=0.05$ for blue trajectory. The
magenta and blue highlighted dots (specified by $T$) are current
values of $\{s,\,q\}$ in the model. The black dot (specified by $F$)
is the stable state of $\{s,\,q\}$ in the future. Also the red dot
(remarked by $\Lambda$CDM) is the value of statefinder $\{s,\,q\}$
in a $\Lambda$CDM scenario.}
\end{figure*}
\begin{figure*}
\flushleft\leftskip0em{
\includegraphics[width=.35\textwidth,origin=c,angle=0]{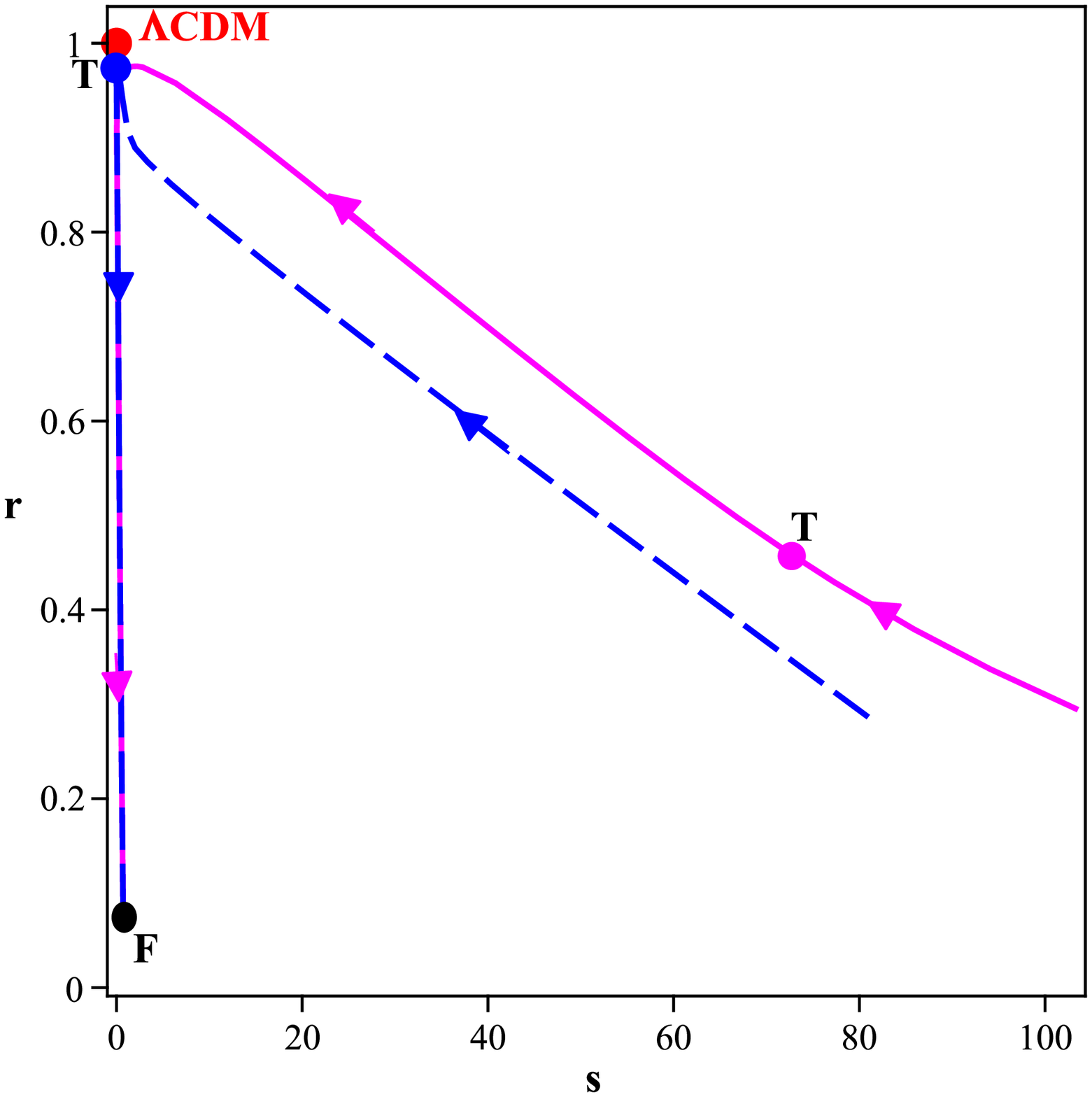}
\hspace{3.6cm}
\includegraphics[width=.35\textwidth,origin=c,angle=0]{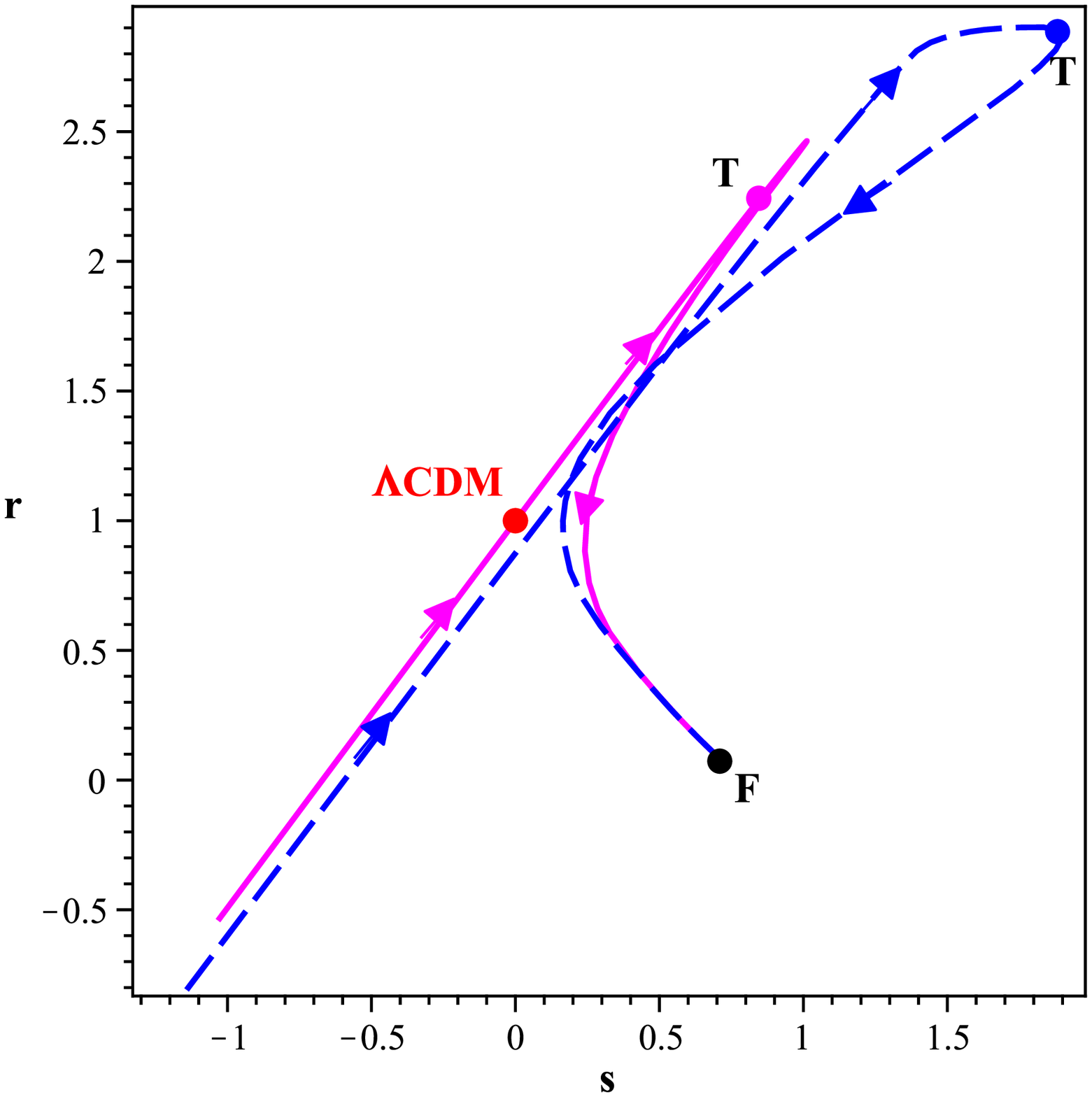}}
\caption{\label{fig11} The trajectories in $\{r,\,s\}$ phase plan
with $\omega=0$ (the left panel, corresponding to dust) and
$\omega=\frac{1}{3}$ (the right panel, corresponding to radiation).
The initial conditions are $x(0)=0.99$ and $y(0)=0.01$ for magenta
trajectory and $x(0)=0.75$ and $y(0)=0.05$ for blue trajectory. The
magenta and blue bold dots (specified by $T$) are current values of
$\{r,\,s\}$ in the model. The black dot (specified by $F$) is the
stable state of $\{r,\,s\}$ in the future. Also the red dot
(remarked by $\Lambda$CDM) is the value of statefinder $\{r,\,s\}$
in a $\Lambda$CDM scenario.}
\end{figure*}

Figure~\ref{fig8} shows the evolution of the effective equation of
state parameter with respect to the cosmic time. As we see form this
figure, the effective equation of state parameter of the model tends
to the value of about $-0.286$. So there is no possibility to
realize the accelerating phase of the universe expansion in a model
with a minimally coupled tachyon field.

\subsection{Statefinder Diagnostic}

In 2003 Sahni et al. have proposed a diagnostic proposal by
introducing a new pair of parameters $\{r,s\}$, called statefinder
parameters, in order to distinguish between different types of dark
energy models ~\cite{Sah03}. These parameters are defined as
\begin{equation}
\label{54}
r=\frac{\dot{}\dot{a}\dot{}}{aH^{3}}=\frac{\ddot{H}}{H^{3}}-3q-2\,,
\end{equation}
\begin{equation}
\label{55} s=\frac{r-1}{3(q-\frac{1}{2})}\,.
\end{equation}
Since statefinder parameters depend on the scale factor and
therefore, on the metric describing the space-time, it is a
``geometrical'' diagnostic. By using the statefinder parameters, we
can study the expansion history of the universe through higher
derivatives of the scale factor (that is,  $\dot{}\dot{a}\dot{}$).
The $\Lambda$CDM scenario corresponds to a fixed point in the
$r$-$s$ diagram with $\{r,\,s\}_{\Lambda CDM}=\{1,\,0\}$. By
plotting the trajectories in the $r$-$s$ phase diagram, the distance
of the model from $\Lambda$CDM can be probed \cite{Ala03}. We can
rewrite the equation \eqref{54} as
\begin{equation}
\label{56}
r=\frac{d}{d\tau}\left(\frac{\dot{H}}{H^{2}}\right)+2\,\left(\frac{\dot{H}}{H^{2}}\right)^{2}
+3\,\left(\frac{\dot{H}}{H^{2}}\right) +1\,,
\end{equation}
which in our setup with a minimally coupled tachyon field takes the
following form
\begin{eqnarray}
r=\frac{-3yy'x^{2}-3y^{2}xx'}{\sqrt{1-x^{2}}}+\frac{3}{2}\frac{y^{2}x^{3}x'}{\big(1-x^{2}\big)^{\frac{3}{2}}}
-3\,d\,
d'(1+\omega)\hspace{0.1cm}\nonumber\\
+\frac{9}{2}\left(\frac{y^{2}x^{2}}{\sqrt{1-x^{2}}}+d^{2}(1+\omega)\right)^{2}
\hspace{3cm}\nonumber\\
-\frac{9}{2}\left(\frac{y^{2}x^{2}}{\sqrt{1-x^{2}}}+d^{2}(1+\omega_{m})\right)+1\,.\hspace{1cm}
\label{57}
\end{eqnarray}
The other parameter, $s$, can be obtained by substituting equation
\eqref{57} in equation \eqref{55} that we avoid to rewrite it here.

Now, we proceed to numerical analysis of the statefinder diagnostic.
Figure~\ref{fig9} shows the trajectories of $\{r,\,q\}$ phase plane
for two initial values of the parameters. The left panel is
$\{r,\,q\}$ diagram with $\omega=0$ (corresponding to the dust) and
the right panel is trajectories of $\{r,\,q\}$ phase plane with
$\omega=\frac{1}{3}$ (corresponding to the radiation). For both
panels, independent of the initial conditions, the trajectories
evolve to a fixed point in the future. The magenta and blue
highlighted dots show the current values of $r$ and $q$. $\{s,\,q\}$
phase plan for both $\omega=0$ (left panel) and $\omega=\frac{1}{3}$
(right panel) is shown in figure~\ref{fig10}. Here also, independent
of the initial conditions, the trajectories tend to a fixed point in
the future. Finally, we can see the $\{r,\,s\}$ diagram in
figure~\ref{fig11}. As the evolution of $q$ shows, this model can
not explain the late time speed-up expansion of the universe. There
is no stable point which could explain the late time acceleration.

After a detailed study of a minimally coupled tachyon field and its
cosmological dynamics, now we extend our analysis to the more
general case of a non-minimally coupled tachyon field.

\section{Non-minimally coupled tachyon field }
\subsection{Inflation}

The 4-dimensional action for a non-minimally coupled tachyon field
is given by the following expression
\begin{equation}
\label{3:1}
S=\int\sqrt{-g}\Bigg[\frac{1}{\kappa^{2}}R-\frac{1}{2}f(\phi)R-V(\phi)\sqrt{1-\partial^{\mu}\phi\partial_{\mu}\phi}\Bigg]d^{4}x\,,
\end{equation}
where $f(\phi)$ shows an explicit non-minimal coupling of the scalar
field with the Ricci scalar. Einstein's field equations calculated
from action \eqref{3:1} are given by
\begin{equation}
\label{3:2}
\Big(1-\kappa^{2}f(\phi)\Big)G_{\mu\nu}=\kappa^{2}T_{\mu\nu}\,,
\end{equation}
where $T_{\mu\nu}$ is the energy momentum tensor of a non-minimally
coupled tachyon field, given by
\begin{eqnarray}
\label{3:3}
T_{\mu\nu}=-\frac{V{(\phi)}}{\sqrt{1-\partial_{\alpha}\phi\partial^{\alpha}\phi}}\,\partial_{\mu}\phi\partial_{\nu}\phi
+\Big(g_{\mu\nu}\Box-\nabla_{\mu}\nabla_{\nu}\Big)f(\phi)\hspace{0cm}\nonumber\\
+g_{\mu\nu}\Big(-V(\phi)\sqrt{1-\partial_{\alpha}\phi\partial^{\alpha}\phi}\Big)
\,.\hspace{1cm}
\end{eqnarray}
The energy-momentum tensor \eqref{3:3} leads to the following energy
density and pressure
\begin{equation}
\label{3:4}
\rho_{\phi}=\frac{V(\phi)}{\sqrt{1-\dot{\phi}^{2}}}+3f'H\dot{\phi}\,,
\end{equation}
and
\begin{equation}
\label{3:5}
p_{\phi}=-V(\phi)\sqrt{1-\dot{\phi}^{2}}-f'\ddot{\phi}-2f'H\dot{\phi}-2f''\dot{\phi}^{2}\,.
\end{equation}

By using the line element \eqref{6}, we find the Friedmann equation
of this model as follows
\begin{equation}
\label{3:6}
H^{2}=\frac{\kappa^{2}}{3-3\kappa^{2}f}\Bigg[\frac{V(\phi)}{\sqrt{1-\dot{\phi}^{2}}}+3f'H\dot{\phi}\Bigg]\,.
\end{equation}

By varying the action \eqref{3:1} with respect to the scalar field,
the equation of motion of the non-minimally coupled tachyon field is
obtained as
\begin{equation}
\label{3:7}
\frac{\ddot{\phi}}{1-\dot{\phi}^{2}}+3H\dot{\phi}+\frac{V'}{V}+\frac{\sqrt{1-\dot{\phi}^{2}}}{2V}f'R=0\,,
\end{equation}
where, as before, a prime denotes derivative with respect to the
tachyon field and a dot refers to derivative with respect to the
time.

In the presence of a non-minimally coupled tachyon field, the energy
conservation equation takes the following form
\begin{equation}
\label{3:8}
\dot{\rho}_{\phi}+3H(\rho_{\phi}+p_{\phi})=-3f'H^{2}\dot{\phi}\,.
\end{equation}
In contrast to equation \eqref{9}, the right hand side of equation
\eqref{3:8} is not zero due to the presence of the non-minimal
coupling between the tachyon field and the Ricci scalar.

Now, we apply the slow-roll approximation to the main equations of
the setup. The energy density and the equation of motion of the
tachyon field, in the slow-roll limit, are given respectively as
\begin{equation}
\label{3:9} \rho_{\phi}=V(\phi)+3f'H\dot{\phi}\,,
\end{equation}
and
\begin{equation}
\label{3:10} 3H\dot{\phi}+\frac{V'}{V}+\frac{f'R}{2V}=0\,.
\end{equation}
Also, the Friedmann equation takes the following form
\begin{equation}
\label{3:11}
H^{2}=\frac{\kappa^{2}}{1-\kappa^{2}f}\Bigg[-\frac{f'^{2}R}{V}+\frac{V}{3}-\frac{f'V'}{3V}\Bigg]\,.
\end{equation}

In the presence of the non-minimally coupled tachyon field, the
slow-roll parameters (defined by equations \eqref{13} and
\eqref{14}) are given by the following equations
\begin{widetext}
\begin{eqnarray}
\epsilon=\frac{1}{2\kappa^{2}}\frac{V'^{2}}{V^{2}}\frac{1-\kappa^{2}f}{\Big[-\frac{f'^{2}R}{2V}+V-\frac{f'V'}{V}\Big]^{2}}
\left[\frac{f'R}{2V'^{2}}+\frac{1}{V'}\right]\times
\hspace{8cm}\nonumber\\
\bigg[-f'f''R+\frac{f'^{2}RV'}{2V}+V'V-f''V-f'V''
+\frac{f'^{2}V'^{2}}{V}+
\frac{\kappa^{2}f'}{1-\kappa^{2}f}[V^{2}-\frac{f'^{2}R}{2}-f'V']\bigg]\,,
\label{3:12}
\end{eqnarray}
\end{widetext}

\begin{widetext}
\begin{eqnarray}
\label{3:13}
\eta=\frac{\kappa^{2}f'}{3H^{2}}\Big(\frac{f'R}{2V}+\frac{V'}{V}\Big)
-\frac{1-\kappa^{2}f}{3X}\left[3X-V'-3f'H^{2}\right]
-\frac{\dot{B}}{\frac{\kappa^{2}X}{3H(1-\kappa^{2}f)}\big(\frac{f'R}{2V}-\frac{V'}{V}\big)}\,,
\end{eqnarray}
\end{widetext}
where $B$ and $X$ are defined as
\begin{equation}
\label{3:14} B=\frac{\kappa^{2}}{H}\left[X-f'H^{2}\right]\,,
\end{equation}
and
\begin{eqnarray}
\label{3:15} X=-\frac{f'f''R}{3V}+\frac{f'^{2}V'R}{6V^{2}}+\frac{V'}{3}\hspace{3.5cm}\nonumber\\
-\frac{f''V'}{3V}
+\frac{f'^{2}V'^{2}}{3V^{2}}-\frac{f'V''}{3V}+f'H^{2}\,.\hspace{1cm}
\end{eqnarray}
The number of e-folds, which is defined by equation \eqref{17}, in
the slow-roll approximation is given by
\begin{equation}
\label{3:16}
N\simeq\int_{\phi_{hc}}^{\phi_{f}}\Bigg(\frac{3V'}{V}\Bigg)\Bigg(\frac{\kappa^{2}\Big[\frac{2VV'}{3}-
\frac{f'^{2}V'R}{3V}-\frac{f'V'^{2}}{3V}\Big]}{\big(1-\kappa^{2}f\big)\big(-f'R-2V'\big)}\Bigg)d\phi\,.
\end{equation}

As the minimal case, in which follows we explore the scalar
perturbation of the metric in order to study the spectrum of
perturbations.

\subsection{Perturbations}

Similar to the minimal case, we use the perturbed metric \eqref{20}
which is written in the longitudinal gauge. With this perturbed
metric, we obtain the following perturbed Einstein's field equations
\begin{equation}
\label{3:17}
-3H(H\Phi+\dot{\Psi})-\frac{k^{2}}{a^{2}}=\frac{\kappa_{4}^{2}}{2}\delta
\rho_{eff}\,,
\end{equation}
\begin{equation}
\label{3:18}
\ddot{\Psi}+3H(H\Phi+\dot{\Psi})+H\dot{\Phi}+2\dot{H}\Phi+\frac{1}{3a^{2}}k^{2}(\Phi-\Psi)=
\frac{\kappa_{4}^{2}}{2}\delta p_{eff}\,,
\end{equation}
\begin{eqnarray}
\label{3:19} \dot{\Psi}+H\Phi=\hspace{6cm}\nonumber\\
-\frac{\kappa^{2}V(\phi)}{\sqrt{1-\dot{\phi}^{2}}}\frac{\dot{\phi}\delta\phi}{2\big(1-\kappa^{2}f\big)}
+\frac{\kappa^{2}\int\big(\delta
T_{i}^{0}\big)^{(nmc)}dx^{i}}{2\big(1-\kappa^{2}f\big)}\,,\hspace{0.5cm}
\end{eqnarray}
\begin{equation}
\label{3:20} \Psi-\Phi=-\frac{\kappa^{2}\delta f}{1-\kappa^{2}f}\,.
\end{equation}
By comparing equations \eqref{3:17}-\eqref{3:20} with equations
\eqref{21}-\eqref{24}, we can see the effects of the non-minimal
coupling between tachyon field and Ricci scalar in perturbed field
equations. In equations \eqref{3:17} and \eqref{3:18}, $\rho_{eff}$
and $p_{eff}$ are the effective energy density and pressure which
contain the effects of non-minimal coupling. We obtain these
effective parameters from the standard Friedmann equation
$H^{2}=\frac{\kappa_{4}^{2}}{3}\rho_{eff}$\, and effective
conservation equation $\dot{\rho}_{eff}+3H(\rho_{eff}+p_{eff})=0$ as
follows
\begin{equation}
\label{3:21} \rho_{eff}=\frac{\rho_{\phi}}{1-\kappa^{2}f}\,,
\end{equation}
and
\begin{equation}
\label{3:22} p_{eff}=\frac{p_{\phi}}{1-\kappa^{2}f}\,.
\end{equation}
where $\rho_{\phi}$ and $p_{\phi}$ are defined by equations
\eqref{3:4} and \eqref{3:5} respectively. By perturbing the
equations \eqref{3:21} and \eqref{3:22}, we find
\begin{eqnarray}
\label{3:23} \delta\rho_{eff}=\frac{1}{1-\kappa^{2}f}\Bigg[\frac{V'\,\delta\phi}{\sqrt{1-\dot{\phi}^{2}}}\hspace{3cm}\nonumber\\
-V\,\frac{\dot{\phi}\,\delta\dot{\phi}
+\dot{\phi}^{2}\,\Phi}{\big(1-\dot{\phi}^{2}\big)^{\frac{3}{2}}}+\frac{\kappa^{2}\rho_{\phi}\delta
f}{1-\kappa^{2}f}+\delta\rho^{(nmc)}\Bigg]\,,
\end{eqnarray}
and
\begin{eqnarray}
\label{3:24} \delta
p=\frac{1}{1-\kappa^{2}f}\Bigg[-V'\,\sqrt{1-\dot{\phi}^{2}}\,\delta\phi\hspace{2cm}\nonumber\\
-V\,\frac{\dot{\phi}\,\delta\dot{\phi}
+\dot{\phi}^{2}\,\Phi}{\sqrt{1-\dot{\phi}^{2}}}+\frac{\kappa^{2}p_{\phi}\delta
f}{1-\kappa^{2}f}+\delta p^{(nmc)}\Bigg]\,.
\end{eqnarray}
$\delta\rho^{(nmc)}$ and $\delta p^{(nmc)}$ are the perturbed energy
density and perturbed pressure corresponding to the non-minimal
sector of the energy-momentum tensor which is given by the following
expression
\begin{equation}
\label{3:25}
T_{\mu\nu}^{(nmc)}=\Big(g_{\mu\nu}\Box-\nabla_{\mu}\nabla_{\nu}\Big)f\,.
\end{equation}
By perturbing the $(0,\,0)$ and $(i,\,i)$ (with $i=1,...,3$)
components of the equation \eqref{3:25}, we obtain the perturbed
non-minimal sector of the energy density and pressure respectively
as follows
\begin{equation}
\label{3:26}
\delta\rho^{(nmc)}=2\Phi\Big[\big(\Box+\nabla_{0}\nabla^{0}\big)f\Big]+\Big[-2\Phi\Box
f-\big(\Box+\nabla_{0}\nabla^{0}\big)\delta f\Big]\,,
\end{equation}
and
\begin{eqnarray}
\label{3:27} \delta p^{(nmc)}=\delta_{j}^{b}\Bigg[\delta
g^{aj}\big(g_{ab}\Box-\nabla_{a}\nabla_{b}\big)f+g^{aj}\delta g_{ab}
\Box f\hspace{0.2cm}\nonumber\\
+g^{aj}\big(g_{ab}\Box-\nabla_{a}\nabla_{b}\big)\delta
f\Bigg]\,.\hspace{1cm}
\end{eqnarray}
The effects of the non-minimal coupling is seen by the presence of
$\big(\delta T_{i}^{0}\big)^{(nmc)}$ in equation \eqref{3:19} and
also by the presence of parameter $f$ in equations \eqref{3:19} and
\eqref{3:20}. Especially, the non-minimal coupling of the tachyon
field and the Ricci scalar causes that the two metric perturbations
($\Psi$ and $\Phi$) are not equal anymore. The equation \eqref{3:20}
shows this property clearly.

Variation of the scalar field's equation of motion in the presence
of the non-minimal coupling is given by the following expression
\begin{widetext}
\begin{eqnarray}
\delta\ddot{\phi}+3H\delta\dot{\phi}+2\,\dot{\phi}\,\ddot{\phi}\,\frac{\delta\dot{\phi}+\dot{\phi}^{2}\Phi}{1-\dot{\phi}^{2}}
+\Bigg[\,\frac{\dot{\phi}\delta\dot{\phi}+\ddot{\phi}^{2}}{2V\sqrt{1-\dot{\phi}^{2}}}
-\frac{V'\delta\phi\sqrt{1-\dot{\phi}^{2}}}{2V}\,\Bigg]f'R+\frac{\sqrt{1-\dot{\phi}^{2}}}{2V}f''R\delta\phi
\hspace{3cm}\nonumber\\
+\frac{\sqrt{1-\dot{\phi}^{2}}}{V}f\,\Bigg[\Big(\frac{ka^{2}}{a^{2}}-3\dot{H}\Big)\Phi
-\frac{2k^{2}}{a^{2}}\Psi-3\Big(\ddot{\Psi}+4H\dot{\Psi}+H\dot{\Phi}+\dot{H}\Phi+4H^{2}\Phi\Big)\Bigg]
\hspace{3cm}\nonumber\\
=\dot{\phi}\Big(\dot{\Phi}+3\dot{\Psi}\Big)+\Bigg(6H\dot{\phi}^{3}-\frac{2V'}{V}(1-\dot{\phi}^{2})\Bigg)\Phi
+(1-\dot{\phi}^{2})\delta\phi\Bigg(\frac{V''}{V}-\frac{V'^{2}}{V^{2}}\Bigg)\,.\hspace{1cm}
\label{3:28}
\end{eqnarray}
\end{widetext}
If we set $f=0$, the above perturbed scalar field's equation of
motion simplifies to the equation \eqref{27}.

Now, we study variation of the curvature perturbation on a
uniform-density hypersurface. In this regard, we rewrite the
gauge-invariant primordial curvature perturbation $\zeta$, given by
equation \eqref{28}, as follows
\begin{equation}
\label{3:29}
\zeta=\Psi-\frac{H}{\dot{\rho}_{eff}}\delta\rho_{eff}\,.
\end{equation}
From this equation, the time evolution of the primordial curvature
perturbation is given by the following equation
\begin{equation}
\label{3:30} \dot{\zeta}=H\left(\frac{\delta
p_{nad}}{\rho_{eff}+p_{eff}}\right)
\end{equation}
Similar to the minimal coupling case, we split the pressure
perturbations into adiabatic and non-adiabatic parts as follows
\begin{equation}
\label{3:31} \delta
p_{eff}=c_{s}^{2}\delta\rho_{eff}+\dot{p}_{eff}\Gamma\,.
\end{equation}
In equation \eqref{3:31}, the non-adiabatic part is $\delta
p_{nad}=\dot{p}_{eff}\Gamma$\,. In the presence of the non-minimal
coupling between the tachyon field and Ricci scalar, $\delta
p_{nad}$ is not zero anymore. From equation \eqref{3:31}, we can
find the $\delta p_{nad}$ as follows
\begin{widetext}
\begin{eqnarray}
\delta
p_{nad}=\frac{2}{\kappa^{2}(1-\kappa^{2}f)}\Bigg[1-\dot{\phi}^{2}
-\frac{\Big(-V'\dot{\phi}\sqrt{1-\dot{\phi}^{2}}+
\frac{V\dot{\phi}\ddot{\phi}}{\sqrt{1-\dot{\phi}^{2}}}++\dot{p}_{nmc}\Big)(1-\kappa^{2}f)+p\kappa^{2}f'\dot{\phi}}
{\Big(-3H\big(\rho+p\big)-3f'H^{2}\dot{\phi}\Big)\Big(1-\kappa^{2}f\Big)+\rho\kappa^{2}f'\dot{\phi}}\Bigg]
\hspace{2cm}\nonumber\\
\times\Bigg[-3H\Big(H\Phi+\dot{\Psi}\Big)-\frac{k^{2}}{a^{2}}\Bigg]
-\frac{1}{1-\kappa^{2}f}\Bigg[2V'\delta\phi\sqrt{1-\dot{\phi}^{2}}+\delta\rho_{nmc}(1-\dot{\phi}^{2})-\delta
p_{nmc}\Bigg]\,.
 \label{3:32}
\end{eqnarray}
\end{widetext}
This non-vanishing, non-adiabatic pressure, leads to the
non-vanishing time evolution of the primordial curvature
perturbation as follows
\begin{widetext}
\begin{eqnarray}
\dot{\zeta}=-H\Bigg[2V'\delta\phi\sqrt{1-\dot{\phi}^{2}}+\delta\rho_{nmc}(1-\dot{\phi}^{2})-\delta
p_{nmc}\Bigg]+2\Bigg[-3H^{2}\Big(H\Phi+\dot{\Psi}\Big)-H\frac{k^{2}}{a^{2}}\Bigg]\times
\hspace{3cm}\nonumber\\
\left[\frac{\Big(1-\dot{\phi}^{2}\Big)^{\frac{3}{2}}}{\kappa^{2}V'\dot{\phi}^{2}}
+\frac{\Big(V'\dot{\phi}\sqrt{1-\dot{\phi}^{2}}-
\frac{V\dot{\phi}\ddot{\phi}}{\sqrt{1-\dot{\phi}^{2}}}-\dot{p}_{nmc}\Big)\Big(1-\kappa^{2}f\Big)\Big(\sqrt{1-\dot{\phi}^{2}}\Big)
-p\kappa^{2}f'\dot{\phi}\sqrt{1-\dot{\phi}^{2}}}
{\kappa^{2}V'\dot{\phi}^{2}\Big(-3H\big(\rho+p\big)-3f'H^{2}\dot{\phi}\Big)\Big(1-\kappa^{2}f\Big)
+\rho\kappa^{4}f'V'\dot{\phi}^{3}}\right]. \label{3:33}
\end{eqnarray}
\end{widetext}
We see that, the non-minimal coupling behaves as a second field in
the theory and causes non-adiabatic perturbation. So, the primordial
curvature perturbations attain an explicit time-dependence.

Now, in order to obtain scalar and tensorial perturbations in our
model, we take into account the slow-roll approximation at the large
scales, $k\ll aH$. This is because the scales of cosmological
interest have spent most of their time far outside the Hubble radius
and have re-entered only relatively recently in the Universe
history. In this scale, $\ddot{\Phi}$, $\ddot{\Psi}$, $\dot{\Phi}$
and $\dot{\Psi}$ are negligible (see ~\cite{Amn06,Amn07}). So, at
large scale, the perturbed tachon field's equation of motion takes
the following form
\begin{eqnarray}
\label{3:34}
3H\delta\dot{\phi}+\Bigg(\frac{f''R-V'f'R}{2V}-\frac{V''}{V}+\frac{V'^{2}}{V^{2}}\Bigg)\delta\phi
\hspace{1cm}\nonumber\\=\Bigg(-\frac{f'R}{V}-\frac{2V'}{V}\Bigg)\Phi\hspace{1cm}
\end{eqnarray}
Also, from equation \eqref{3:19} and by using the relation $\int
\delta(T_{i}^{0})_{nmc}\,dx^{i}=2f\Big(H\Phi+\dot{\Psi}\Big)$, we
obtain
\begin{equation}
\label{3:35}
\Phi=-\frac{\kappa^{2}V\dot{\phi}\delta\phi}{2H\Big(1-2\kappa^{2}f\Big)\sqrt{1-\dot{\phi}^{2}}}\,.
\end{equation}
So, equation \eqref{3:34} can be rewritten as follows
\begin{eqnarray}
\label{3:36}
3H\delta\dot{\phi}=\Bigg(-\frac{f'R}{V}-\frac{2V'}{V}\Bigg)\Bigg(\frac{\kappa^{2}\,\dot{\phi}\,\delta\phi}
{2H\Big(1-2\kappa^{2}f\Big)\sqrt{1-\dot{\phi}^{2}}}\Bigg)\hspace{0.1cm}\nonumber\\
-\Bigg(\frac{f''R-V'f'R}{2V}-\frac{V''}{V}+\frac{V'^{2}}{V^{2}}\Bigg)\delta\phi\,.\hspace{1cm}
\end{eqnarray}
Here we introduce the function ${\cal{F}}$ as
\begin{equation}
\label{3:37} {\cal{F}}\equiv\frac{V\delta\varphi}{V'}\,,
\end{equation}
by which we rewrite equation \eqref{3:36} as follows
\begin{eqnarray}
\label{3:38}
\frac{{\cal{F}}'}{{\cal{F}}}=\frac{\kappa^{2}\Big(f'R+2VV'\Big)}{6H^{2}V\Big(1-\kappa^{2}f\Big)}\hspace{0.1cm}\nonumber\\
-\frac{V'f'R-f''R-2V''+\frac{2V'^{2}}{V}}{-f'R-2V'}-\frac{V''}{V'}+\frac{V'}{V}\,.\hspace{1cm}
\end{eqnarray}
A solution of this equation is given by the following expression
\begin{equation}
\label{3:39}
{\cal{F}}={\cal{C}}\exp\bigg(\int\frac{{\cal{F}}'}{{\cal{F}}}d\varphi\bigg)\,,
\end{equation}
where ${\cal{C}}$ is an integration constant. So, from equation
\eqref{3:37} we find
\begin{widetext}
\begin{equation}
\label{3:40} \delta\phi=\frac{{\cal{C}}V'}{V}\exp
\Bigg[\int\Bigg(\frac{\kappa^{2}\Big(f'R+2VV'\Big)}{6H^{2}V\Big(1-\kappa^{2}f\Big)}
-\frac{V'f'R-f''R-2V''+\frac{2V'^{2}}{V}}{-f'R-2V'}-\frac{V''}{V'}+\frac{V'}{V}\Bigg)d\phi\Bigg]\,.
\end{equation}
\end{widetext}
So, the density perturbation amplitude is given by
\begin{widetext}
\begin{equation}
\label{3:41}
A_{s}^{2}=\frac{k^{3}{\cal{C}}}{2\pi^{2}}\frac{V'^{2}}{V^{2}} \exp
\Bigg[2\int\Bigg(\frac{\kappa^{2}\Big(f'R+2VV'\Big)}{6H^{2}V\Big(1-\kappa^{2}f\Big)}
-\frac{V'f'R-f''R-2V''+\frac{2V'^{2}}{V}}{-f'R-2V'}-\frac{V''}{V'}+\frac{V'}{V}\Bigg)d\phi\Bigg]\,.
\end{equation}
\end{widetext}
By using this equation, the scalar spectral index (defined by
equation \eqref{37}) becomes
\begin{widetext}
\begin{equation}
\label{3:42}
n_{s}-1=\Bigg[\frac{\kappa^{2}\Big(f'R+2VV'\Big)}{3H^{2}V\Big(1-\kappa^{2}f\Big)}
-\frac{2V'f'R-2f''R-4V''+\frac{4V'^{2}}{V}}{-f'R-2V'}\Bigg]
\Bigg[\frac{\Big(1-\kappa^{2}f\Big)\Big(-f'R-2V'\Big)}{2V^{2}-f'^{2}R-2f'V'}\Bigg]\,.
\end{equation}
\end{widetext}

In the non-minimal case, the tensor perturbations amplitude of a
given mode when leaving the Hubble radius (equation \eqref{40}) are
given by
\begin{equation}
\label{3:43}
A_{T}^{2}=\frac{4\kappa^4}{75\pi\big(1-\kappa^{2}f\big)}\Bigg[V-\frac{f'^{2}R}{2V}-\frac{f'V'}{V}\Bigg]\,.
\end{equation}
So, the tensor spectral index (defined by equation \eqref{42}) takes
the following form
\begin{widetext}
\begin{equation}
n_{T}=\Bigg(\frac{-f'R-2V'}{6H^{2}V}\Bigg)\Bigg(\frac{\kappa^{2}}{9\big(1-\kappa^{2}f\big)H^{3}}\Bigg)
\Bigg(-\frac{f'^{2}R}{2V}-\frac{f'V'}{2V}\Bigg)
\times\Bigg(V'-\frac{f'f''R}{V}+\frac{f'^{2}RV'}{2V^{2}}
-\frac{f''V'}{V}+\frac{f'V''}{V}+\frac{f'V'^{2}}{V^{2}}+3\kappa^{2}H^{2}\Bigg)\,.\label{3:44}
\end{equation}
\end{widetext}
Also, the tensor-to-scalar ratio in the presence of the non-minimal
coupling is given by
\begin{widetext}
\begin{eqnarray}
r_{t-s}\equiv\frac{A_{T}^{2}}{A_{s}^{2}}\simeq\frac{8\pi}{25k^{3}}\Bigg[\frac{V^{3}-f'^{2}RV-f'V'V}{V'^{2}{\cal{C}}}\Bigg]
\hspace{9cm}\nonumber\\ \times \exp
\Bigg[-2\int\Bigg(\frac{\kappa^{2}\Big(f'R+2VV'\Big)}{6H^{2}V\Big(1-\kappa^{2}f\Big)}
-\frac{V'f'R-f''R-2V''+\frac{2V'^{2}}{V}}{-f'R-2V'}-\frac{V''}{V'}+\frac{V'}{V}\Bigg)d\phi\Bigg]\,.\label{3:45}
\end{eqnarray}
\end{widetext}

\begin{figure*}
\flushleft\leftskip1em{
\includegraphics[width=.35\textwidth,origin=c,angle=0]{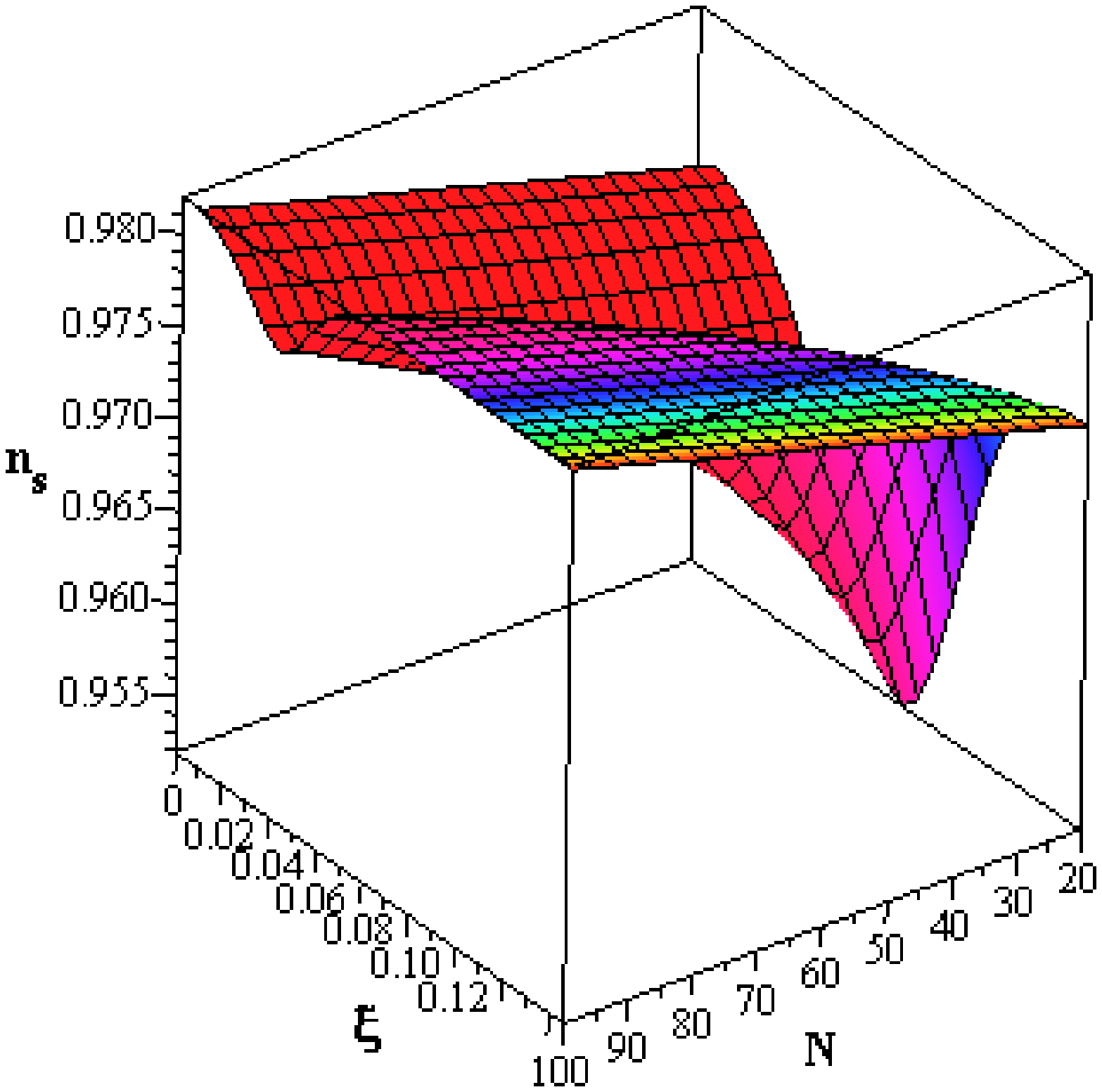}}
\hspace{3.6cm}
\includegraphics[width=.35\textwidth,origin=c,angle=0]{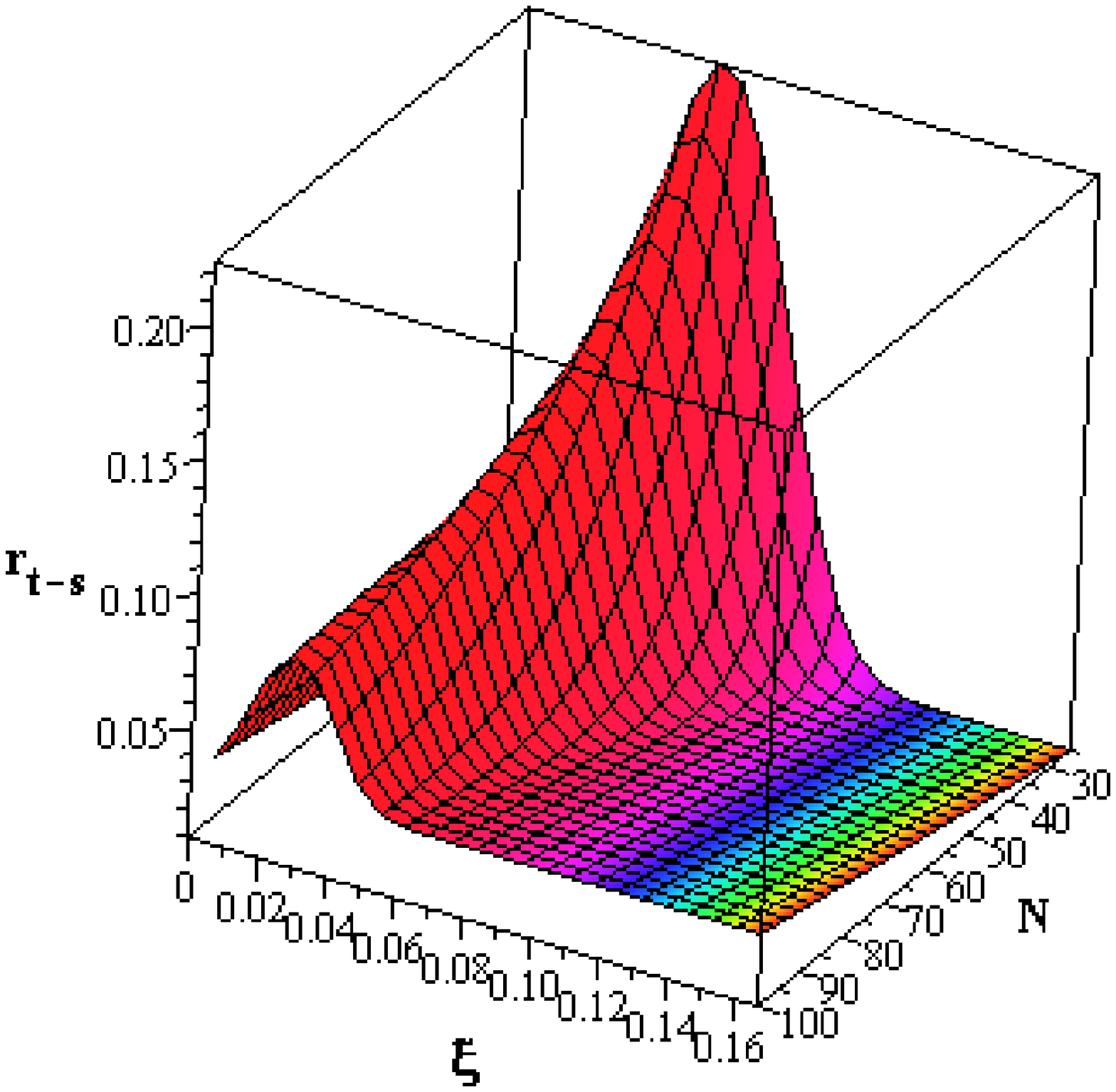}
\caption{\label{fig12} Evolution of the scalar spectral index (left
panel) and the tensor to scalar ratio (right panel) versus the
non-minimal coupling and number of e-folds parameter.}
\end{figure*}
\begin{figure}
\flushleft\leftskip0em{
\includegraphics[width=.40\textwidth,origin=c,angle=0]{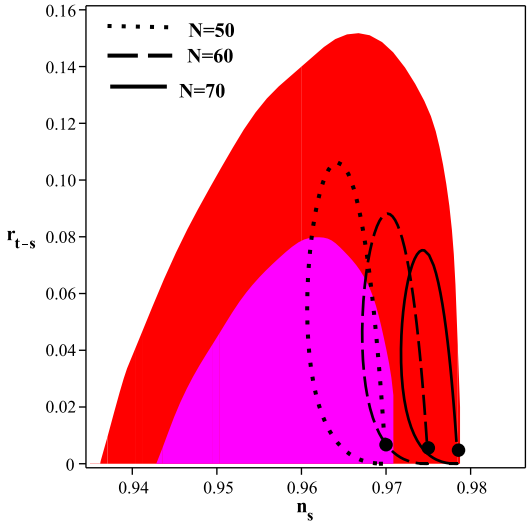}}
\caption{\label{fig13} Behavior of the tensor to scalar ratio with
respect to the scalar spectral index for various $\xi$ and for
quadratic potential in the background of WMAP9+eCMB+BAO+H$_{0}$
data. The two contours are corresponding to the 68$\%$ and 95$\%$
levels of confidence.}
\end{figure}

\subsubsection{$V=\frac{1}{2}\sigma\phi^{2}$}

Similar to the minimal coupling case, for numerical analysis of the
model's parameter space, we firstly solve the integral of equation
\eqref{3:16} with the quadratic tachyon potential and
$f(\phi)=\xi\phi^{2}$. By solving this integral we obtain
\begin{eqnarray}
N=\frac{3}{2}\frac {\sigma^{2}\big(\phi_{f}^{2}-\phi_{hc}^{2}\big)}{
\left(12\,\xi\,R+12\,\sigma
\right)\xi}-\frac{\xi\,R}{\xi\,R+\sigma}\ln
\left(\frac{-1+{\kappa}^{2}\xi\,\phi_{f}^{2}}{
-1+{\kappa}^{2}\xi\,\phi_{hc}^{2}}\right) \hspace{0.1cm}\nonumber\\
+\frac{{\sigma}^{2}}{{\kappa}^{2}\left( 8\, \xi\,R+8\,\sigma \right)
{\xi}^{2}}\ln\left(\frac{-1+{\kappa}^{2}\xi\,\phi_{f}^{2}}{-1+{\kappa}^{2}\xi\,\phi_{hc}^{2}}\right)\hspace{1cm}\nonumber\\
-\frac{\sigma}{\xi\,R+\sigma}\ln\left(\frac{-1+{\kappa}^{2}\xi\,\phi_{f}^{2}}{-1+{\kappa}^{2}\xi\,\phi_{hc}^{2}}\right)
\hspace{1cm} \label{3:46}
\end{eqnarray}
We assume $\phi_{hc}\gg\phi_{f}$ and find $\phi_{hc}$ from equation
\eqref{3:46}. By substituting the obtained parameter in equation
\eqref{3:42}, we plot the evolution of the scalar spectral index
with respect to the number of e-folds and the non-minimal coupling
parameter. One can see the result in the left panel of
figure~\ref{fig12}. As figure shows, for all values of $N$, the
scalar spectral index firstly decreases by increasing the
non-minimal coupling parameter and then increases by more increment
of $\xi$. Also, in this case the scalar spectral index is
red-tilted. The right panel of figure~\ref{fig12} shows the behavior
of the tensor to scalar ratio with respect to the number of e-folds
and the non-minimal coupling parameter. This figure confirms that
for all values of $N$, tensor to scalar ratio firstly decreases by
increasing the non-minimal coupling parameter and then increases by
more increment of $\xi$.

The behavior of the tensor to scalar perturbation amplitudes with
respect to the scalar spectral index, at the horizon crossing and in
the background of WMAP9+eCMB+BAO+H$_{0}$ data, can be seen in
figure~\ref{fig13}. This figure has been plotted for $\xi\geq 0$ and
for $N=50$, $N=60$ and $N=70$. This figure also shows that as $\xi$
increases, the scalar spectral index decreases whereas tensor to
scalar ratio increases. The reduction of $n_{s}$ and increment of
$r_{t-s}$ stop at some values of $\xi$ and after that, the scalar
spectral index increases by increment of $\xi$, while $r_{t-s}$
decreases. Note that in the case of quadratic potential, for all
values of $\xi$, the values of $r_{t-s}$ and $n_{s}$ are well in the
region that is compatible with observation. The dots in the figure
show $\xi=0$ and are compatible with corresponding points in
figure~\ref{fig2}.
\begin{figure*}
\flushleft\leftskip1em{
\includegraphics[width=.35\textwidth,origin=c,angle=0]{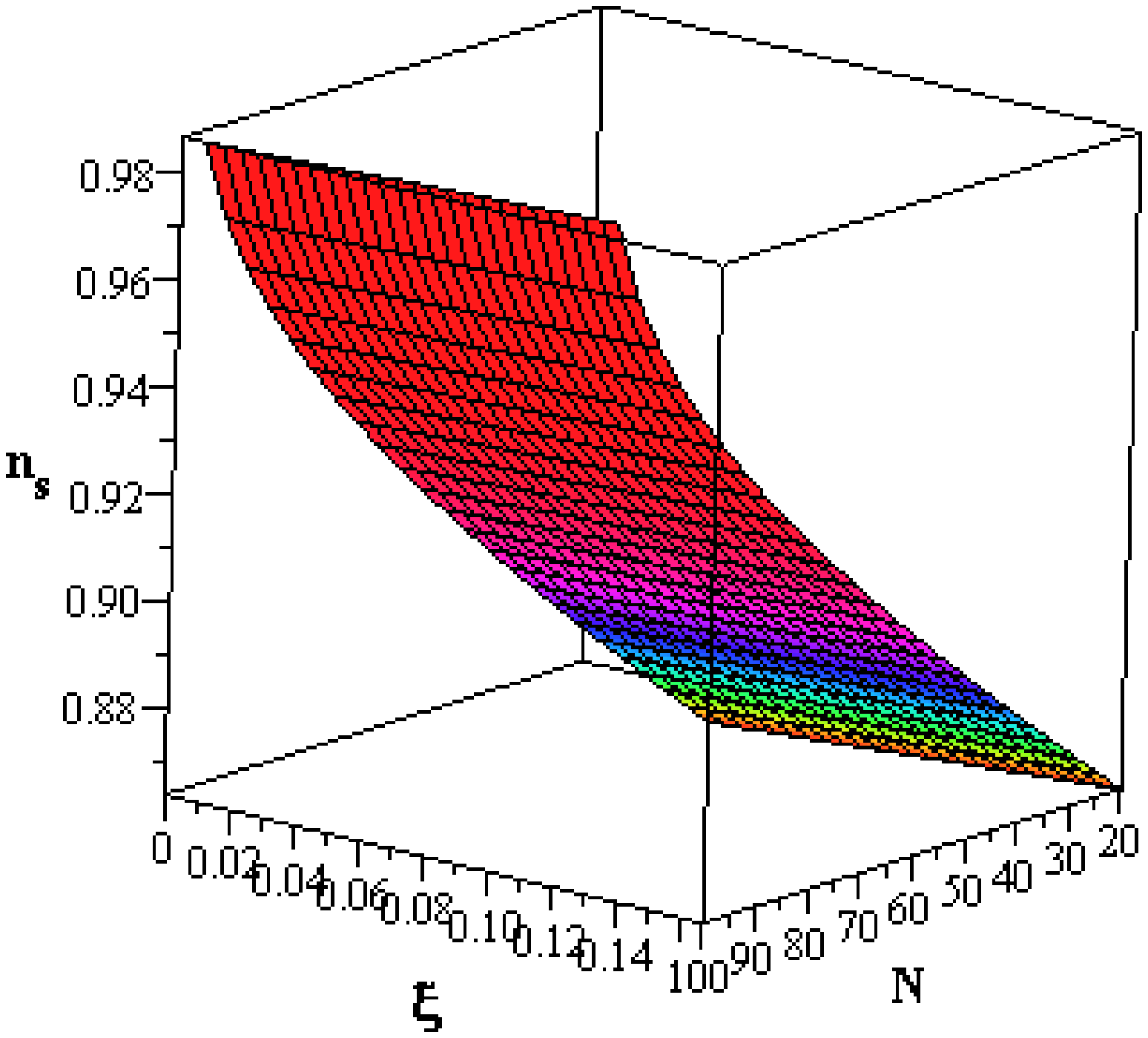}
\hspace{3.6cm}
\includegraphics[width=.35\textwidth,origin=c,angle=0]{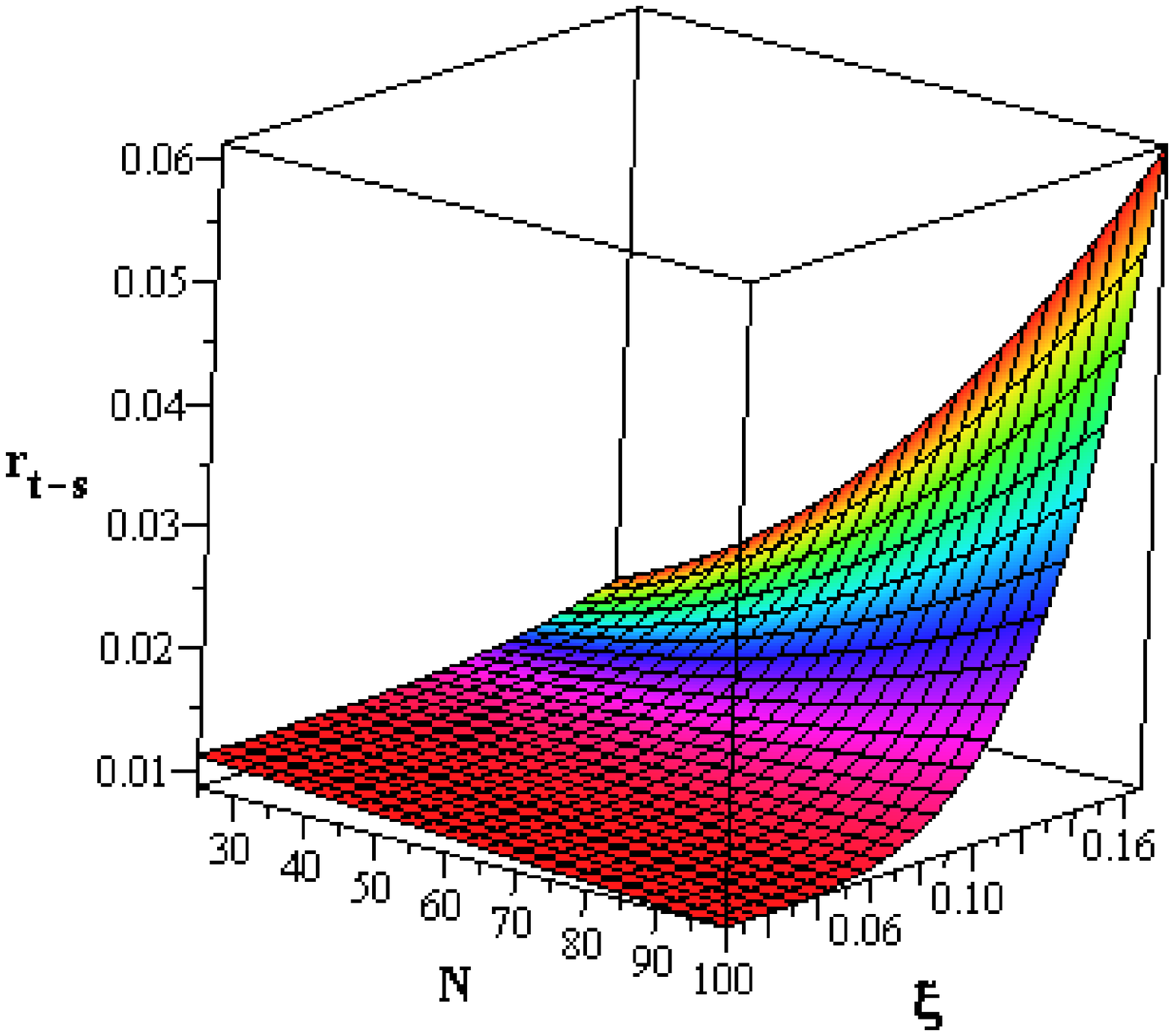}}
\caption{\label{fig14} Evolution of the scalar spectral index (left
panel) and the tensor to scalar ratio (right panel) with respect to
the non-minimal coupling and number of e-folds parameters. }
\end{figure*}
\begin{figure}
\flushleft\leftskip0em{
\includegraphics[width=.40\textwidth,origin=c,angle=0]{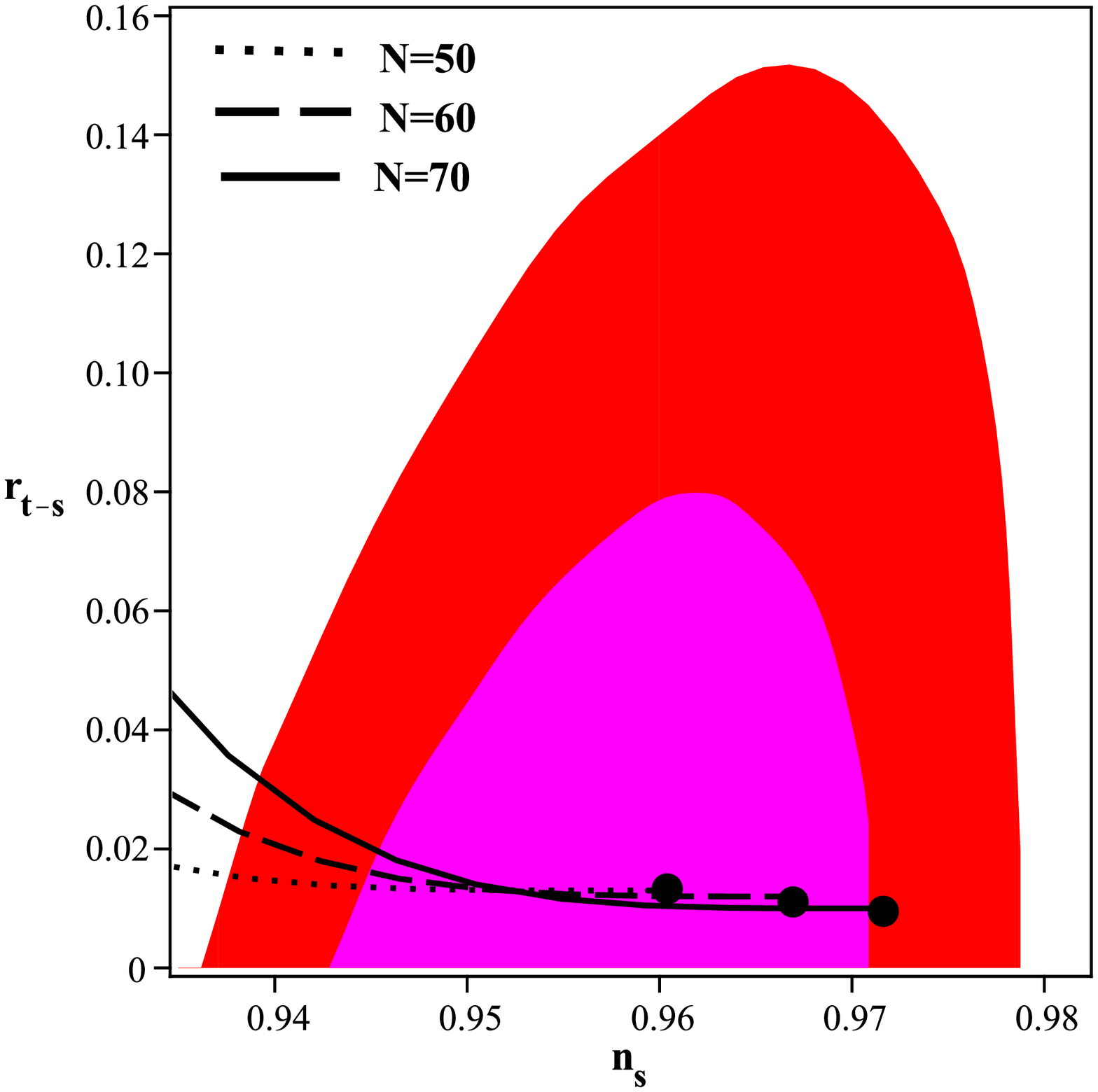}}
\caption{\label{fig15} Behavior of the tensor to scalar ratio with
respect to the scalar spectral index for various $\xi$ and for
exponential potential in the background of WMAP9+eCMB+BAO+H$_{0}$
data. The two contours are corresponding to the 68$\%$ and 95$\%$
levels of confidence.}
\end{figure}

\begin{table*}
\caption{\label{tab4} The range of the non-minimal coupling
parameter which leads to observationally viable scalar spectral
index and tensor to scalar ratio for exponential potential and for
three different values of $N$.}
\begin{ruledtabular}
\begin{tabular}{cccc}
 & $N=50$ & $N=60$ & $N=70$\\
\hline $\xi$ & $0<\xi<0.02$ & $0<\xi<0.033$ & $0<\xi<0.05$ \\
\end{tabular}
\end{ruledtabular}
\end{table*}

\subsubsection{$V=V_{0}\,e^{-\kappa\sigma\phi}$}
Now, we consider an exponential potential and solve the integral of
equation \eqref{3:16}. Solving this integral gives the following
expression for number of e-folds parameter
\begin{eqnarray}
N=\left(-R+2\frac{\kappa^{2}v\alpha^{
2}}{\xi}-\frac{\kappa^{4}v^{2}\alpha^{4}}{\xi^{2}R}+
\kappa^{2}\alpha^{2}\right){\cal{M}}\hspace{1cm}\nonumber\\+
\left(\sqrt {\xi}R+2\frac{vR}{\sqrt{\xi}}-\frac
{\kappa^{2}v^{2}}{\sqrt{\xi}}-\frac
{\kappa^{2}v^{2}\alpha^{2}}{\xi^{\frac{3}{2}}}
\right){\cal{P}}+\bigg(\frac{1}{2}\kappa^{2}v^{2}R
\hspace{0.1cm}\nonumber\\
-\xi R^{2}+\frac{1}{2}\frac{\kappa^
{2}v^{2}R\alpha^{2}}{\xi}+\frac{1}{2}\kappa^{4}v^{2}\alpha^{2}-
\frac {\kappa^{4}v^{3}\alpha^{2}}{\xi} \bigg){\cal{N}}
\,,\hspace{1cm} \label{3:46-2}
\end{eqnarray}
where
\begin{equation}
\label{3:46-3} {\cal{M}}=\frac {\kappa^{2}v^{2} \ln \left(\frac{
\xi\,\phi_{f}\,R-v\kappa\,\alpha}{\xi\,\phi_{hc}\,R-v\kappa\,\alpha}
\right) }{\xi\,{R}^{2 }-{\kappa}^{4}{v}^{2}{\alpha}^{2}}\,,
\end{equation}
\begin{equation}
\label{3:46-4} {\cal{N}}=\frac {\kappa^{2}v\alpha \ln  \left(\frac{
-1+{\kappa}^{2}\xi\,\phi_{f}^{2}}{-1+{\kappa}^{2}\xi\,\phi_{hc}^{2}}
\right) }{\xi\,{R}^ {2}-{\kappa}^{4}{v}^{2}{\alpha}^{2}}\,,
\end{equation}
and
\begin{equation}
\label{3:46-5} {\cal{P}}=\frac {\arctan \left( \kappa\,\sqrt
{\xi}\phi_{f} \right)-\arctan \left( \kappa\,\sqrt {\xi}\phi_{hc}
\right) }{\xi\,{R }^{2}-{\kappa}^{4}{v}^{2}{\alpha}^{2}}\,.
\end{equation}

By setting equation~\eqref{3:12} equal to unity and finding
$\phi_{f}$, we can deduce $\phi_{hc}$ from equation~\eqref{3:46-2}.
As before, if we substitute expression of $\phi_{hc}$ in
equation~\eqref{3:42}, we can explore the evolution of the scalar
spectral index in terms of the number of e-folds and non-minimal
coupling parameter (see left panel of figure~\ref{fig14}). We see
from this figure that for all values of $N$, the scalar spectral
index decreases as $\xi$ increases. Also, for all values of $\xi$,
the scalar spectral index is red-tilted. The behavior of the tensor
to scalar ratio with respect to the number of e-folds and the
non-minimal coupling parameter is shown in the left panel of
figure~\ref{fig14}. As figure shows, for all values of the number of
e-folds parameter, $r_{t-s}$ increases by increment of $\xi$.

In order to see which values of $\xi$ lead to observationally viable
scalar spectral index and tensor to scalar ratio in this model, we
show the evolution of $r_{t-s}$ with respect to $n_{s}$ for various
$\xi$ in figure~\ref{fig15}. We explore this evolution for $N=50$,
$N=60$ and $N=70$. As before, dots in this figure are corresponding
to minimal case with $\xi=0$. This figure shows also that depending
on the value of $N$, some values of $\xi$ are observationally
viable. We have found the range of $\xi$ for mentioned values of $N$
and the results are summarized in table~\ref{tab4}.

\begin{figure*}
\flushleft\leftskip1em{
\includegraphics[width=.35\textwidth,origin=c,angle=0]{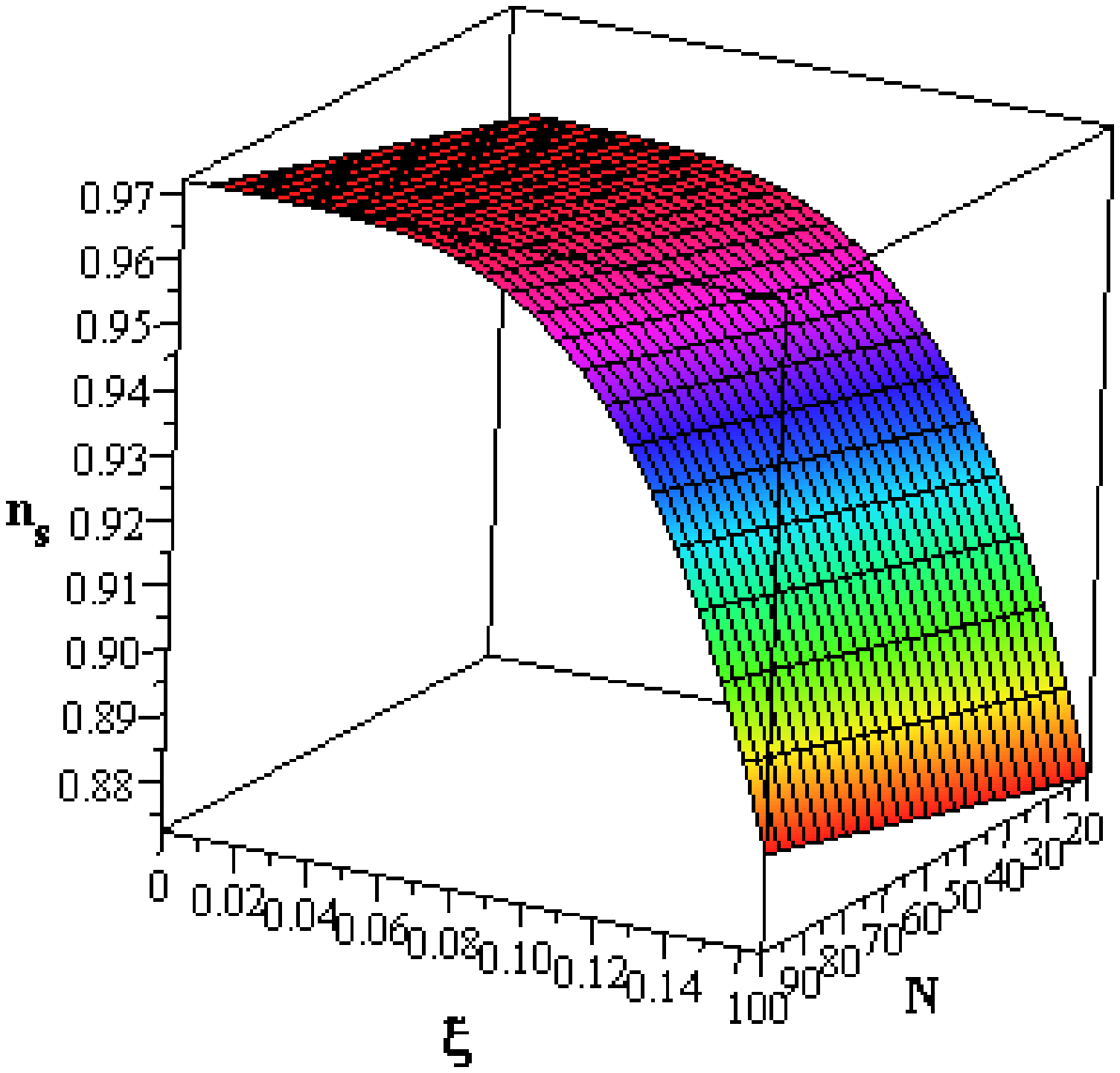}
\hspace{3.6cm}
\includegraphics[width=.35\textwidth,origin=c,angle=0]{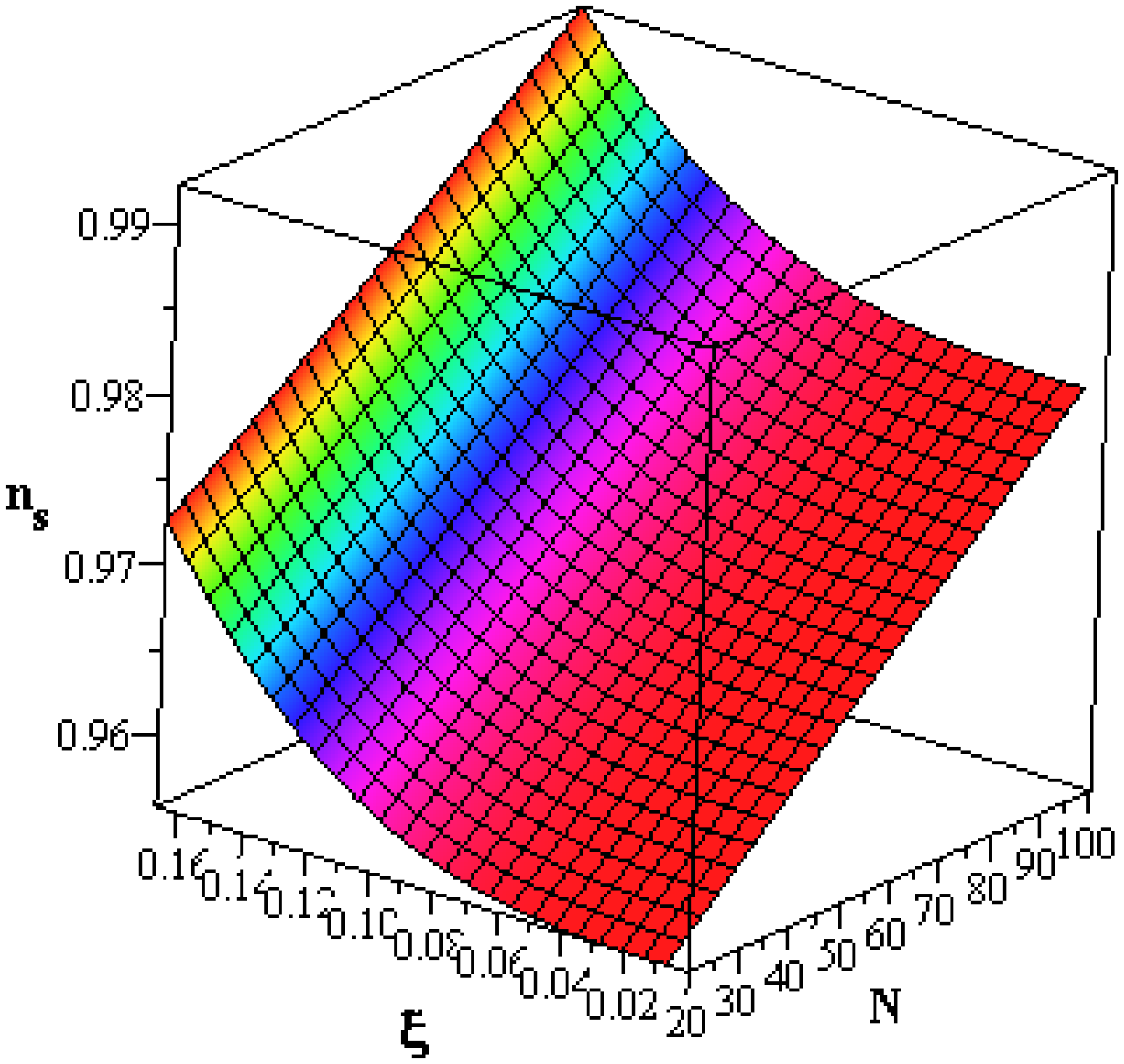}}
\caption{\label{fig16} Evolution of the scalar spectral index with
$l=0.4$ (left panel) and $l=0.87$ (right panel) with respect to the
non-minimal coupling and number of e-folds parameter. }
\end{figure*}
\begin{figure*}
\flushleft\leftskip1em{
\includegraphics[width=.35\textwidth,origin=c,angle=0]{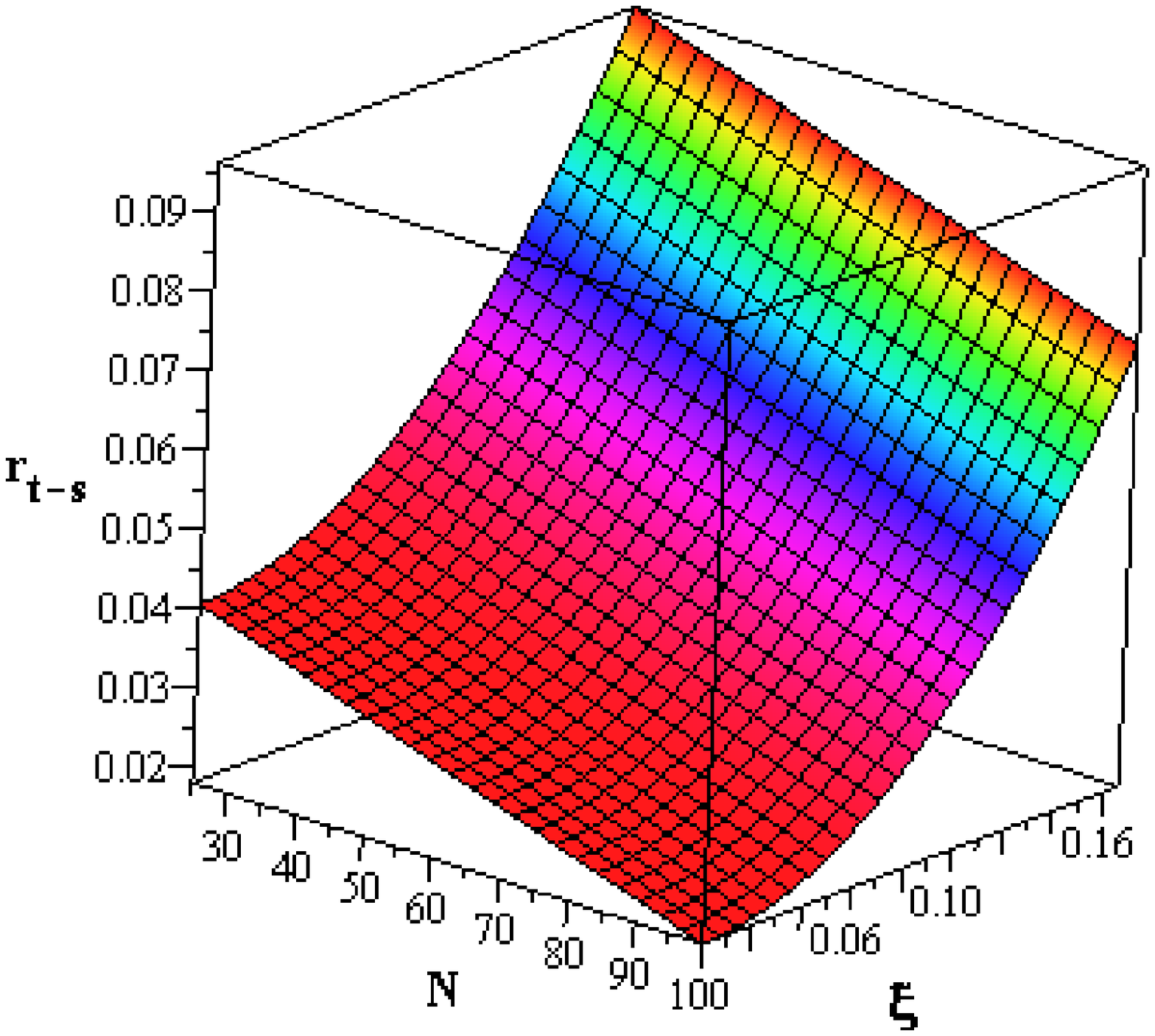}
\hfill
\includegraphics[width=.35\textwidth,origin=c,angle=0]{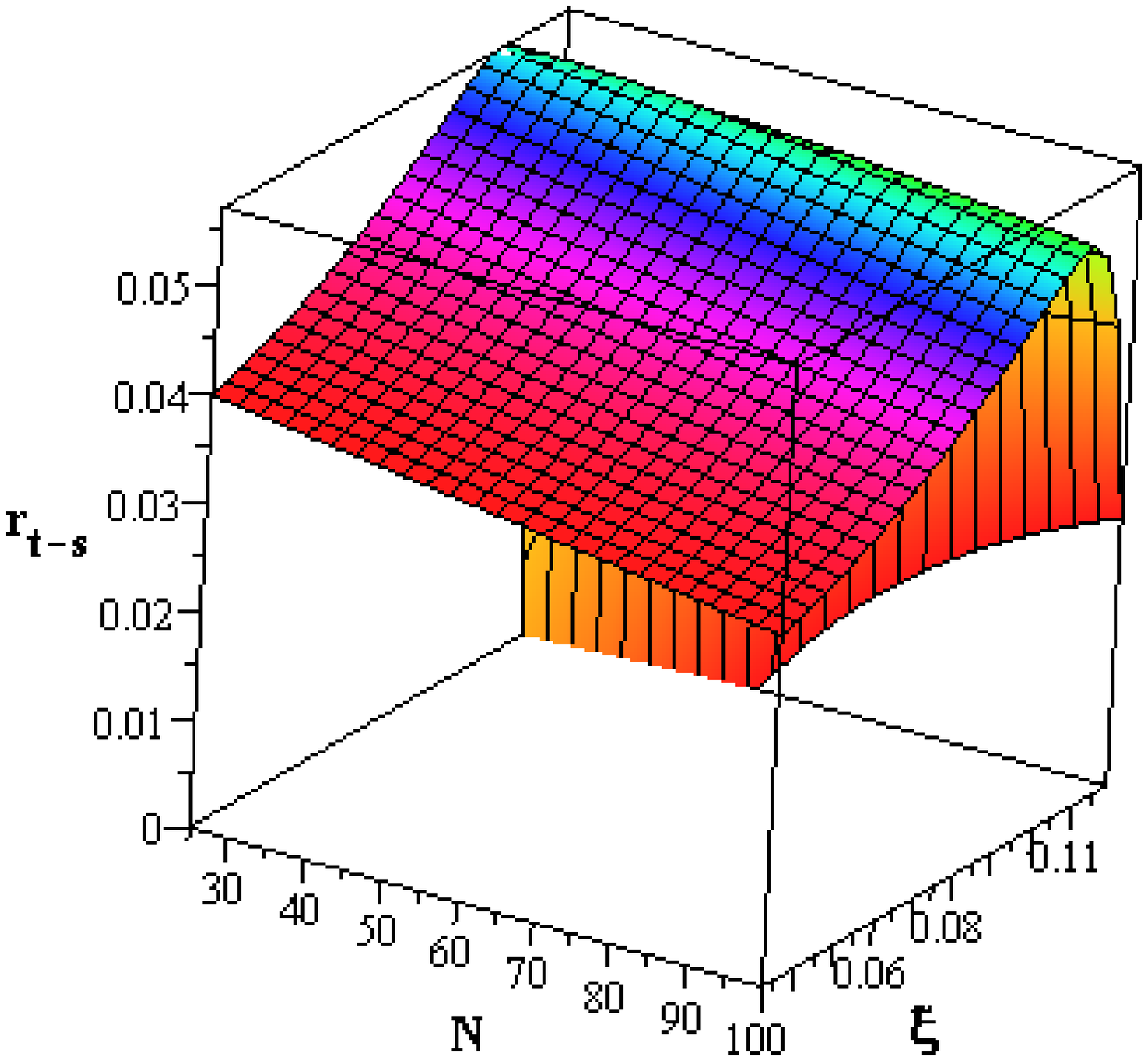}}
\caption{\label{fig17} Evolution of the tensor to scalar ratio with
$l=0.4$ (left panel) and $l=0.87$ (right panel) with respect to the
non-minimal coupling and the number of e-folds parameter. }
\end{figure*}
\begin{figure*}
\flushleft\leftskip1em{
\includegraphics[width=.35\textwidth,origin=c,angle=0]{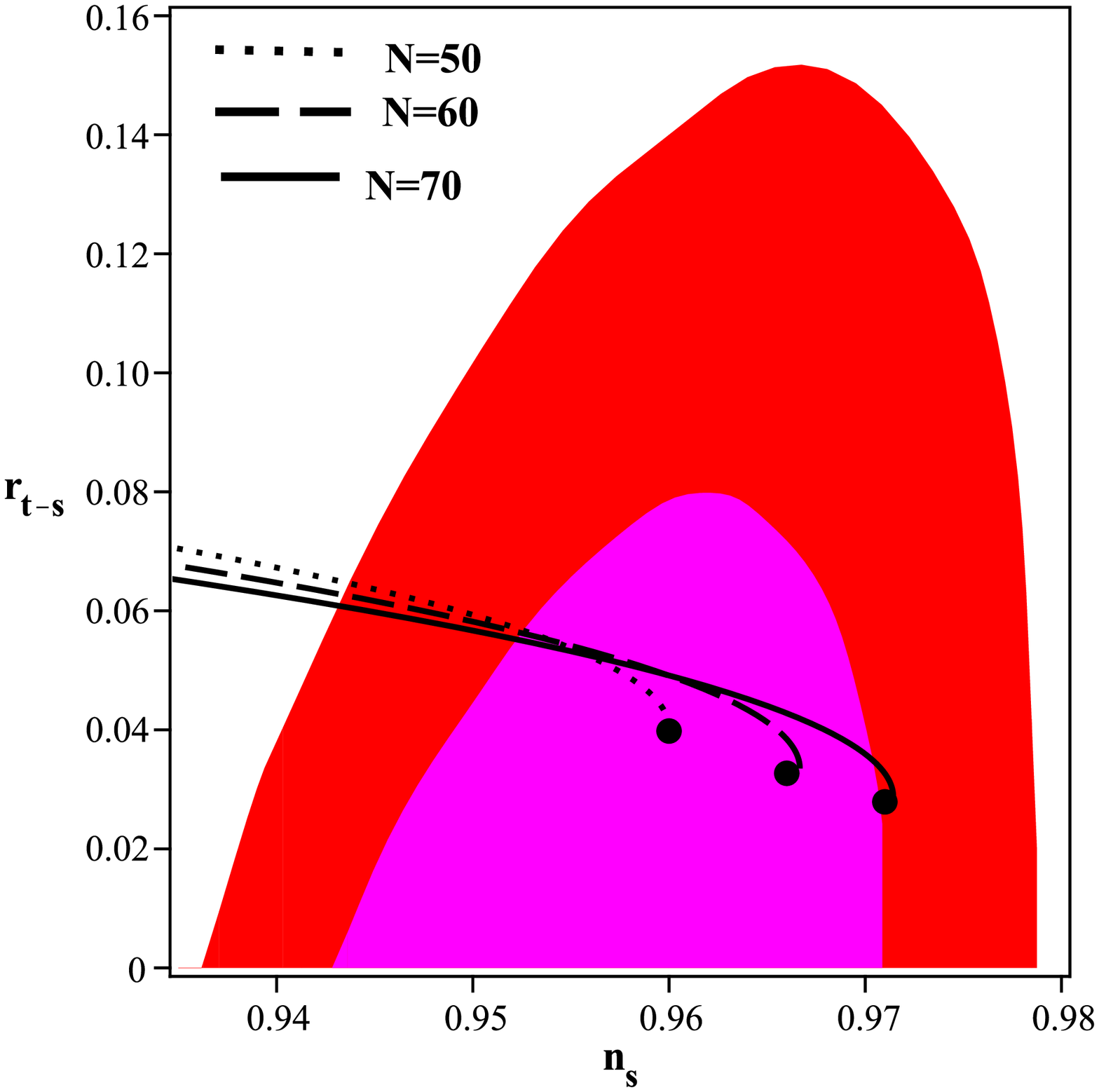}
\hfill
\includegraphics[width=.35\textwidth,origin=c,angle=0]{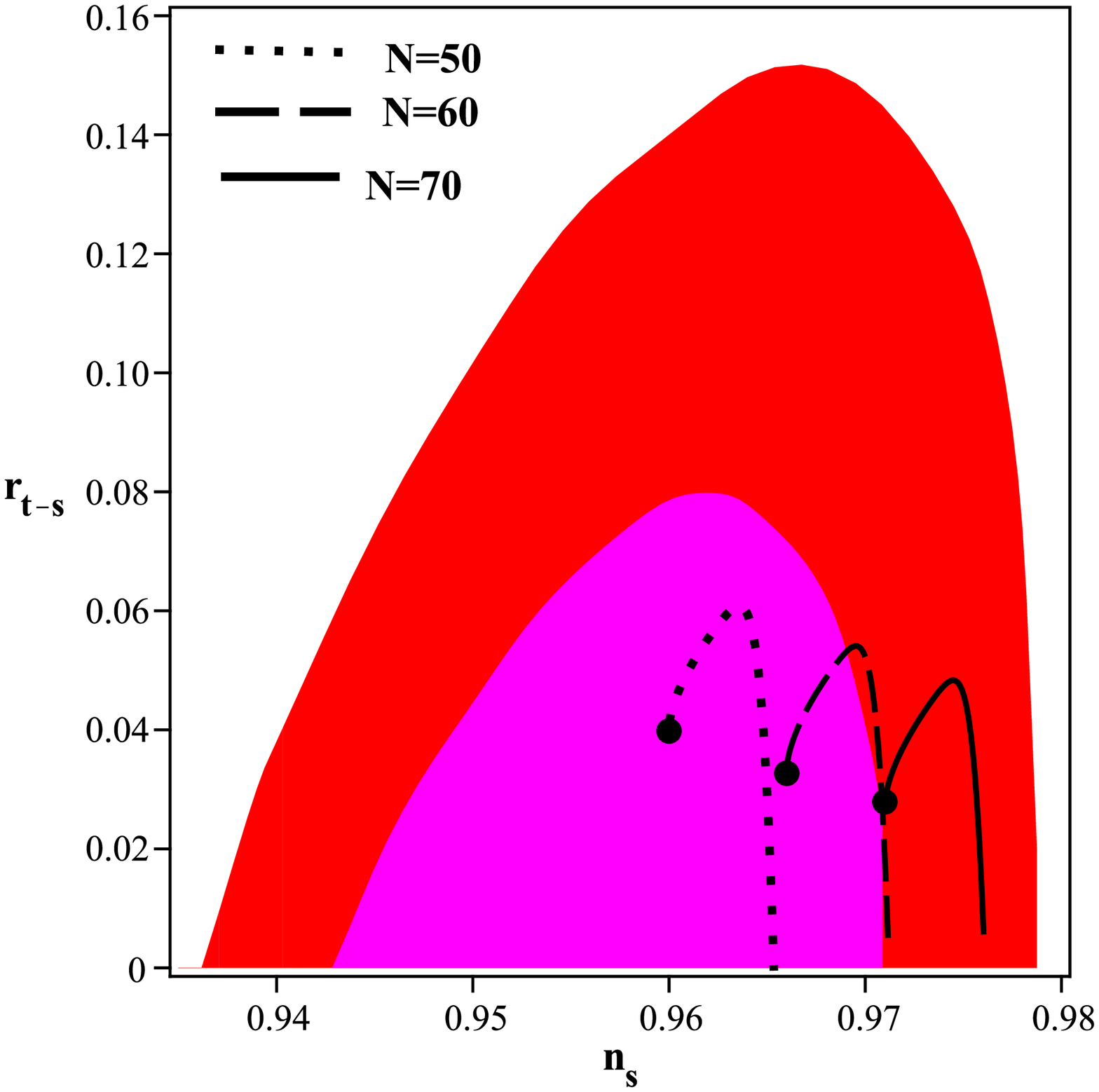}}
\caption{\label{fig18} Behavior of the tensor to scalar ratio with
respect to the scalar spectral index for various $\xi$ and for
exponential potential in the background of WMAP9+eCMB+BAO+H$_{0}$
data. The two contours are corresponding to the 68$\%$ and 95$\%$
levels of confidence. Left panel is corresponding to $l=0.4$ and
right panel with $l=0.87$.}
\end{figure*}
\begin{table*}
 \caption{\label{tab5} The range of the non-minimal
coupling parameter which leads to viable scalar spectral index and
tensor to scalar ratio for intermediate potential with $l=0.4$.}
\begin{ruledtabular}
\begin{tabular}{cccc}
 & $N=50$ & $N=60$ & $N=70$\\
\hline $\xi$ & $0<\xi<0.149$ & $0<\xi<0.14$ & $0<\xi<0.131$ \\
\end{tabular}
\end{ruledtabular}
\end{table*}

\subsubsection{Intermediate inflation}
In this subsection, we firstly obtain the intermediate potential in
the presence of the non-minimal coupling between the tachyon field
and Ricci scalar. As the minimal coupling case, we start with scale
factor in the form of $a=a_{0}\exp(\vartheta t^{l})$ and by using
the equations~\eqref{3:11} and ~\eqref{3:12} we obtain
\begin{eqnarray}
\label{3:46-6} V=3l^{2}\vartheta^{2}\left( {\frac{b}{{\phi}^{2}}}
\right)^{\beta}\left(1
-{\kappa}^{2}\xi\phi^{2}\right)\kappa^{-2}-{\frac{\kappa^{2
}\xi^{2}{\phi}^{2+2\beta}}{1-{\kappa}^{2}\xi\phi^{2}}}\hspace{0.1cm}\nonumber\\+\xi\phi
\vartheta l{b}^{\beta}\sqrt{{\frac
{\kappa^{4}\xi^{2}\phi^{2+2\beta}}{\vartheta^{2}{
l}^{2}{b}^{\beta}\left(1-{\kappa}^{2}\xi\phi^{2} \right) ^{2}}}+6{
\frac {l-1}{\vartheta l}}}\,,\hspace{1cm}
\end{eqnarray}
where
\begin{equation}
\label{3:46-7}b=\frac {3 \left( -2+f \right) ^{2}\vartheta
l}{8\left(1-l\right)}\,,
\end{equation}
and
\begin{equation}
\label{3:46-8}\beta=\frac {2l-2}{-2+l}\,.
\end{equation}

Now, we solve the integral of equation~\eqref{3:16} to obtain the
number of e-folds parameter. The result is so complicated and we
present it in Appendix~\ref{A}. As other cases, by finding
$\phi_{hc}$ from $N$ and substituting it in~\eqref{3:42}, we can
explore the evolution of the scalar spectral index versus the number
of e-folds and the non-minimal coupling parameter
(figure~\ref{fig16}). The left panel of figure~\ref{fig16} has been
plotted with $l=0.4$ and the right panel with $l=0.87$. This figure
shows that both with $l=0.4$ and $l=0.87$, for all values of $N$,
the scalar spectral index decreases as the non-minimal coupling
parameter increases. The evolution of the tensor to scalar ratio
versus the number of e-folds and the non-minimal coupling parameter
is shown in figure~\ref{fig17}. With $l=0.4$, for all values of $N$,
the tensor to scalar ratio increases as $\xi$ increases. With
$l=0.87$, for all values of $N$, this ratio first increases by
increment of $\xi$ and then decreases by more increment of $\xi$.
Note that for all $l\leq0.63$ the behavior of the scalar spectral
index and tensor to scalar ratio are more or less similar to the
left panel of figures~\ref{fig16} and~\ref{fig17} and for all
$l>0.63$ the behavior of $n_{s}$ and $r_{t-s}$ are similar to the
right panel of figures~\ref{fig16} and~\ref{fig17} respectively.

Figure~\ref{fig18} shows the evolution of the tensor to scalar ratio
with respect to the scalar spectral index in the background of
WMAP9+eCMB+BAO+H$_{0}$ data for $N=50$, $N=60$ and $N=70$. As figure
shows, with $l=0.87$ (the right panel), for all values of $\xi$, the
tensor to scalar ratio and scalar spectral index are compatible with
observational data. With $l=0.4$, only in some range of the
non-minimal coupling parameter, $n_{s}$ and $r_{t-s}$ lie in the
region that observational data shows. One can see the corresponding
range of $\xi$ for three values of $N$ in table~\ref{tab5}.

\subsection{Non-Gaussianity}

To study the non-gaussianity in this model we should keep the
gradient term in the equation of motion of the scalar field. So, in
the presence of the non-minimal coupling we have
\begin{equation}
\label{3:46-9}
3H\dot{\phi}=a^{-2}\,\nabla^{2}\phi-\frac{V'}{V}+\frac{f'R}{2V}+\Xi\,.
\end{equation}
By expanding equation~\eqref{3:46-9} up to the second order in the
fluctuations of the tachyon field
($\delta\phi=\delta\phi^{(1)}+\delta\phi^{(2)}$) and in a time
interval $t_{n}-t_{n-1}=\frac{1}{H}$, we obtain the following
expression for the first and second order fluctuations
\begin{eqnarray}
\label{3:46-10}
\frac{d}{dt}\Big(\delta\phi^{(1)}(\textbf{k},t)\Big)=\frac{1}{3H}\Bigg\{\bigg[-k^{2}-\frac{2V''+f''R}{2V}
+\frac{V'^{2}}{V^{2}}\hspace{0.1cm}\nonumber\\
-\frac{f'RV'}{2V^{2}}\bigg]\delta\phi^{(1)}+\Xi\Bigg\}\,,\hspace{1cm}
\end{eqnarray}
and
\begin{eqnarray}
\frac{d}{dt}\Big(\delta\phi^{(2)}(\textbf{k},t)\Big)=\frac{1}{3H}\Bigg\{\bigg[-k^{2}-\frac{2V''+f''R}{2V}
+\frac{V'^{2}}{V^{2}}\hspace{0.1cm}\nonumber\\
-\frac{f'RV'}{2V^{2}}\bigg]\delta\phi^{(2)}+\bigg[\frac{3V'V''}{V^{2}}-\frac{V'''}{V}
-\frac{2V'^{3}}{V^{3}}-\frac{f'''R}{2V}
\hspace{0.1cm}\nonumber\\
+\frac{f''V'R}{V^{2}}+\frac{f'RV''}{2V^{2}}-\frac{f'RV'^{2}}{V^{3}}
\bigg]
\Big(\delta\phi^{(1)}\Big)^{2}\Bigg\}\,.\hspace{1cm}\label{3:46-11}
\end{eqnarray}
The solution of the equations~\eqref{3:46-10} and~\eqref{3:46-11}
are given by
\begin{eqnarray}
\delta\phi^{(1)}(\textbf{k},t)=\Theta(k,t-t_{n-1})\int_{t_{n-1}}^{t}dt'\frac{\Xi}{3H}\Theta^{-1}(k,t-t_{n-1})
\hspace{0.1cm}\nonumber\\
+\Theta(k,t-t_{n-1})\delta\phi^{(1)}(\textbf{k}e^{-H(t_{n}-t_{n-1})},t_{n-1})\,,\hspace{1cm}
\label{3:46-12}
\end{eqnarray}
and
\begin{widetext}
\begin{eqnarray}
\delta\phi^{(2)}(\textbf{k},t)=\Theta(k,t-t_{n-1})\int_{t_{n-1}}^{t}dt'\chi(k,t')
\Bigg[\int\frac{dp^{3}}{(2\pi)^{3}}\delta\phi^{(1)}(\textbf{p},t')\delta\phi^{(1)}(\textbf{k}-\textbf{p},t')\Bigg]
\Theta^{-1}(k,t'-t_{n-1})\hspace{1cm}\nonumber\\
+\Theta(k,t-t_{n-1})\delta\phi^{(2)}(\textbf{k}e^{-H(t_{n}-t_{n-1})},t_{n-1})\,,\hspace{1cm}
\label{3:46-13}
\end{eqnarray}
\end{widetext}
where
\begin{eqnarray}
\label{3:46-14}
\Theta=\exp\Bigg[-\int_{t_{0}}^{t}\bigg(\frac{k^{2}}{3H}+\frac{2V''+f''R}{6HV}
-\frac{V'^{2}}{3HV^{2}}\hspace{1cm}\nonumber\\-\frac{f'RV'}{6HV^{2}}\bigg)dt'\Bigg]\,,\hspace{1cm}
\end{eqnarray}
and
\begin{eqnarray}
\label{3:46-15}\chi=\frac{2V'V''}{HV^{2}}-\frac{2V'''}{3HV}
-\frac{4V'^{3}}{3HV^{3}}
-\frac{f'''R}{3HV}+\frac{2f''V'R}{3HV^{2}}\hspace{1cm}\nonumber\\
+\frac{f'RV''}{3HV^{2}}-\frac{2f'RV'^{2}}{3HV^{3}}\,.\hspace{1cm}
\end{eqnarray}
In the presence of the non-minimal coupling between the tachyon
field and the Ricci scalar, the freeze–out momentum $k_{F}$ is
expressed by the following condition
\begin{equation}
\label{3:46-16} \frac{k_{F}^{2}}{3H^{2}}+\frac{2V''+f''R}{6H^{2}V}
-\frac{V'^{2}}{3H^{2}V^{2}}-\frac{f'RV'}{6H^{2}V^{2}}=1\,.
\end{equation}
So, we find
\begin{equation}
\label{3:46-17}k_{F}=\sqrt{\frac{6H^{2}V^{2}-(2V''+f''R)V
+2V'^{2}+f'RV'}{2V^{2}}}\,.
\end{equation}
Now, we compute the three-point correlation function of the tachyon
fluctuations at large scale, at the time about 50 e-folds before the
end of inflation, by using equations~\eqref{3:46-12}
and~\eqref{3:46-13}. The result is as follows
\begin{eqnarray}
\langle\delta\phi(\textbf{k}_{1},t)\,\delta\phi(\textbf{k}_{2},t)\,\delta\phi(\textbf{k}_{3},t)\rangle=
\Theta(k_{3},t-t_{50}-\frac{1}{H})\times
\hspace{0.2cm}\nonumber\\
\int_{t_{50}-\frac{1}{H}}^{t_{50}}dt'
\Theta^{-1}(k_{3},t'-t_{50}-\frac{1}{H})\chi(k_{3},t')
\hspace{2cm}\nonumber\\
\Bigg[\int\frac{d^{3}p}{(2\pi)^{3}}\langle\delta\phi^{(1)}(\textbf{k}_{1},t_{1})\,
\delta\phi^{(1)}(\textbf{p},t')\rangle\hspace{2cm}\nonumber\\
\langle\delta\phi^{(1)}(\textbf{k}_{2},t_{2})\delta\phi^{(1)}(\textbf{k}_{3}-\textbf{p},t')\rangle\Bigg]
\hspace{2cm}\nonumber\\
+\Theta(k_{3},t-t_{50}-\frac{1}{H})\langle\delta\phi^{(1)}(\textbf{k}_{1},t_{50})
\,\delta\phi^{(1)}(\textbf{k}_{2},t_{50})\,\hspace{0.2cm}\nonumber\\
\delta\phi^{(1)}(\textbf{k}_{3}e^{-1},t_{50}-\frac{1}{H})\rangle
+(\textbf{k}_{1}\leftrightarrow
\textbf{k}_{3})+(\textbf{k}_{2}\leftrightarrow
\textbf{k}_{3})\hspace{1cm} \label{3:46-18}
\end{eqnarray}
On the scales with $k<k_{F}$, we can take $\Theta$ about unity and
$\chi$ as a constant. So we have
\begin{eqnarray}
\langle\delta\phi(\textbf{k}_{1},t)\,\delta\phi(\textbf{k}_{2},t)\,\delta\phi(\textbf{k}_{3},t)\rangle\approx
\chi(k_{F},t_{F})\frac{1}{H}\ln\Big(\frac{k_{F}}{H}\Big)\times
\hspace{0.2cm}\nonumber\\
\Bigg[\int\frac{d^{3}p}{(2\pi)^{3}}\langle\delta\phi^{(1)}(\textbf{k}_{1},t_{1})\,
\delta\phi^{(1)}(\textbf{p},t')\rangle\hspace{2cm}\nonumber\\
\langle\delta\phi^{(1)}(\textbf{k}_{2},t_{2})\delta\phi^{(1)}(\textbf{k}_{3}-\textbf{p},t')\rangle
\hspace{2cm}\nonumber\\
+(\textbf{k}_{1}\leftrightarrow
\textbf{k}_{3})+(\textbf{k}_{2}\leftrightarrow
\textbf{k}_{3})\Bigg]\hspace{1cm} \label{3:46-19}
\end{eqnarray}
The bispectrum for the single field inflation models in the
slow-roll limit is given by the following expression
\begin{eqnarray}
\label{3:46-20}
\langle\Phi(\textbf{k}_{1})\,\Phi(\textbf{k}_{2})\,\Phi(\textbf{k}_{3})\rangle=
2f_{NL}(2\pi)^{3}
\delta^{3}(\textbf{k}_{1}+\textbf{k}_{2}+\textbf{k}_{3})\hspace{0.2cm}\nonumber\\
\big[P_{\Phi}(\textbf{k}_{1})+P_{\Phi}(\textbf{k}_{2})+
permutations\big].\hspace{1cm}
\end{eqnarray}
On the other hand, the gravitational field potential, in the
presence of the non-minimal coupling is given by
\begin{equation}
\label{3:46-20-2}
\Phi=-\frac{1}{2}\frac{H}{\dot{\phi}}\frac{\left[\Sigma+\frac{\kappa^{2}f'\Upsilon}
{1-\kappa^{2}f}\right]^{-1}}{(1-\frac{f'R}{6HV})}\delta\phi
\end{equation}
where
\begin{equation}
\label{3:46-20-3}
\Sigma=\frac{1}{3}-\frac{2f'f''RV-f'^{2}RV'}{6V'V^{2}}-\frac{f''V+f'\frac{V''V}{V'}-f'V'}{3V^{2}}\,,
\end{equation}
and
\begin{equation}
\label{3:46-20-4}
\Upsilon=\frac{V}{3V'}-\frac{f'^{2}R}{6VV'}-\frac{f'}{3V}
\end{equation}

So, from equations~\eqref{3:46-20-2},~\eqref{3:46-19}
and~\eqref{3:46-20} we find
\begin{eqnarray}
f_{NL}=2\,\frac{\dot{\phi}}{H}\left(1-\frac{f'R}{6HV}\right)\left(\Sigma+\frac{\kappa^{2}f'\Upsilon}
{1-\kappa^{2}f}\right)\hspace{1cm}
\hspace{0.2cm}\nonumber\\
\label{3:46-21}\times\Bigg[\frac{1}{H}
\ln\bigg(\frac{k_{F}}{H}\bigg)\bigg(\frac{2V'V''}{HV^{2}}-\frac{2V'''}{3HV}
-\frac{4V'^{3}}{3HV^{3}}\hspace{1cm}\nonumber\\
-\frac{f'''R}{3HV}+\frac{2f''V'R}{3HV^{2}}+\frac{f'RV''}{3HV^{2}}-\frac{2f'RV'^{2}}{3HV^{3}}
\bigg)\Bigg]\hspace{1cm}
\end{eqnarray}
\begin{figure*}
\flushleft\leftskip0em{
\includegraphics[width=.31\textwidth,angle=0]{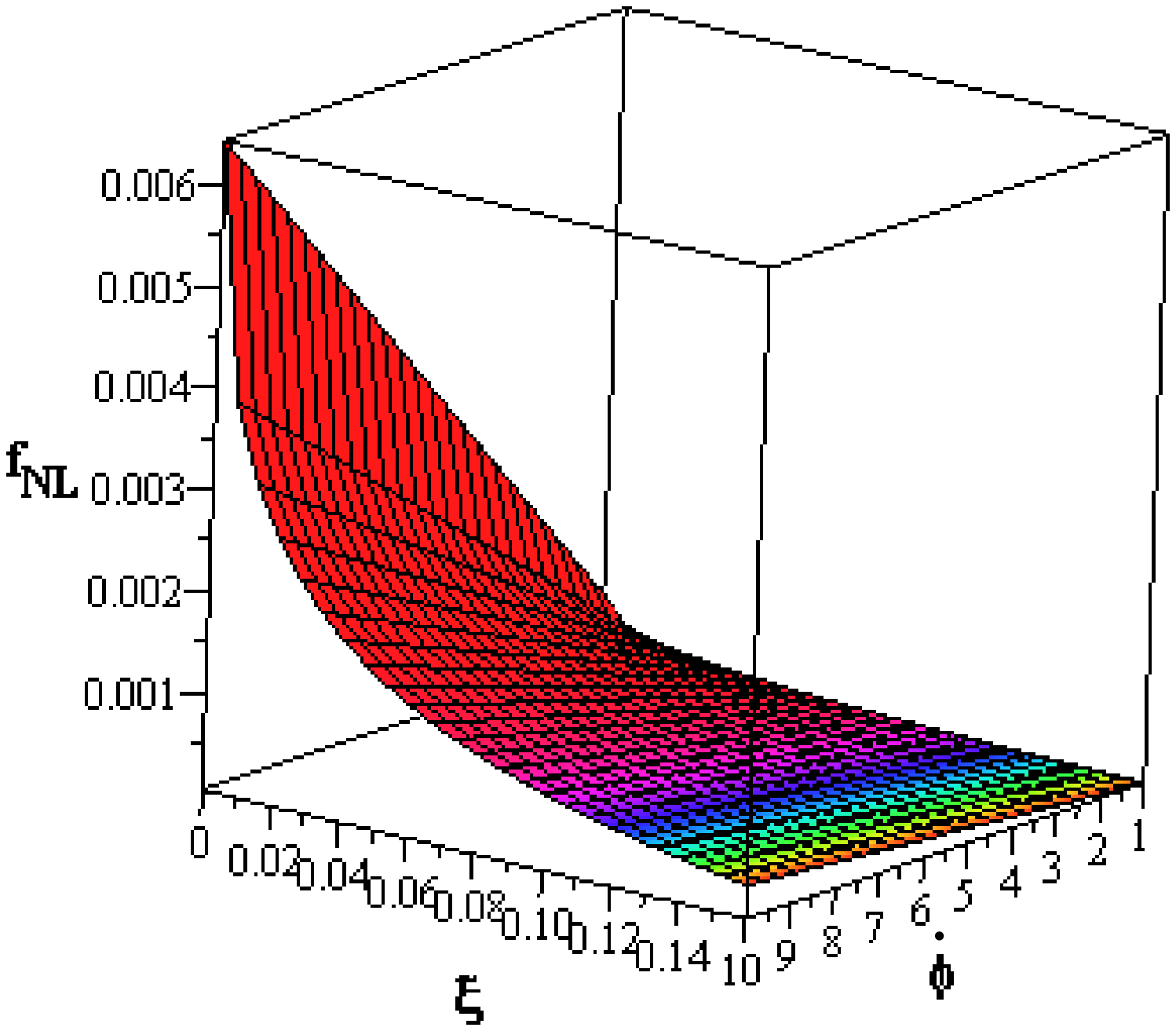}
\includegraphics[width=.36\textwidth,angle=0]{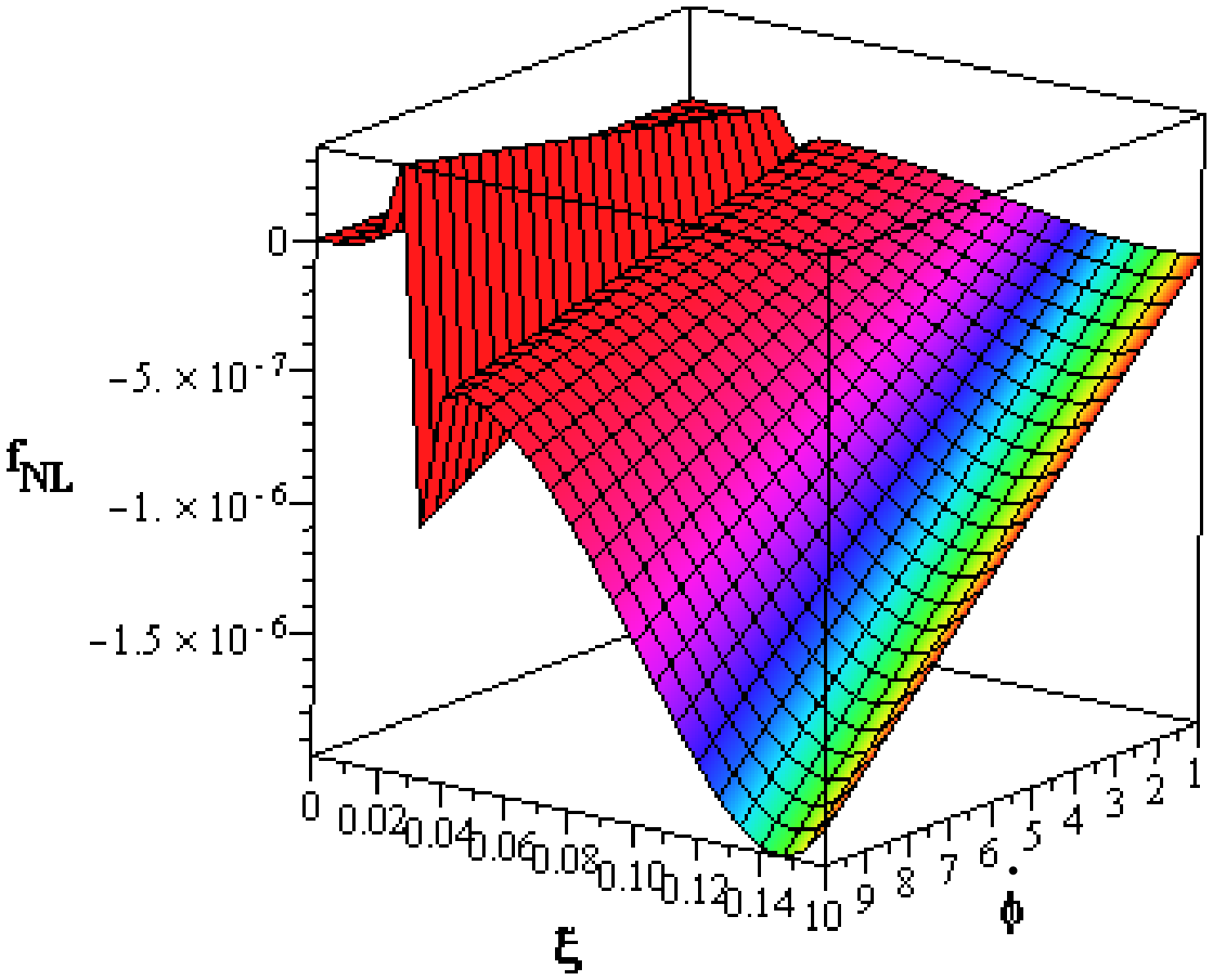}
\includegraphics[width=.31\textwidth,angle=0]{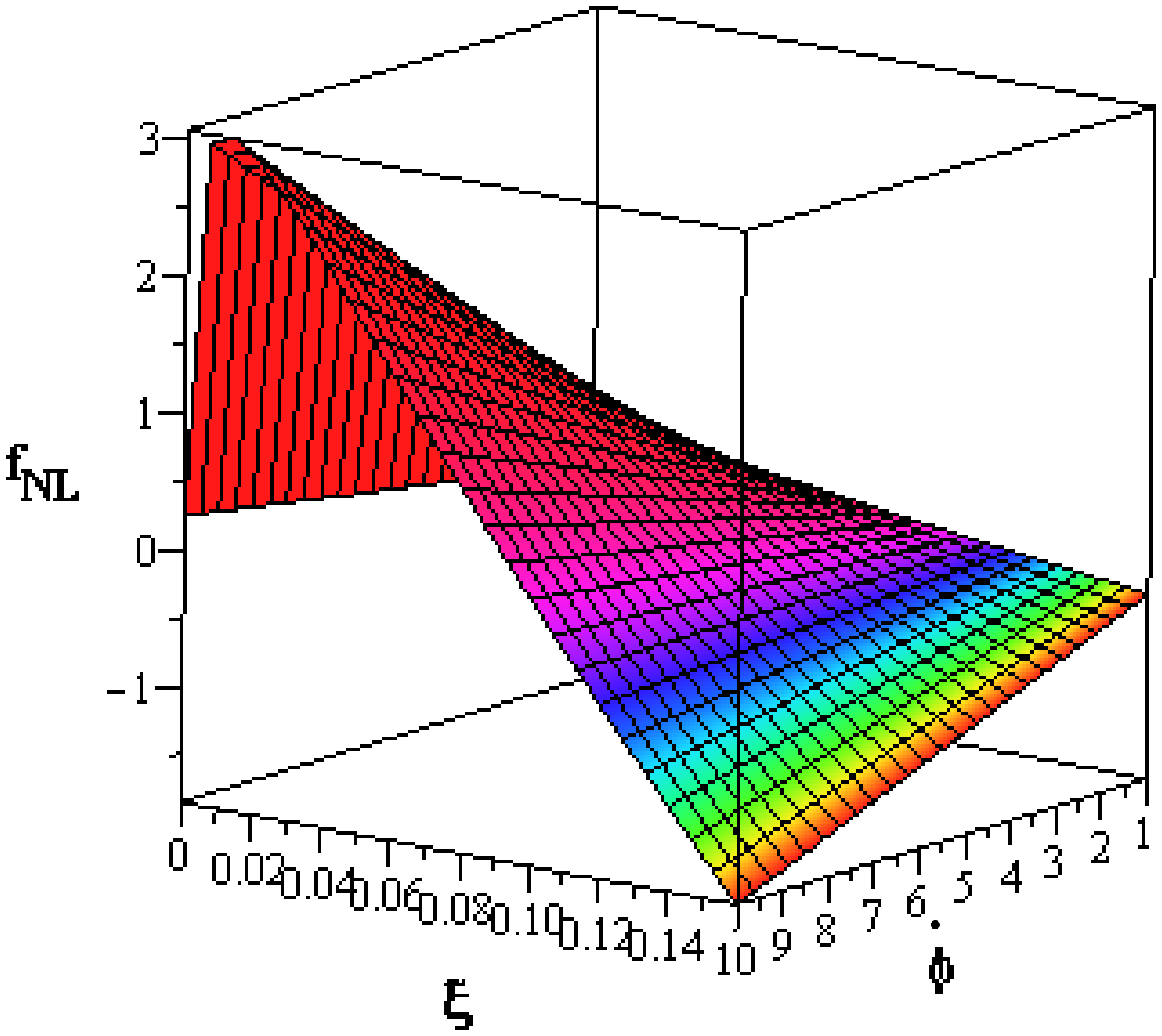}}
\caption{\label{fig3-3-1} Non-gaussianity versus $\xi$ and
$\dot{\phi}$ for $N=50$ for quadratic (left panel), exponential
(middle panel) and intermediate potential (right panel) in the
non-minimal coupling setup.}
\end{figure*}
In figure~\ref{fig3-3-1} we have plotted the behavior of $f_{NL}$
versus $\dot{\phi}$ and $\xi$ for $N=50$ and for three types of
potentials: quadratic, exponential and intermediate potential. As
the figure shows, for quadratic potential, similar to the minimal
case, $f_{NL}$ is positive. In this case, the value of
non-gaussianity parameter decreases as the non-minimal coupling
increases. With an exponential potential, the non-gaussianity
parameter can be either positive or negative, depending on the value
of $\xi$. In this case, $f_{NL}$ decreases as the non-minimal
coupling parameter increases. But, at some value of $\xi$ it reaches
a minimum and then increases by more increment of the non-minimal
coupling. With an intermediate potential, depending on the value of
the non-minimal coupling, the non-gaussianity parameter is either
positive or negative (note that in the minimal coupling case,
$f_{NL}$ for an intermediate potential is always positive). With
this potential, $f_{NL}$ increases as the non-minimal coupling
parameter increases until it reaches a maximum and then decreases as
$\xi$ increases. For intermediate potential we have set $l=0.4$.

\subsection{Cosmological dynamics and late time acceleration}

In this section, similar to the minimal coupling case, we study
cosmological dynamics of a non-minimally coupled tachyon field by
using the dynamical system technique. In the presence of a
non-minimally coupled tachyon field and ordinary component(s), the
Friedmann equation of our setup takes the following form
\begin{equation}
\label{3:47}
\big(1-\kappa^{2}f\big)H^{2}=\frac{\kappa^{2}}{3}\Bigg[\frac{V(\phi)}{\sqrt{1-\dot{\phi}^{2}}}+3f'H\dot{\phi}+\rho_{m}\Bigg]\,.
\end{equation}
Now, as usual we should define some dimensionless variables in order
to translate our cosmological equations in the language of the
autonomous dynamical system. In this non-minimal coupling framework,
we use the dimensionless parameters defined in \eqref{47} and in
addition to two new parameters defined as follows
\begin{equation}
\label{3:48} z=\kappa\sqrt{\xi}\phi\,\,,\quad\quad
u=\frac{\kappa\sqrt{\xi}}{H}
\end{equation}
With these dimensionless parameters, the Friedmann equation
\eqref{3:47} can be written as follows
\begin{equation}
\label{3:49}
1=\frac{y^{2}}{\sqrt{1-x^{2}}}+2\,u\,z\,x+z^{2}+d^{2}\,,
\end{equation}
which gives a constraint on the parameters of the model. As usual
this constraint can be used to express one of the dimensionless
parameters in terms of the others.

In the presence of a non-minimally coupled tachyon field and in
terms of the new variables, the acceleration equation takes the
following form
\begin{eqnarray}
\frac{\dot{H}}{H^{2}}=\Bigg[-\frac{3}{2}\frac{y^{2}x^{2}}{\sqrt{1-x^{2}}}-uzx+\frac{3}{2}y^{2}\sqrt{1-x^{2}}
\hspace{1.4cm}\nonumber\\-\big(1-x^{2}\big)\big(3uzx+\sqrt{3}uzy\alpha+4u^{2}z^{2}y^{2}\big)+u^{2}x^{2}
\hspace{1cm}\nonumber\\
-\frac{3}{2}d^{2}\big(1+\omega\big)\Bigg]\times
\Bigg[1-z^{2}+2\frac{u^{2}z^{2}}{2}\big(1-x^{2}\big)^{\frac{3}{2}}\Bigg]^{-1}\,,\hspace{1cm}
\label{3:50}
\end{eqnarray}
and the scalar field's equation of motion becomes
\begin{equation}
\label{3:51}
\frac{\ddot{\phi}}{H}=\bigg(1-x^{2}\bigg)\bigg(-3x-\sqrt{3}\alpha
y\bigg)-\frac{2uz}{y^{2}}\bigg(1-x^{2}\bigg)^{\frac{3}{2}}\bigg(\frac{\dot{H}}{H^{2}}+2\bigg).
\end{equation}
Also, the effective equation of state parameter in terms of the
dimensionless parameters is given by the following expression
\begin{widetext}
\begin{equation}
\label{3:52} \omega_{eff}=\frac{\frac{\sqrt{3}\,y^{2}\alpha
x}{\sqrt{1-x^{2}}}-\frac{y^{2}x}{(1-x^{2})^{\frac{3}{2}}}\frac{\ddot{\phi}}{H}-2u^{2}x^{2}-2uz\frac{\dot{H}}{H^{2}}
-2uz\frac{\ddot{\phi}}{H}-2uzx-2z^{2}\frac{\dot{H}}{H^{2}}}
{\frac{3y^{2}}{\sqrt{1-x^{2}}}+6uzx+3z^{2}}-1\,.
\end{equation}
\end{widetext}

In the next step, we obtain the autonomous equations with a
non-minimally coupled tachyon field
\begin{equation}
\label{3:53}
\frac{dx}{d\tau}=\bigg(1-x^{2}\bigg)\bigg(-3x-\sqrt{3}\alpha
y\bigg)-\frac{2uz\big(1-x^{2}\big)^{\frac{3}{2}}}{y^{2}}\bigg(\frac{\dot{H}}{H^{2}}+2\bigg),
\end{equation}
\begin{equation}
\label{3:54} \frac{dy}{d\tau}=\frac{\sqrt{3}}{2}y^{2}\alpha\,
x-y\,\frac{\dot{H}}{H^{2}}\,,
\end{equation}
\begin{equation}
\label{3:55} \frac{dz}{d\tau}=u\,z\,,
\end{equation}
and
\begin{equation}
\label{3:56} \frac{du}{d\tau}=-u\,\frac{\dot{H}}{H^{2}}\,,
\end{equation}

\begin{table*}
\caption{\label{tab6} Location, Eigenvalues and Dynamical Characters
of the Critical Lines.}
\begin{ruledtabular}
\begin{tabular}{cccccc}
Line & ($x$,$y$,$z$,$u$) &eigenvalues & $\omega_{eff}$&stability\\
\hline ${\cal{L}}_{1}$ &
($x_{{\cal{L}}_{1}}$,$y_{{\cal{L}}_{1}}$,$z$,0) &($0$,\,
$\lambda_{2{\cal{L}}_{1}}$,\,$\lambda_{3{\cal{L}}_{1}}$,\,$\lambda_{4{\cal{L}}_{1}}$)& $\omega_{{\cal{L}}_{1}}$& stable\\
\hline ${\cal{L}}_{2}$ & (0,0,$z$,0) &($0$,\,
$\lambda_{{\cal{L}}}$,\,$\lambda_{{\cal{L}}}$,\,$\lambda_{{\cal{L}}}$)& $\omega$ & unstable \\
\hline ${\cal{L}}_{3}$ & (1,0,$z$,0) &($0$,\,
$6$,\,$\lambda_{{\cal{L}}}$,\,$\lambda_{{\cal{L}}}$)& $\omega$ & unstable \\
\hline ${\cal{L}}_{4}$ & (-1,0,$z$,0) &($0$,\,
$6$,\,$\lambda_{{\cal{L}}}$,\,$\lambda_{{\cal{L}}}$)& $\omega$ & unstable \\
\hline
 ${\cal{L}}_{5}$ &
($-x_{{\cal{L}}_{1}}$,$-y_{{\cal{L}}_{1}}$,$z$,0) &($0$,\,
$\lambda_{2{\cal{L}}_{1}}$,\,$\lambda_{3{\cal{L}}_{1}}$,\,$\lambda_{4{\cal{L}}_{1}}$)& $\omega_{{\cal{L}}_{1}}$& stable\\
\end{tabular}
\end{ruledtabular}
\end{table*}
where $\frac{\dot{H}}{H^{2}}$ is given by equation \eqref{3:50}. By
finding the roots of the system of equations
\eqref{3:53}-\eqref{3:56}, we find 5 critical lines denoted as
${\cal{L}}_{1}$, ${\cal{L}}_{2}$, ${\cal{L}}_{3}$, ${\cal{L}}_{4}$
and ${\cal{L}}_{5}$. The characters of these critical lines are
shown in table~\ref{tab6}. The eigenvalues
$\lambda_{2{\cal{L}}_{1}}$,\,$\lambda_{3{\cal{L}}_{1}}$ and
$\lambda_{4{\cal{L}}_{1}}$ in table 4, are defined in
Appendix~\ref{B} and $\lambda_{\cal{L}}$ is given by the following
expression
\begin{equation}
\label{3:57}
\lambda_{{\cal{L}}}=\frac{3}{2}\,\omega+\frac{3}{2}-\frac{3}{2}\,{z}^{2}-\frac{3}{2}\,{z}^{2}\omega\,.
\end{equation}
Also, we have
\begin{eqnarray}
\label{3:58} x_{{\cal{L}}_{1}}=\frac{2\sqrt{3}}{9}
\bigg(-6-6z^{4}+12z^{2}-3z^{2}\sqrt
{4z^{4}-8z^{2}+13}\hspace{0.01cm}\nonumber\\+3\,\sqrt
{4\,{z}^{4}-8\,{z}^{2}+13}\bigg)^{\frac{1}{2}}\,,\hspace{1cm}
\end{eqnarray}

and
\begin{eqnarray}
\label{3:59} y_{{\cal{L}}_{1}}=\frac{1}{3}
\bigg(-6-6z^{4}+12z^{2}-3z^{2}\sqrt{4z^{4}-8z^{2}+13}\hspace{0.01cm}\nonumber\\
+3\sqrt{4z^{4}-8z^{2}+13}\bigg)^{\frac{1}{2}}\,.\hspace{1cm}
\end{eqnarray}

\begin{figure}
\flushleft\leftskip0em{
\includegraphics[width=.40\textwidth,origin=c,angle=0]{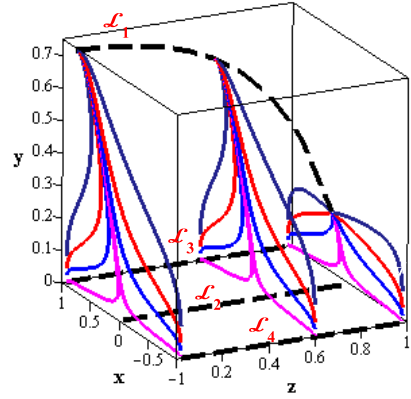}}
\caption{\label{fig19} The phase space trajectories of a model with
ordinary component(s) and a non-minimally coupled tachyon field. We
assumed that, the ordinary component is dust with $\omega=0$. There
is one stable critical line, ${\cal{L}}_{1}$, and 3 unstable
critical lines (${\cal{L}}_{2}$, ${\cal{L}}_{3}$ and
${\cal{L}}_{4}$). The trajectories are plotted for just three values
of $z$: $z(0)=0.02$, $z(0)=0.6$ and $z(0)=0.98$. We have not plotted
the critical line ${\cal{L}}_{5}$ to avoid complexity in the
figure.}
\end{figure}

The critical line ${\cal{L}}_{2}$, which is effectively ordinary
component dominated, is a saddle line. This means that if the
universe during its evolution reaches this state, evolves to other
states. Critical lines ${\cal{L}}_{3}$ and ${\cal{L}}_{4}$ are
unstable (repeller) lines, because their eigenvalues are positive.
These lines are effectively ordinary component dominated solutions
and can be relevant to the early time. The critical lines
${\cal{L}}_{1}$ and ${\cal{L}}_{5}$, which are effectively dark
energy dominated, are stable critical lines, meaning that if the
universe reaches this state, remains there forever. This can be
corresponding to the late time in the history of the universe
expansion. Figure~\ref{fig19} shows the phase space trajectories of
the parameters space in 3 dimensions for $\omega=0$.

As we have seen, for the minimal coupling case the stable point was
not able to explain the late time accelerating phase of the universe
expansion. However, in the non-minimal coupling case the stable
lines ${\cal{L}}_{1}$ and ${\cal{L}}_{5}$ have the capability of
explaining the accelerating phase of the universe expansion. This is
because for all values of $z$, the effective equation of state
parameter, say for ${\cal{L}}_{1}$, which is defined as
\begin{equation}
\label{3:60}
\omega_{{\cal{L}}_{1}}=\frac{1}{9}\,\frac{-4\,{A}^{\frac{3}{2}}{z}^{2}+4\,{A}^{\frac{3}{2}}+{z}^{2}{B}^{2}A-9\,{z}^
{2}A+9\,A}{ \left( z-1 \right)  \left( z+1 \right)  \left(
A+3\,{z}^{2 }B \right) }\,,
\end{equation}
where
\begin{eqnarray}
\label{3:61} A=-6-6\,{z}^{4}+12\,{z}^{2}-3\,{z}^{2}\sqrt
{4\,{z}^{4}-8\,{z}^{2}+13}\hspace{0.2cm}\nonumber\\+3 \,\sqrt
{4\,{z}^{4}-8\,{z}^{2}+13}\,,\hspace{1cm}
\end{eqnarray}
and
\begin{eqnarray}
\label{3:62} B=\bigg(17+8\,{z}^{4}-16z^{2}+4z^{2}\sqrt {4z^{4}-8z^{2
}+13}\hspace{0.2cm}\nonumber\\-4\sqrt{4z^{4}-8z^{2}+13}\bigg)^{\frac{1}{2}}\hspace{1cm}\nonumber\\
\end{eqnarray}
is always below $-\frac{1}{3}$. So, the universe in this case
experiences the late time accelerating phase of expansion. It is
interesting to note that the effective equation of state parameter
of the model shows effective phantom behavior in some subspaces of
the model parameter space.

\begin{figure*}
\flushleft\leftskip1em{
\includegraphics[width=.35\textwidth,origin=c,angle=0]{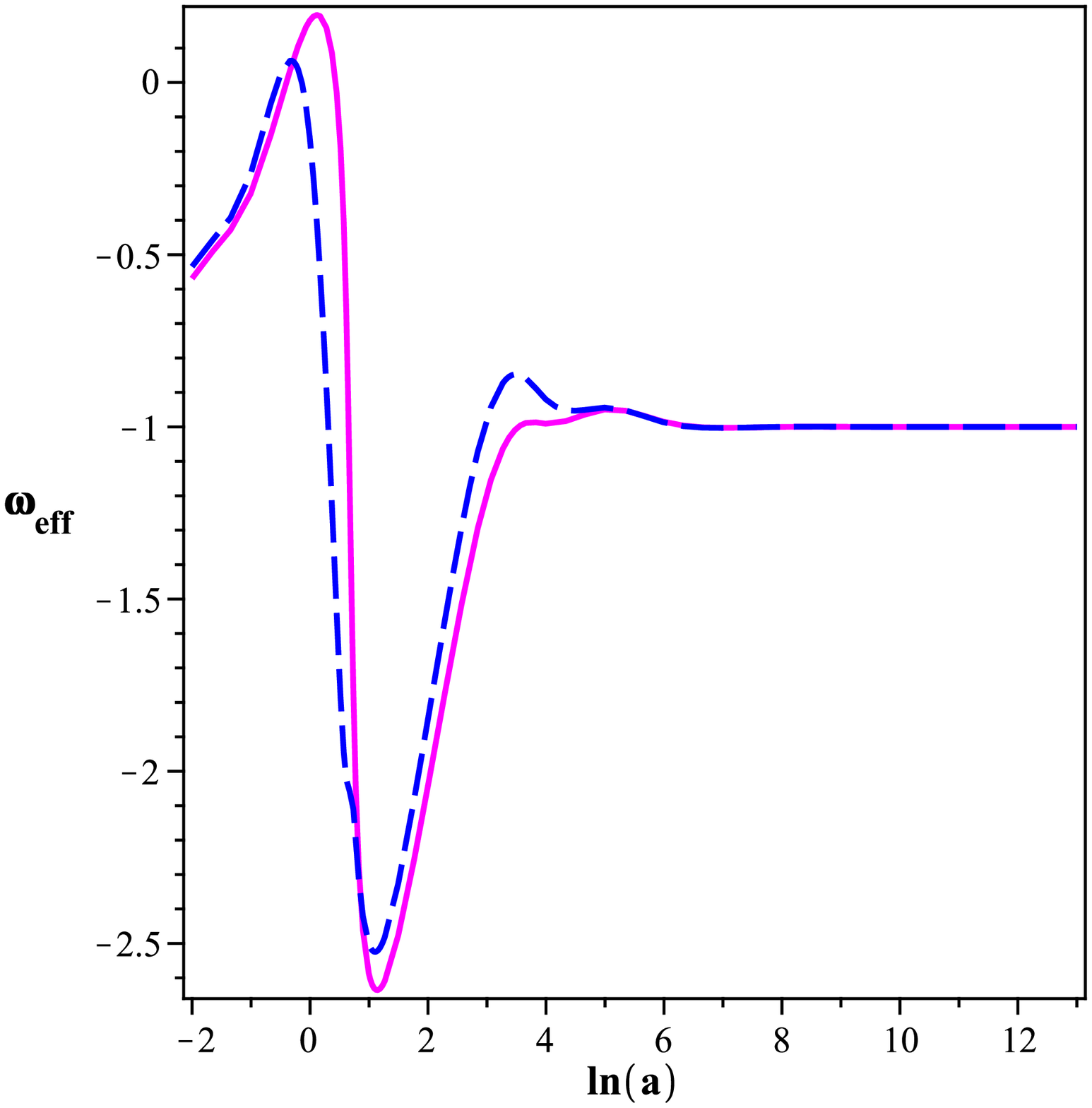}
\hspace{3.6cm}
\includegraphics[width=.35\textwidth,origin=c,angle=0]{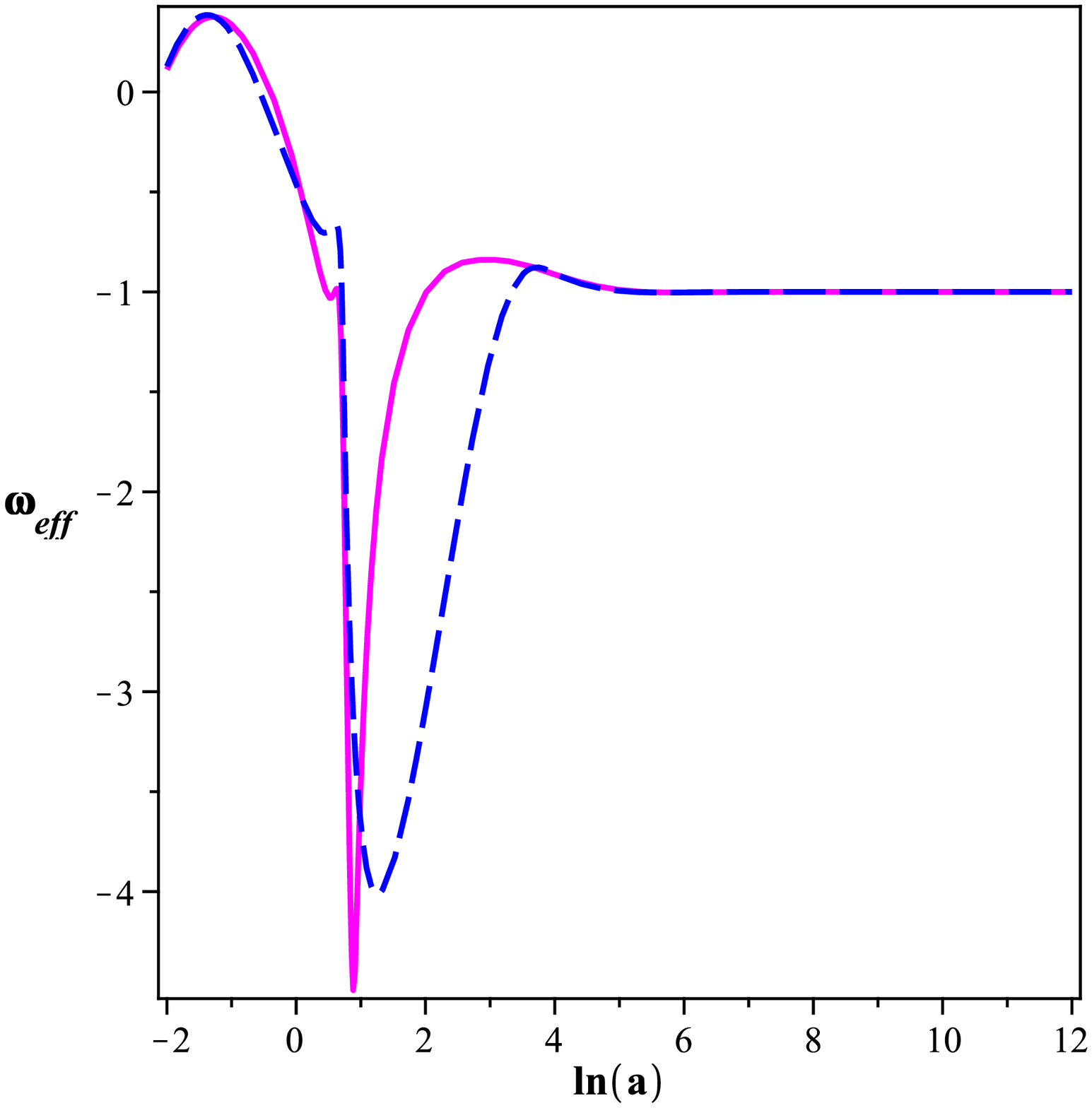}}
\caption{\label{fig20} The behavior of the effective equation of
state parameter in the non-minimal coupling case. The left panel is
corresponding to $\omega=0$ (dust) and the right one is
corresponding to $\omega=\frac{1}{3}$ (radiation). As the figure
shows, $\omega_{eff}$ evolves and tends to the stable state with
$\omega_{eff}=-1$. The figure is plotted with $x(0)=0.99$,
$y(0)=0.01$, $z(0)=0.2$ and $u(0)=0.1$ for magenta curve and
$x(0)=0.75$, $y(0)=0.05$, $z(0)=0.2$ and $u(0)=0.1$ for blue curve.}
\end{figure*}

Figure~\ref{fig20} shows the evolution of the effective equation of
state parameter in the presence of ordinary component(s) and a
non-minimally coupled tachyon field. The effective equation of state
parameter tends to a stable state which is corresponding to the
$\Lambda$CDM scenario. So, the universe with a non-minimally coupled
tachyon field experiences the accelerating phase of expansion in its
history.

Now, we analyze the statefinder parameters diagnostic. As for the
minimal coupling case, we rewrite equation \eqref{56} in terms of
the newly introduced dimensionless parameters as follows
\begin{eqnarray}
\label{3:63}
r=\frac{{\cal{D}}}{{\cal{B}}}-\frac{{\cal{A}}{\cal{K}}}{{\cal{B}}^{2}}+2\frac{{\cal{A}}^{2}}{{\cal{B}}^{2}}
+3\frac{{\cal{A}}}{{\cal{B}}}+1\,,
\end{eqnarray}
where ${\cal{A}}$, ${\cal{B}}$, ${\cal{D}}$ and ${\cal{K}}$ are
defined as
\begin{eqnarray}
\label{3:64}
{\cal{A}}=-\frac{3}{2}\frac{y^{2}x^{2}}{\sqrt{1-x^{2}}}-uzx+\frac{3}{2}y^{2}\sqrt{1-x^{2}}
-\big(1-x^{2}\big)\big(3uzx \hspace{0.2cm}\nonumber\\
+\sqrt{3}uzy\alpha+4u^{2}z^{2}y^{2}\big)+u^{2}x^{2}
-\frac{3}{2}d^{2}\big(1+\omega\big)\,,\hspace{1cm}
\end{eqnarray}

\begin{eqnarray}
\label{3:65}
{\cal{B}}=1-z^{2}+2\frac{u^{2}z^{2}}{2}\big(1-x^{2}\big)^{\frac{3}{2}}\,,
\end{eqnarray}

\begin{eqnarray}
{\cal{D}}=-\frac{3y\frac{dy}{dN}}{\sqrt{1-x^{2}}}-\frac{3}{2}\frac{y^{2}x\frac{dx}{dN}}{(1-x^{2})^{\frac{3}{2}}}
-zx\frac{du}{dN}-ux\frac{dz}{dN}\hspace{1cm}\nonumber\\
-uz\frac{dx}{dN}+3yy'\sqrt{1-x^{2}}-\frac{3}{2}\frac{y^{2}x\frac{dx}{dN}}{\sqrt{1-x^{2}}}
+2ux^{2}\frac{du}{dN}
\hspace{1cm}\nonumber\\
+2x\frac{dx}{dN}\Big(3uzx+\sqrt{3}uzy\alpha+4u^{2}z^{2}y^{2}\Big)
-\Big(1-x^{2}\Big)\hspace{1cm}\nonumber\\
\Big(3zx\frac{du}{dN}+3ux\frac{dz}{dN}
+3uz\frac{dx}{dN}+\sqrt{3}zy\alpha \frac{du}{dN}
\hspace{1.5cm}\nonumber\\
+\sqrt{3}uy\alpha
\frac{dz}{dN}+\sqrt{3}uz\alpha\frac{dy}{dN}+8u\frac{du}{dN}z^{2}y^{2}+8z\frac{dz}{dN}u^{2}y^{2}
\hspace{0.1cm}\nonumber\\
+y\frac{dy}{dN}u^{2}z^{2}\Big)+2u^{2}x\frac{dx}{dN}-3d\frac{d\,d}{dN}(1+\omega)\,,\hspace{1cm}
\label{3:66}
\end{eqnarray}

\begin{eqnarray}
\label{3:67}
{\cal{K}}=+\Big(1-x^{2}\Big)^{\frac{3}{2}}\Big(\frac{4u\frac{du}{dN}z^{2}+4u^{2}z\frac{dz}{dN}}{y^{2}}
-\frac{4u^{2}z^{2}\frac{dy}{dN}}{y^{3}}\Big)\hspace{0.1cm}\nonumber\\
-2z\frac{dz}{dN}-\frac{6u^{2}z^{2}x\frac{dx}{dN}}{y^{2}}\Big(1-x^{2}\Big)^{\frac{1}{2}}\,.
\end{eqnarray}
The parameter $s$, can be obtained by substituting equation
\eqref{3:63} into equation \eqref{55}.

\begin{figure*}
\flushleft\leftskip1em{
\includegraphics[width=.35\textwidth,origin=c,angle=0]{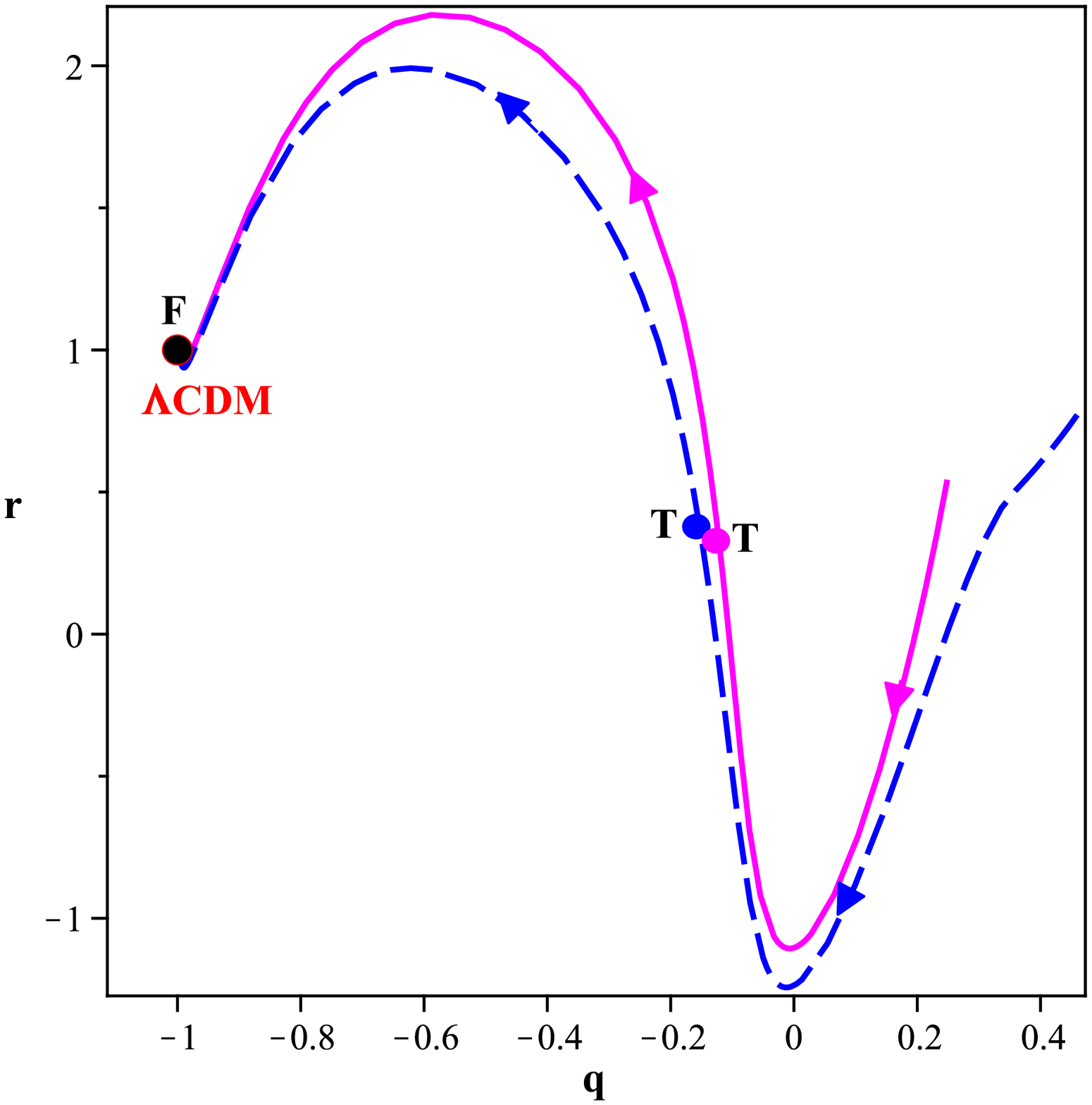}
\hspace{3.6cm}
\includegraphics[width=.35\textwidth,origin=c,angle=0]{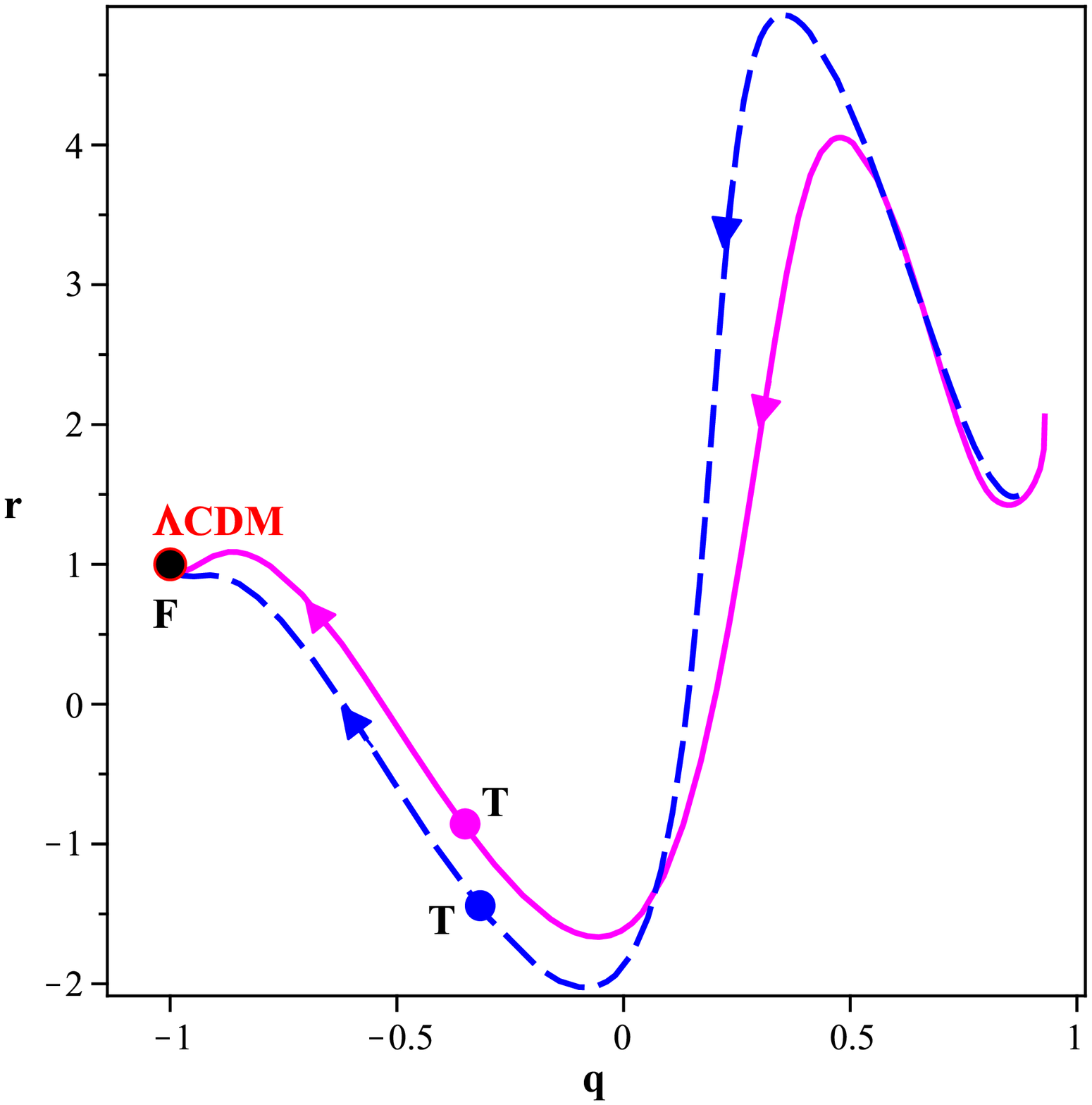}}
\caption{\label{fig21} The trajectories in $\{r,\,q\}$ phase plan
with $\omega=0$ (left panel) and $\omega=\frac{1}{3}$ (right panel).
The initial conditions are $x(0)=0.99$, $y(0)=0.01$, $z(0)=0.2$ and
$u(0)=0.1$ for magenta trajectory and $x(0)=0.75$, $y(0)=0.05$,
$z(0)=0.2$ and $u(0)=0.1$ for blue trajectory. The magenta and blue
highlighted dots (specified by $T$) are current values of
$\{r,\,q\}$ in the model with non-minimally coupled tachyon field.
The black dot (specified by $F$) is the stable state of $\{r,\,q\}$
in future. Also the red dot (marked by $\Lambda$CDM) is the value of
statefinder $\{r,\,q\}$ in a $\Lambda$CDM scenario.}
\end{figure*}
\begin{figure*}
\flushleft\leftskip1em{
\includegraphics[width=.35\textwidth,origin=c,angle=0]{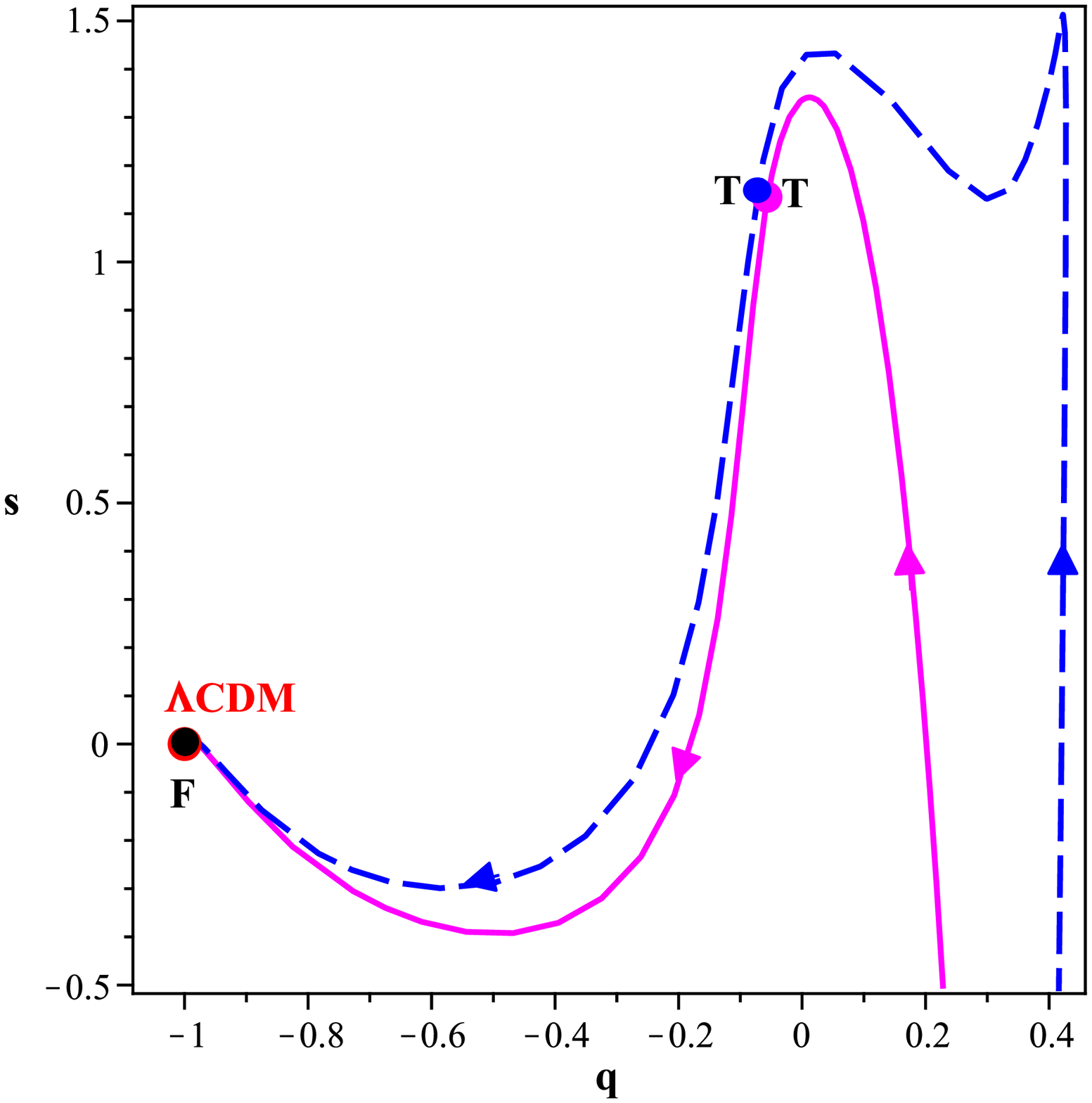}
\hspace{3.6cm}
\includegraphics[width=.35\textwidth,origin=c,angle=0]{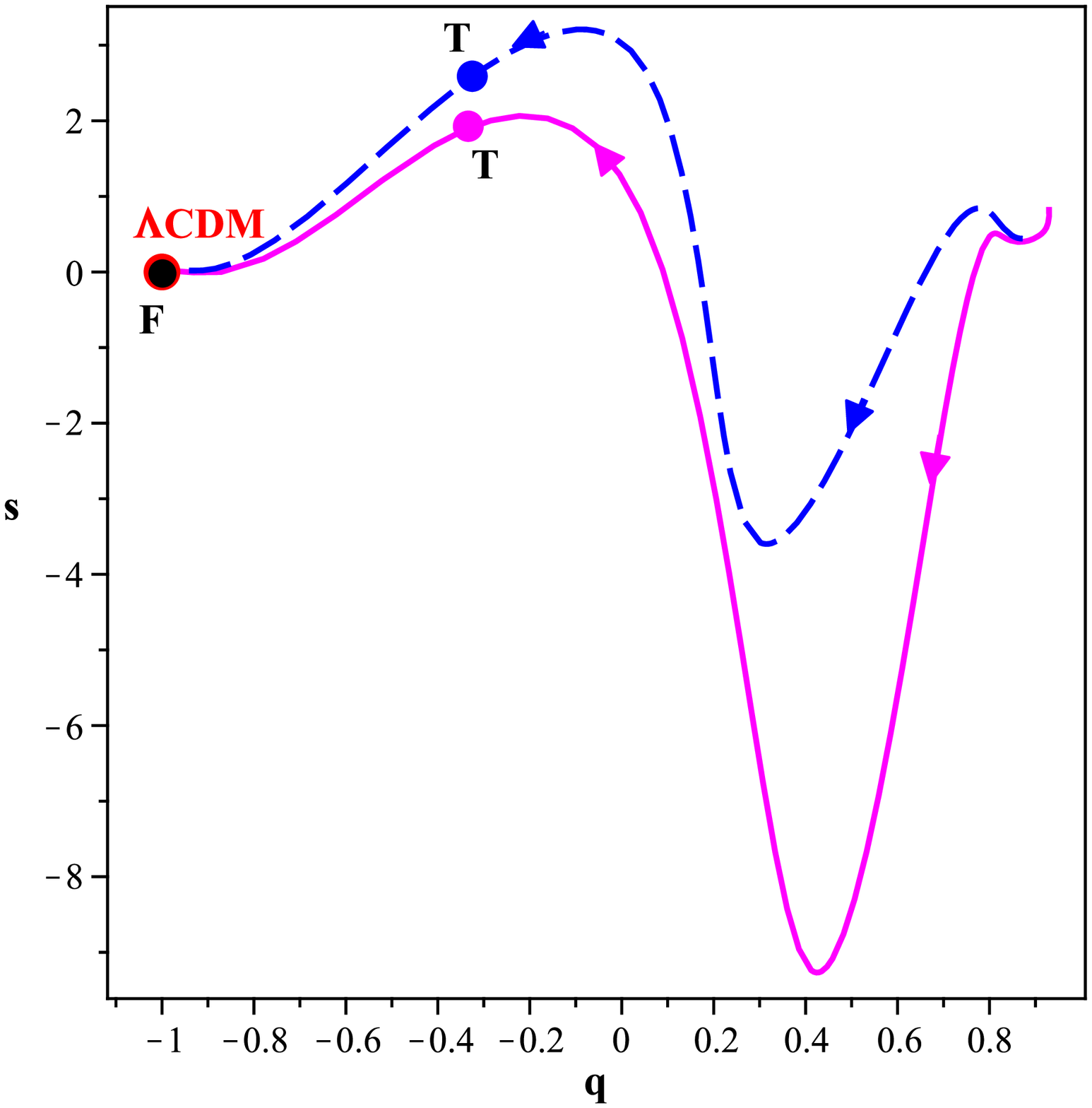}}
\caption{\label{fig22} The trajectories in $\{s,\,q\}$ phase plan
with $\omega=0$ (left panel) and $\omega=\frac{1}{3}$ (right panel).
The initial conditions are $x(0)=0.99$, $y(0)=0.01$, $z(0)=0.2$ and
$u(0)=0.1$ for magenta trajectory and $x(0)=0.75$, $y(0)=0.05$,
$z(0)=0.2$ and $u(0)=0.1$ for blue trajectory. The magenta and blue
dots (specified by $T$) are current values of $\{s,\,q\}$ in the
model with non-minimally coupled tachyon field. The black dot ($F$)
is the stable state of $\{s,\,q\}$ in the future. Also the red dot
(remarked by $\Lambda$CDM) is the value of statefinder $\{s,\,q\}$
in a $\Lambda$CDM scenario.}
\end{figure*}
\begin{figure*}
\flushleft\leftskip1em{
\includegraphics[width=.35\textwidth,origin=c,angle=0]{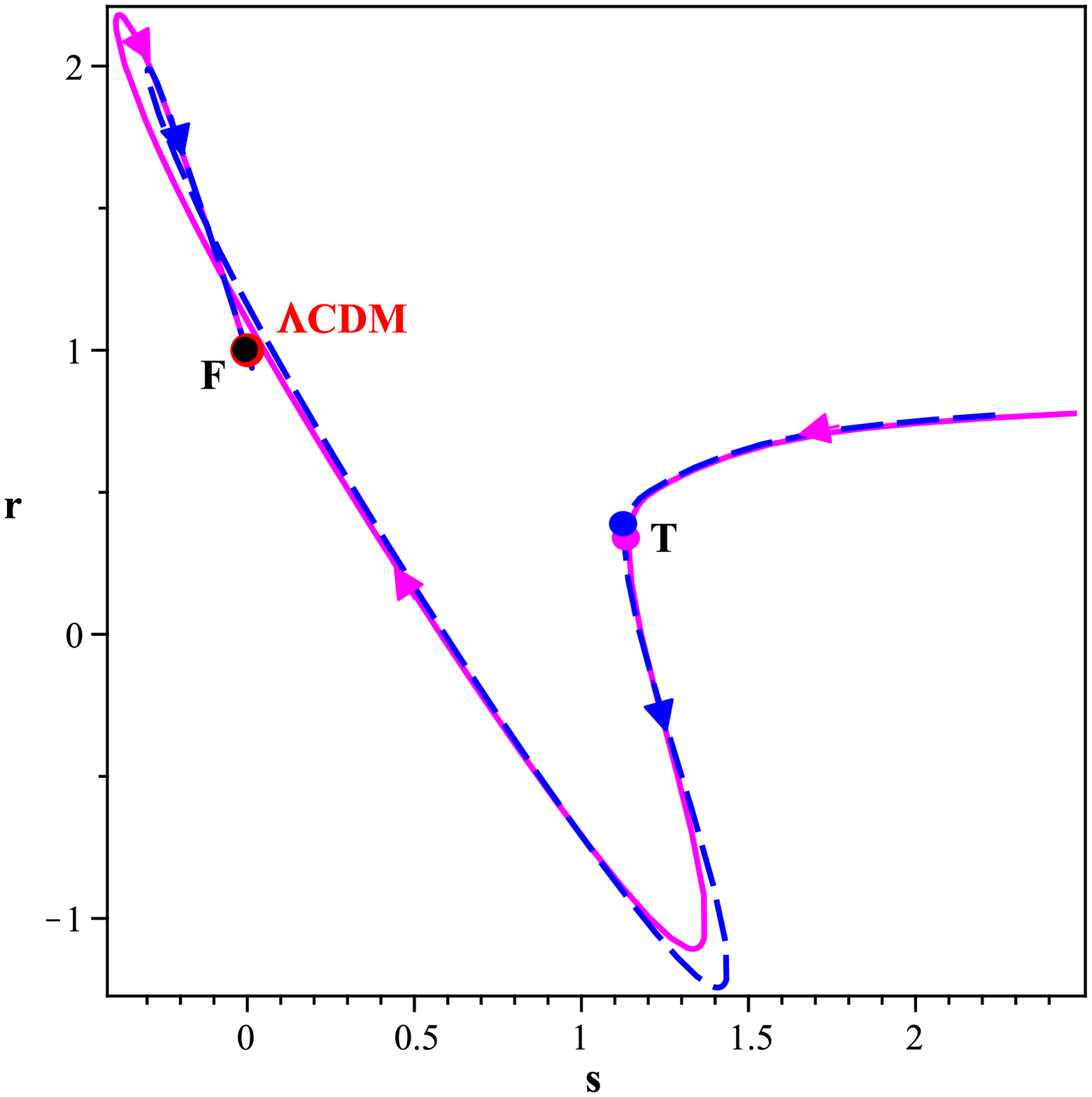}
\hspace{3.6cm}
\includegraphics[width=.35\textwidth,origin=c,angle=0]{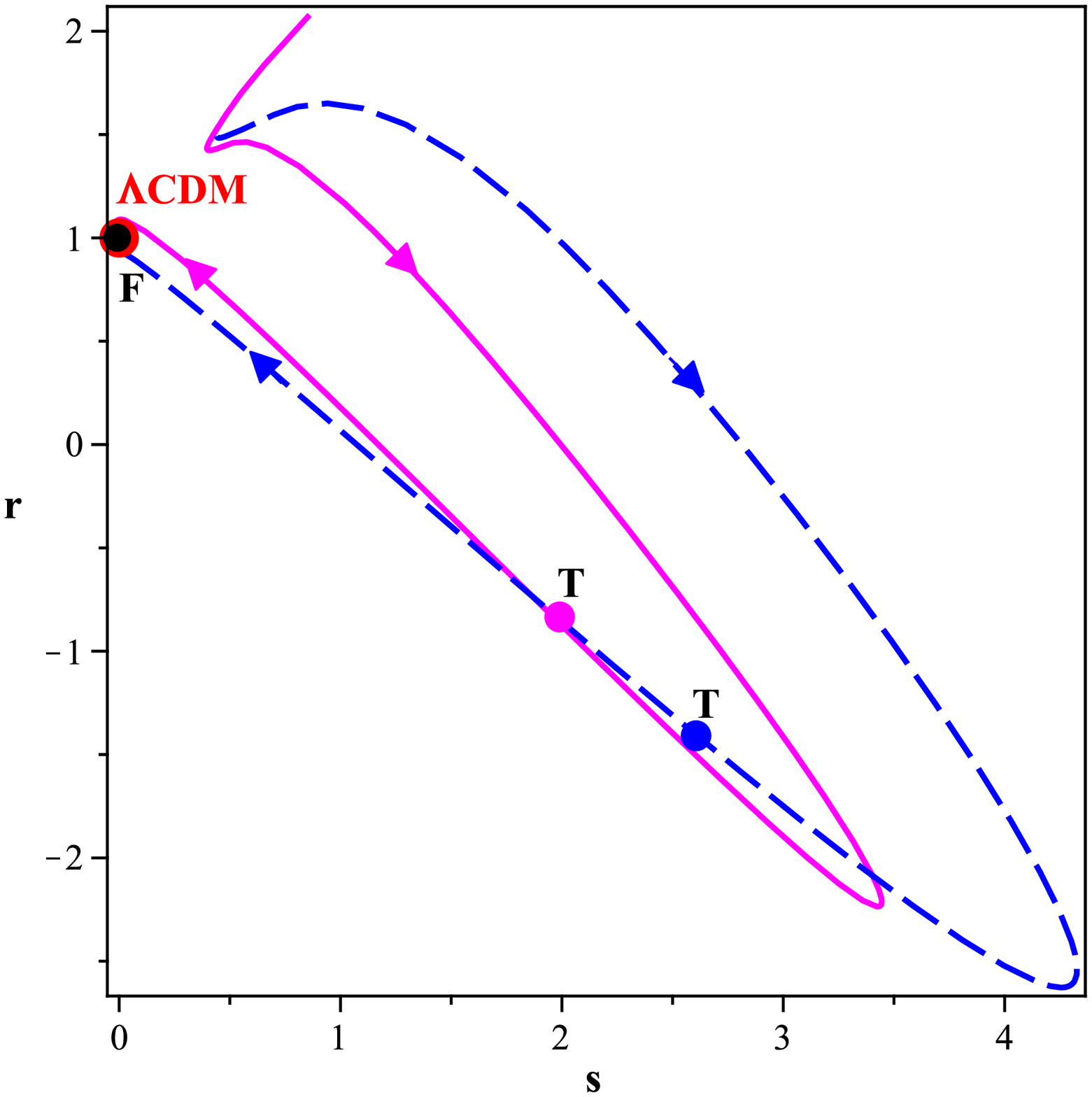}}
\caption{\label{fig23} The trajectories in $\{r,\,s\}$ phase plan
with $\omega=0$ (left panel) and $\omega=\frac{1}{3}$ (right panel).
The initial conditions are $x(0)=0.99$, $y(0)=0.01$, $z(0)=0.2$ and
$u(0)=0.1$ for magenta trajectory and $x(0)=0.75$, $y(0)=0.05$,
$z(0)=0.2$ and $u(0)=0.1$ for blue trajectory. The magenta and blue
highlighted dots (specified by $T$) are current values of
$\{r,\,s\}$ in the model with non-minimally coupled tachyon field.
The black dot ($F$) is the stable state of $\{r,\,s\}$ in the
future. Also red dot (marked by $\Lambda$CDM) is the value of
statefinder $\{r,\,s\}$ in a $\Lambda$CDM scenario.}
\end{figure*}

For numerical analysis of the statefinder diagnostic, we study the
phase plane of the pairs $\{r,\,q\}$, $\{s,\,q\}$ and $\{r,\,s\}$.
The left panel of figure~\ref{fig21} shows the trajectories of the
phase plane $\{r,\,q\}$ for $\omega=0$. Also, the right panel of
this figure shows the trajectories in the phase plan $\{r,\,q\}$ for
$\omega=\frac{1}{3}$. As this figure shows, in a model with ordinary
components (dust or radiation) and a non-minimally coupled tachyon
field, the trajectories of $\{r,\,q\}$ plan terminate in a stable
point in future, which is corresponding to the $\Lambda$CDM model
(with $\{r,\,q\}=\{0,\,-1\}$). As before, magenta and blue
highlighted dots show the current values of the parameters. The
trajectories of phase plane $\{s,\,q\}$ is shown in
figure~\ref{fig22} for $\omega=0$ (left panel) and
$\omega=\frac{1}{3}$ (right panel). In this phase plane also, the
trajectories evolve to a stable point in future which is
corresponding to the $\Lambda$CDM scenario (with
$\{s,\,q\}=\{1,\,-1\}$). Another important pair is $\{r,\,s\}$,
which its phase plane has been plotted in figure~\ref{fig23}. As
other cases, the left panel is corresponding to $\omega=0$ and the
right panel is corresponding to $\omega=\frac{1}{3}$. The
trajectories in phase plan $\{r,\,s\}$ reach a point corresponding
to $\Lambda$CDM model.

\section{Summary}

We have studied cosmological dynamics of a tachyon field as a field
responsible for initial inflation and late time cosmic acceleration.
Firstly, we have considered a tachyon field which is minimally
coupled to gravity and we have explored its dynamic both in the
inflationary stage and late time in the history of the universe. In
the inflationary stage, we have calculated the inflation parameters
and perturbations with details. If there is only one field in
inflation period as responsible for inflation, the adiabatic
perturbations are generated. On the other hand, if there is more
than one scalar field in a model, or a scalar field interacts with
other fields such as the Ricci scalar, the isocurvature
perturbations can be generated. So, in the minimal coupling case,
the perturbation are adiabatic (meaning that $\delta p_{nad}=0$ and
so $\dot{\zeta}=0$). By adopting three types of potential, a
quadratic tachyon potential, an exponential potential and an
intermediate potential, we have performed a numerical analysis of
the model's parameter space and the results have been shown in
figures and tables. Also, we have compared the minimally coupled
tachyon model with observational data by plotting the evolution of
the tensor to scalar ratio with respect to the scalar spectral index
in the background of the WMAP9+eCMB+BAO+H$_{0}$ data. By this
comparison we have found the range of intermediate parameter, $l$,
for which $n_{s}$ and $r_{t-s}$ are compatible with recent
observational data. We have also studied the non-gaussianity of
perturbations for a minimally coupled tachyon field model. As we
have shown, in the minimal coupling case, both with quadratic and
intermediate potentials $f_{NL}$ is positive whereas with the
exponential potential $f_{NL}$ is either positive or negative,
depending on the value of $N$.

Within a dynamical system approach, the cosmological dynamics and
especially late time cosmic evolution of a model with ordinary
component(s) and the minimally coupled tachyon field has been
explored. By finding the roots of the autonomous equations, we have
obtained the critical points of the phase space. We have found three
unstable critical points (which are ordinary component(s) dominated
solutions) and one stable fixed points (which is tachyon field
dominated) in the minimal setup. The effective equation of state
parameter corresponding to the stable point is larger then
$-\frac{1}{3}$. This means that, the minimally coupled tachyon field
model cannot explain the late time acceleration of the universe.
Also, the statefinder diagnostic of the minimal case has been
studied. By plotting the phase plan of the pairs $\{r,q\}$,
$\{s,q\}$ and $\{r,s\}$ we have seen that the trajectories of the
minimal case tend to the state which has distance from $\Lambda$CDM
model.

In the next step, we have considered a tachyon field which is
non-minimally coupled to the Ricci scalar. we have obtained the
inflation parameters and perturbations with details. We have seen
that due to the presence of the non-minimal coupling between the
tachyon field and Ricci scalar, the perturbations are non-adiabatic
and there are isocurvature perturbations in the setup. Actually, in
the presence of the non-minimal coupling there is non-vanishing,
non-adiabatic pressure leading to the non-vanishing time evolution
of the primordial curvature perturbations. By adopting
$f(\phi)=\xi\phi^{2}$, we have numerically analyzed the model's
parameter space in the inflationary stage for three types of
potential. The results have been shown in figures. Note that we have
obtained the intermediate potential in the presence of the
non-minimal coupling of the tachyon field and Ricci scalar, which is
more complicated than the corresponding parameter in the minimal
coupling case. In the non-minimal case, we have plotted the behavior
of the tensor to scalar ratio versus the scalar spectral index in
the background of the WMAP9+eCMB+BAO+H$_{0}$ data, for $\xi\geq 0$.
With a quadratic potential and with $N=50$, $N=60$ and $N=70$, for
all values of $\xi$, the tensor to scalar ratio and scalar spectral
index are compatible with observational data. With an exponential
potential, $n_{s}$ and $r_{t-s}$ are compatible with observational
data in some range of the non-minimal coupling parameter. In the
case of intermediate potential, there are two states. If $l>0.63$,
scalar spectral index and tensor to scalar ratio, for all values of
$\xi$, lie in the region compatible with observational data. If
$l\leq0.63$, the scalar spectral index and tensor to scalar ratio
lie in the viable domain only in some range of the non-minimal
coupling parameter. We have studied the non-gaussianity of
perturbations in this model in the presence of the non-minimal
coupling. In the non-minimal coupling case, both with exponential
and intermediate potentials $f_{NL}$ is either positive or negative,
depending on the value of $\xi$. But, with a quadratic potential the
non-gaussianity parameter is positive and well in the range
supported by recent observations. Similar to the minimal case, the
cosmological dynamics of the setup has been considered with details
in a dynamical system technique. We have found that, there are three
unstable critical lines in the model with ordinary component(s) and
the non-minimally coupled tachyon field. These unstable lines are
effectively ordinary component(s) dominated solutions. In the
non-minimal coupling case, there are two stable critical lines which
are effectively dark energy dominated. In contrast to the minimal
coupling case, a model with a non-minimally coupled tachyon field
has capability to explain the late time accelerating phase of the
universe expansion. By studying the statefinder pairs in phase
plans, we have found that the non-minimal model reaches a stable
state which is corresponding to the $\Lambda$CDM scenario. This
behavior is independent of the initial conditions.

\appendix
\section{Number of e-folds parameter for an intermediate potential in the presence of the non-minimal coupling}
\label{A}
\begin{widetext}
\begin{eqnarray*}
N=\sqrt {\frac{l-1}{\vartheta
l}}\Bigg[\Bigg(\frac{\sqrt{6}\kappa^{2}b^{\beta}(1+\frac{1}{4\beta})}{6\vartheta
l\left(-\kappa^{2}\xi \right) ^{\frac{1}{2}+\beta}}+{\frac
{\sqrt{6}\kappa^{2}(1+\frac{1}{\beta}+b^{\beta})}{144\vartheta^{3}l^{3}\beta\left(-{\kappa}^{2}\xi
\right)^{\frac{1}{2}+\beta}b^{\beta} }}\frac {1}{\left( {\frac
{{b}^{-\beta}}{3\vartheta l \left( l-1 \right) }}+1
\right)}\Bigg)\Bigg(
\frac{1}{2}\grave{{\cal{B}}}\grave{{\cal{O}}}-{\frac
{\grave{{\cal{B}}}}{1+2\,\beta}}\Bigg)
\hspace{2.5cm}\nonumber\\
+\frac{\sqrt{6}\kappa^{4}(1+\frac{1}{16b^{2\beta}})}{36\vartheta^{2}l^{2}\left(l-1\right)
\left(-\kappa^{2}\xi \right) ^{\frac{1}{2}+\beta}}\Bigg( {\frac
{\grave{{\cal{\vartheta}}}\left(
5{\kappa}^{2}\xi{\phi}^{2}+4\,\beta{\kappa}^{2}
\xi{\phi}^{2}-4\beta-3 \right) }{ \left( -1+{\kappa}^{2}\xi{\phi}^{2
} \right) ^{2}}}+\frac{\left(16\beta+3+16 \beta^{2}
\right)\grave{{\cal{\vartheta}}}}{2\vartheta^{2}l^{2}\beta}\grave{{\cal{O}}}
\Bigg)\Bigg]-
\hspace{2.5cm}\nonumber\\
\frac{\kappa^{2}}{6b^{2}l^{2}\left(-\kappa^{2}\xi
\right)^{2\beta}b^{\beta}} \Bigg({\frac {\grave{{\cal{C}}}
\left(\kappa^{2}\xi\phi^{2}-1-2\beta \right)}{\beta
\left(-1+\kappa^{2}\xi\phi^{2}\right)}}-\grave{{\cal{C}}}\left(
2+4\beta\right)\grave{{\cal{O}}}\Bigg)-\frac{\kappa^{2}\phi^{4\beta}(1-2b^{\beta})}{24\vartheta^{2}l^{2}\beta{b}^{\beta}}
\Bigg((6\beta+4\beta^{2}+2)\grave{{\cal{O}}}
\hspace{2.5cm}\nonumber\\
-{\frac { 1-2{\kappa}^{2}\xi\phi^{2}+3\beta+{\kappa}^{4}{\xi}^{2}{
\phi}^{4}-2\beta^{2}{\kappa}^{2}\xi\phi^{2}+2\beta^{2}-4\beta\kappa^{2}\xi\phi^{2}
}{\beta \left(-1+{\kappa}^{2}\xi\phi^{2}\right)^{2}}}\Bigg)-\frac
{{\kappa}^{4}\xi\phi^{8\beta}(\frac{8\beta+8}{\beta})}{54b^{\beta}\vartheta^{4}l^{4}\beta^{-1}}\Bigg((6\beta+1+8\beta^{2}
)\grave{{\cal{O}}}
\hspace{2.5cm}\nonumber\\
+\frac { \left( 1-3\kappa^{2}\xi\phi^{2}+6\beta+8\beta
\phi^{4}\kappa^{4}\xi^{2}+8\beta^{2}\kappa^{4}\xi^{2}\phi^{
4}-16\kappa^{2}\xi\phi^{2}\beta^{2}+3\kappa^{4}\xi^{2}
\phi^{4}+8\beta^{2}-14\beta\kappa^{2}\xi\phi^{2} \right) }{\beta
\left(-1+{\kappa}^{2}\xi\phi^{2}\right)^{3}}\Bigg)
\hspace{2.5cm}\nonumber\\
+{\frac{{\kappa}^{4}\xi\phi^{8\beta}{\vartheta}^{-4}l^{-4}{\beta}^{-1}}{54
\left(4+4\beta \right)\left( \beta+1 \right)
b^{\beta}}}\Bigg({\frac{12\kappa^{2}
\xi\phi^{2}-3+32\beta^{3}{\kappa}^{6}{\xi}^{3}{\phi}^{6}-96\beta^{3}
\kappa^{4}\xi^{2}\phi^{4}+96\beta^{3}\kappa^{2}\xi\phi^{2}
-18\kappa^{4}\xi^{2}\phi^{4}}{\left(-1+{\kappa}^{2}\xi\phi^{2}
\right) ^{4}}}
\hspace{2.5cm}\nonumber\\
+\frac{76\beta
\kappa^{2}\xi\phi^{2}-22\beta-48\beta^{2}-32\beta^{3}-90\beta\phi^{4}\kappa^{4}\xi^{2}-160
\beta^{2}\kappa^{4}\xi^{2}\phi^{4}+152\kappa^{2}\xi\phi^{2}
\beta^{2}+56\beta^{2}\kappa^{6}\xi^{3}\phi^{6}}{ \left(
-1+{\kappa}^{2}\xi\phi^{2} \right)^{4}}
\hspace{2.5cm}\nonumber\\
+\frac{36\beta\phi^{6}\kappa^{6}\xi^{3}+12\kappa^{6}
\xi^{3}\phi^{6}}{ \left( -1+{\kappa}^{2}\xi\phi^{2}
\right)^{4}}+4\beta (\beta+1 )(3+22\beta+48\beta^{2}+32\beta^{3})
\grave{{\cal{O}}}\Bigg)+\Bigg[\frac{\sqrt
{6}\xi{\phi}^{1+8\beta}(7+8\beta)}{2592\vartheta^{4}l^{4}\beta(l-1)b^{2\beta}
}
\hspace{2.5cm}\nonumber\\
\Bigg(\frac{73\kappa^{4}{\xi}^{
2}{\phi}^{4}-55\kappa^{2}\xi\phi^{2}-56\beta-448\beta^{2}-512\beta^{3}+232\beta{\kappa}^{2}\xi\phi
^{2}+1472\kappa^{2}\xi\phi^{2}{\beta}^{2}-360\beta\phi^{4}\kappa^{4}\xi^{2}}{16
\left(7/2+4\beta \right)  \left( -1+{\kappa}^{2}\xi{\phi}^{2}
\right) ^{4}{\kappa}^{6}{\xi}^{3}}
\hspace{2.5cm}\nonumber\\
+\frac{1536\beta^{3}\kappa^{2}\xi\phi ^{2}+15
\kappa^{6}\xi^{3}\phi^{6}-1600\beta^{2}\kappa^{4}\xi^{2}\phi^{4}-1536\beta^
{3}\kappa^{4}\xi^{2}\phi^{4}+576\beta^{2}\kappa^{6}
\xi^{3}\phi^{6}+512\beta^{3}\kappa^{6}\xi^{3}\phi^{6}}{ \left(
7/2+4\beta \right) \left( -1+{\kappa}^{2}\xi\,{\phi}^{2} \right)
^{4}{\kappa}^{6}{\xi}^{3}}
\hspace{2.5cm}\nonumber\\
+\frac{15+184\beta\phi^{6}\kappa^{6}\xi^{3}}{ \left( 7/2+4\beta
\right) ( -1+{\kappa}^{2}\xi\,{\phi}^{2})
^{4}{\kappa}^{6}{\xi}^{3}}+\frac{1}{32}{\frac {\left(
15+184\,\beta+576\,{\beta} ^{2}+512\,{\beta}^{3} \right) (
-1+8\beta)\grave{{\cal{O}}} }{ \left( 7/2+4\,\beta \right)
{\kappa}^{6}{\xi}^{3}}} \Bigg)+\Bigg(\frac{3}{2}\grave{{\cal{D}}}
\hspace{2.5cm}\nonumber\\
(1+8\beta+12\beta^{2}) \grave{{\cal{O}}}-{ \frac { (
5{\kappa}^{2}\xi\phi^{2}+6\beta{\kappa}^{2}\xi\phi^{2}-6\beta-3)\grave{{\cal{D}}}}{
\left( -1+{\kappa}^ {2}\xi\,{\phi}^{2} \right)
^{2}}}\Bigg){\frac{\sqrt
{6}{\kappa}^{4}\xi}{72\vartheta^{3}l^{3}\beta b^{\beta}
\left(-\kappa^{2}\xi \right)^{\frac{1}{2}+3\beta}}}
\Bigg]\sqrt{\frac{\vartheta l}{l-1}}
+\prod+\bigwedge\,,\hspace{1.5cm}
\end{eqnarray*}
\end{widetext}

where
\begin{widetext}
\begin{eqnarray*}
\prod=-\Bigg[{\frac {\sqrt
{6}{\kappa}^{4}\xi}{216b^{5}l^{5}\beta(-\kappa^{2}\xi)^{\frac{1}{2}+3\beta}b^{\beta}}}
\Bigg( { \frac {( 5{\kappa}^{2}\xi\phi^{2}+6\beta{\kappa}^{2}\xi
\phi^{2}-6\beta-3)\grave{{\cal{D}}}}{(-1+{\kappa}^
{2}\xi\phi^{2})^{2}}} +\frac{3}{2}\grave{{\cal{D}}}(
1+8\beta+12\beta^{2}) \grave{{\cal{O}}}\Bigg)+\Bigg(\frac{3}{2}
\hspace{2.5cm}\nonumber\\
(8\beta+1+12\beta^{2})
(6\beta-1)\grave{{\cal{O}}}+\frac{8\kappa^{2}\xi
\phi^{2}+3\kappa^{4}\xi^{2}\phi^{4}+36\beta^{2}-36\beta\kappa^{2}\xi\phi^{2}-72\beta^{2}\kappa^{2}\xi\phi^{2}+24
\kappa^{4}\xi^{2}\phi^{4}\beta}{ \left( -1+{\kappa}^{2}
\xi{\phi}^{2} \right) ^{3}}
\hspace{2.5cm}\nonumber\\
+\frac{36\beta^{2}\kappa^{4}\xi^{2}\phi^{4}-3+12\beta}{ \left(
-1+{\kappa}^{2} \xi{\phi}^{2} \right) ^{3}}\Bigg){\frac {\sqrt
{6}\kappa^{4}\xi^{2}\vartheta^{-
2}l^{-2}\beta^{-1}\grave{{\cal{E}}}}{648\left(-\kappa^{2}\xi
\right)^{\frac{1}{2}+3\beta}}}\Bigg] \frac
{\sqrt{\frac{l-1}{\vartheta l}}}{\left( {\frac {{b}
^{-\beta}}{3\vartheta l \left( l-1 \right) }}+1 \right)}+\frac{\sqrt
{{\frac {l-1}{\vartheta l}}}}{\vartheta^{4}l^{4}\left( l-1
\right)}\Bigg[\Bigg(-\frac{1}{2}(15+92\beta
\hspace{2.5cm}\nonumber\\
+64\beta^{3}+144\beta^{2})
\grave{{\cal{O}}}-\frac{-40\kappa^{2}\xi\phi^{2}+33\kappa
^{4}\xi^{2}\phi^{4}+16\beta^{2}+15-72\beta\kappa^{2}\xi\phi^{2
}-32\beta^{2}\kappa^{2}\xi\phi^{2}+40\kappa^{4}\xi^{2}
\phi^{4}\beta}{\left( -1+\kappa^{2}\xi\phi^{2}\right)^{3}}
\hspace{2.5cm}\nonumber\\
+\frac{16\beta^{2}\kappa^{4}\xi^{2}\phi^{4}+32\beta}{\left(
-1+\kappa^{2}\xi\phi^{2}\right)^{3}}\Bigg)\frac
{\sqrt{6}\kappa^{4}\xi\beta^{-1}\grave{{\cal{F}}}}{108\left(-\kappa^{2}\xi
\right)^{\frac{1}{2}+2 \beta}}+\frac
{\sqrt{6}\kappa^{6}\xi^{2}(\frac{\grave{{\cal{G}}}}{\beta}+\grave{{\cal{H}}})}{1296b^{\beta}
\left(-\kappa^{2}\xi
\right)^{\frac{1}{2}+4\beta}}\Bigg(\frac{1}{2}{\cal{G}}\left(
8\beta-1\right)\grave{{\cal{O}}}(15+184\beta+
\hspace{2.5cm}\nonumber\\
576\beta^{2}+512\beta^{3})+\frac{73\kappa^{4}{\xi}^{2}{\phi}^{4}
-448\beta^{2}-512\beta^{3}+15+ 232\beta{\kappa}^{2}\xi\phi^{2}
+1472\beta^{2}{\kappa}^{2}\xi\phi
^{2}-360\kappa^{4}\xi^{2}\phi^{4}\beta}{
\left(-1+{\kappa}^{2}\xi\phi^{2} \right) ^{4}}
\hspace{2.5cm}\nonumber\\
\frac {184\kappa^{6}\xi^{3}\phi^{6 }\beta-1600\beta^{2}\kappa^{4}
\xi^{2}\phi^{4}-1536\beta^{3}\kappa^{4}\xi^{2}\phi^{4}+1536
\beta^{3}\kappa^{2}\xi\phi^{2}+576\beta^{2}\kappa^{6}\xi^{3}\phi^{6}+512\beta^{3}\kappa^{6
}\xi^{3}\phi^{6}}{ \left( -1+{\kappa}^{2}\xi\,{\phi}^{2} \right)
^{4}}
\hspace{2.5cm}\nonumber\\
+\frac{15\kappa^{6}\xi^{3}\phi^{6}-56\beta}{\left(
-1+{\kappa}^{2}\xi\phi^{2} \right) ^{4}}\Bigg)-\frac {\sqrt {6}
{\kappa}^{6}{\xi}^{2}\beta^{-1}\grave{{\cal{K}}}}{6480b^{\beta}\left(-\kappa^{2}\xi
\right)^{\frac{1}{2}+4\beta}}\Bigg(\frac{\grave{{\cal{O}}}}{2}(1408\beta+8192
{\beta}^{3}+4096{\beta}^{4}+105+5504{\beta}^{2}) (1-8\beta )
\hspace{2.5cm}\nonumber\\
- \frac {790{\kappa}^{6}\xi^{3
}\phi^{6}-896\kappa^{4}\xi^{2}\phi^{4}+3584\beta^{2}+7168\beta^{3}4096\beta^{4}+105\kappa^{
8}\xi^{4}\phi^{8}+272\beta+24576\beta^{4}\kappa^{4}\xi^{2}\phi^{4}}{\left(
-1+{\kappa}^{2}\xi\phi^{ 2}\right)^{5}}+
\hspace{2.5cm}\nonumber\\
\frac{1408\kappa^{8}{\xi}^{4}{\phi}^{8}\beta+490\kappa^{2}\xi
\phi^{2}-16384\beta^{4}{\kappa}^{2}\xi\phi^{2}-1456\beta
\kappa^{2}\xi\phi^{2}-16384\beta^{4}\kappa^{6} \xi^{3}\phi^{6}+5504
\beta^{2}\kappa^{8}\xi^{4}\phi^{8}}{\left(
-1+{\kappa}^{2}\xi\,{\phi}^{ 2} \right)^{5}}
\hspace{2.5cm}\nonumber\\
+ \frac {8192\beta^{3}\kappa^{8}\xi
^{4}\phi^{8}-105+4096\beta^{4}\kappa^{8}\xi^{4}\phi^{8}-15744\beta^{2}\kappa^{2}\xi\phi^{2}+
3120\kappa^{4}\xi^{2}\phi^{4}\beta+26240\beta^{2}\kappa^{4}\xi
^{2}\phi^{4}}{\left( -1+{\kappa}^{2}\xi\,{\phi}^{ 2} \right) ^{5}}
\hspace{2.5cm}\nonumber\\
+\frac{46080\beta^{3}\kappa^{4}\xi^{2}\phi^{4}-29696\beta
^{3}\kappa^{2}\xi\phi^{2}-3344\kappa^{6}\xi^{3}\phi^{6}
\beta-19584\beta^{2}\kappa^{6}\xi^{3}\phi^{6}-31744\beta^{3}\kappa
^{6}\xi^{3}\phi^{6}}{\left( -1+{\kappa}^{2}\xi\,{\phi}^{ 2} \right)
^{5}}\Bigg)+\Bigg(4
\hspace{2.5cm}\nonumber\\
(1-4\beta)\beta(48\beta^{2}+32\beta^{3}+ 3+22\beta
)\grave{{\cal{O}}}-\frac{
15\kappa^{2}\xi\phi^{2}+38\beta\kappa^{2}\xi\phi^{
2}-30{\kappa}^{4}{\xi}^{2}{\phi}^{4}+40{\beta}^{2}+160{\beta}^{3}+128
{\beta}^{4}}{ \left( -1+{\kappa}^{2}\xi\,{\phi}^{2} \right) ^{5}}
\hspace{2.5cm}\nonumber\\
\frac {352{\beta}^{2}{\kappa}^{4
}{\xi}^{2}{\phi}^{4}-3-46{\kappa}^{4}{\xi}^{2}{\phi}^{4}\beta-192{\beta}^{2}{\kappa}^{2}\xi{\phi}^{2}-288{\beta}^
{2}{\kappa}^{6}{\xi}^{3}{\phi}^{6}-736{\beta}^{3}{\kappa}^{6}{\xi}^{3}{
\phi}^{6}+1056{\beta}^{3}{\kappa}^{4}{\xi}^{2}{\phi}^{4}}{ \left(
-1+{\kappa}^{2}\xi\,{\phi}^{2} \right) ^{5}}
\hspace{2.5cm}\nonumber\\
+\frac {88{\beta}^{2}{\kappa}^{8}{\xi}^{4}{\phi}
^{8}-672{\beta}^{3}{ \kappa}^{2}\xi{\phi}^{2}-10\beta +6{
\kappa}^{6}{\xi}^{3}{\phi}^{6}\beta+192{\beta}^{3}{\kappa}^{8}{\xi}^{4}{\phi}^{8}+12{\kappa}^{8}{\xi}
^{4}{\phi}^{8}\beta-512\beta^{4}{\kappa}^{2}\xi{\phi}^{2}}{ \left(
-1+{\kappa}^{2}\xi\,{\phi}^{2} \right) ^{5}}+
\hspace{2.5cm}\nonumber\\
\frac{128{\beta}^{4}{\kappa}^{8}{\xi}^{4}{\phi}^{8}+30{\kappa}^{6}{\xi}^{3}{\phi}^{6}-512{\beta}^{
4}{\kappa}^{6}{\xi}^{3}{\phi}^{6}+768{\beta}^{4}{\kappa}^{4}{\xi}^{2}{
\phi}^{4}}{(-1+{\kappa}^{2}\xi\phi^{2})
^{5}}\Bigg)\frac{{\kappa}^{6}{\xi}^{2}\beta^{-1}\vartheta^{-1}l^{-1}\grave{{\cal{I}}}}{6480b^{2\beta}\left(-\kappa^{2}\xi
\right)^{4\beta}}\Bigg]\,,\hspace{2cm}
\end{eqnarray*}
\end{widetext}
\begin{widetext}
\begin{eqnarray*}
\bigwedge=-\frac{{\kappa}^{4}\xi}{2\left(-\kappa^{2}\xi
\right)^{3\beta}}\Bigg(
{\frac { 2-6{\kappa}^{2}\xi{\phi}^{2}+9\beta
+6{\kappa}^{4}{\xi}^{2}{\phi}^{4}+9{\beta}^{2}-21\beta{\kappa}^{2}\xi{
\phi}^{2}+12{\kappa}^{4}{\xi}^{2}{\phi}^{4}\beta+9{\beta}^{2}{\kappa}^{4}{
\xi}^{2}{\phi}^{4}-18\beta^{2}{\kappa}^{2}\xi{\phi}^{2}}{
\left(-1+{\kappa}^{2}\xi{\phi}^{2} \right) ^{3}}}+3\beta
\hspace{2.5cm}\nonumber\\
\left(9\beta+9{\beta}^{2}+2 \right)\grave{{\cal{O}}}\Bigg)
\frac{\left( {\frac {\grave{{\cal{Q}}}}{108}}+{\frac
{\grave{{\cal{Q}}}}{108}}\beta +\frac{\grave{{\cal{J}}}\left(
1+3\vartheta l^{2}{b}^{\beta}-3\,\vartheta l{b}^{\beta}
\right)}{27b^{\beta}}\right)}{\beta l^{4}\vartheta^{4} \left(
1+3\vartheta{l}^{2}{b}^{\beta}-3\vartheta l{b}^{\beta}
\right)}+\frac{{\vartheta}^{-1}{l}^{-1}{\beta
}^{-1}\grave{{\cal{R}}}}{ \left( 1+3\vartheta
l^{2}{b}^{\beta}-3\vartheta l{b}^{\beta} \right)}\frac
{\kappa^{4}\xi^{2}a^{\beta}
}{432\left(-\kappa^{2}\xi\right)^{3\beta}}
\hspace{3.5cm}\nonumber\\
\Bigg( \frac{
2-8\kappa^{2}\xi{\phi}^{2}+12{\kappa}^{4}{\xi}^{2}{\phi}^{4}+3
\beta-18{\beta}^{2}-6\beta{\kappa}^{2}\xi{\phi}^{2}-3{\kappa
}^{4}{\xi}^{2}{\phi}^{4}\beta-72{\beta}^{2}{\kappa}^{4}{\xi}^{2}{\phi}^{4}+
63{\beta}^{2}{\kappa}^{2}\xi{\phi}^{2}}{ \left( -1+{\kappa}^{2}
\xi\,{\phi}^{2} \right) ^{4}}+
\hspace{3.5cm}\nonumber\\
\frac{27{\beta}^{2}{\kappa}^{6}{\xi}^{3
}{\phi}^{6}-27{\beta}^{3}+27{\beta}^{3}{\kappa}^{6}{\xi}^{3}{\phi}^{6}-81{\beta}^{3}{
\kappa}^{4}{\xi}^{2}{\phi}^{4}+81{\beta}^{3}{\kappa}^{2}\xi{\phi}^{2}+
6\beta{\kappa}^{6}{\xi}^{3}{\phi}^{6} }{(-1+{\kappa}^{2}
\xi\phi^{2}) ^{4}}+3(1-3\beta)(2+9{\beta}^{2}
\hspace{3.5cm}\nonumber\\
+ 9\beta)\beta\grave{{\cal{O}}}\Bigg)-\sqrt {{\frac {l-1}{\vartheta
l}}}\Bigg\{\frac{\sqrt{6}{\kappa}^{2}{
b}^{\beta}\beta^{-1}}{6\vartheta
l\left(-\kappa^{2}\xi\right)^{\frac{1}{2}+\beta}}\left( {\frac
{\grave{\mathcal{S}}( 2 {\kappa}^{2}\xi{\phi}^{2}-3-2\beta)}{ \left(
-1+{\kappa}^{2} \xi{\phi}^{2} \right)  \left( 2\beta+1 \right)
}}-\frac{1}{4}\grave{\mathcal{S}} \left( 3+2\beta
\right)\grave{{\cal{O}}}\right)-\Bigg[3(1+\beta)
\hspace{3.5cm}\nonumber\\
\Bigg( {\frac {\grave{{\cal{T}}} \left(
6{\kappa}^{2}\xi{\phi}^{2}\beta+5\kappa^{2}\xi{\phi}^{2}-6\beta-3
\right) }{ \left(-1+{\kappa}^{2}\xi{\phi}^{2} \right)^{2}}}
+\frac{3}{2}\grave{{\cal{T}}}\left( 8\beta +1+12{\beta}^{2} \right)
\grave{{\cal{O}}}\Bigg)+\Bigg(\frac{3}{2}\grave{{\cal{U}}}(
5+46\beta+108\beta^{2}+72\beta^{3})\grave{{\cal{O}}}
\hspace{3.5cm}\nonumber\\
+\frac{\grave{{\cal{U}}}\left(33\kappa
^{4}\xi^{2}\phi^{4}-40\kappa^{2}\xi\phi^{2}+48\beta+15-108\kappa^{2}\xi
\phi^{2}\beta-72\beta^{2}\kappa^{2}\xi\phi^{2}+60\beta\kappa^{4}
\xi^{2}\phi^{4}+36\beta^{2}\kappa^{4}\xi^{2}\phi^{4}\right) }{
\left( -1+{\kappa}^{2}\xi\,{\phi}^{2} \right) ^{3}}
\hspace{3.5cm}\nonumber\\
+\frac{36\beta^{2}\grave{{\cal{U}}}}{(-1+\kappa^{2}\xi\phi^{2})
^{3}}\Bigg) \Bigg]\frac{\sqrt{6} {\kappa}^
{4}\xi}{108\vartheta^{3}l^{3}\beta\left(-{\kappa}^{2}\xi
\right)^{\frac{1}{2}+3\beta}}\Bigg\}+\frac{{\kappa}^{2}\left( l-1
\right) }{6\vartheta^{3}l^{3}\beta\left(-\kappa^{2}\xi
\right)^{2\beta}}\Bigg[
\Bigg(\grave{{\cal{V}}}+\frac{\grave{{\cal{W}}}b^{\beta} \left(
1+\beta \right)}{2\left( 1+3\vartheta l^{2}b^{\beta}-3\vartheta
lb^{\beta} \right)}
\hspace{3.5cm}\nonumber\\
+\vartheta l\beta
\grave{{\cal{Z}}}\Bigg)\left({\frac{\kappa^{2}\xi{\phi}^{2}-1-2\beta
}{\beta \left(-1+{\kappa}^{2}\xi{\phi}^{2} \right)}}-\left(2+4\beta
\right) \grave{{\cal{O}}}\right)+\frac{\xi
b^{\beta}\vartheta^{2}l^{2}\grave{{\cal{X}}}}{6\left( {\frac
{{b}^{-\beta}}{3\vartheta l \left( l-1 \right) }}+1
\right)}\Bigg(\frac{\left( 2\beta{\kappa}^{2}\xi
{\phi}^{2}-2\beta+2{\kappa}^{2}\xi{\phi}^{2}-1 \right)}{ \left( -
1+{\kappa}^{2}\xi{\phi}^{2}\right)^{2}}
\hspace{3.5cm}\nonumber\\
+2\grave{{\cal{X}}}\beta \left( 1+2\beta \right) \grave{{\cal{O}}}
\Bigg)\Bigg]+\frac{1}{4}{\kappa}^{2}{\phi}^{2}+\frac{1}{4}{\frac
{{\kappa}^{2}{\phi}^{2}}{\beta}}+\frac{1}{4}{\frac {\ln  \left(
-1+{\kappa}^{2}\xi{\phi}^{2} \right) }{\xi
\beta}}\,\Bigg|_{hc}^{f}\,,\hspace{4cm}
\end{eqnarray*}
\end{widetext}
\begin{widetext}
\begin{eqnarray*}
\grave{{\cal{\vartheta}}}=\frac{{\phi}^{1+4\,\beta} \left(
-{\kappa}^ {2}\xi \right) ^{5/2+2\,\beta}\left( 5+4\,\beta \right)
}{ 4\left( 5/2+2\,\beta \right) {\kappa}^{4}{\xi}^{2}}\,,\,\quad
\grave{{\cal{B}}}={\frac {{\phi}^{1+2\,\beta} \left( -{\kappa}^{2}
\xi \right) ^{\beta+3/2} \left( -2\,\beta-3 \right) }{ \left(
\beta+3/2 \right) { \kappa}^{2}\xi}}\,,\,\quad
\grave{{\cal{C}}}={\frac {{\phi}^{4\,\beta} \left( -{\kappa}^{2}\xi
\right) ^{2\,\beta} \left( 1+\beta \right) }{2+2\,\beta}}\,,
\end{eqnarray*}
\end{widetext}
\begin{widetext}
\begin{eqnarray*}
\grave{{\cal{D}}}=\frac{{\phi}^{1+6\,\beta} \left( -{\kappa}^ {2}\xi
\right) ^{5/2+3\,\beta} \left( 5+6\,\beta \right) }{4 \left(
5/2+3\,\beta \right)
{\kappa}^{4}{\xi}^{2}}\,,\,\quad
\grave{{\cal{E}}}=\frac{{\phi}^{1+6\,\beta}
\left( -{\kappa}^ {2}\xi \right) ^{5/2+3\,\beta} \left( 5+6\,\beta
\right) }{8 \left( 5/2+3\,\beta \right)
{\kappa}^{4}{\xi}^{2}}\,,\,\quad
\grave{{\cal{F}}}=\frac{{\phi}^{4\,\beta+1} \left( -{\kappa}^ {2}\xi
\right) ^{7/2+2\,\beta} \left( 7+4\,\beta \right) }{ 8\left(
7/2+2\,\beta \right) {\kappa}^{6}{\xi}^{3}}\,,
\end{eqnarray*}
\end{widetext}
\begin{widetext}
\begin{eqnarray*}
\grave{{\cal{G}}}=\frac {{\phi}^{8\,\beta+1} \left( -{\kappa}
^{2}\xi \right) ^{4\,\beta+7/2} \left( 8\,\beta+7 \right) }{16
\left( 4\,\beta+7/2 \right) {\kappa}^{6}{\xi}^{3}}\,,\,\quad
\grave{{\cal{H}}}=\frac {{\phi}^{8\,\beta+1} \left( -{\kappa}
^{2}\xi \right) ^{4\,\beta+7/2} \left( 8\,\beta+7 \right) }{16
\left( 4\,\beta+7/2 \right) {\kappa}^{6}{\xi}^{3}}\,,\,\quad
\grave{{\cal{K}}}=\frac{{\phi}^{8\,\beta+1} \left( -{\kappa} ^{2}\xi
\right) ^{9/2+4\,\beta} \left( 9+8\,\beta \right) }{32 \left(
9/2+4\,\beta \right) {\kappa}^{8}{\xi}^{4}}\,,
\end{eqnarray*}
\end{widetext}
\begin{widetext}
\begin{eqnarray*}
\grave{{\cal{O}}}=\,{\it LerchPhi}
\Big(\kappa^{2}\xi\phi^{2},1,\beta+\frac{1}{2}\Big)\,,\,\quad
\grave{{\cal{I}}}=\frac
{8{\phi}^{8\,\beta} \left( -{\kappa}^{2} \xi \right) ^{4\,\beta}
\left( \beta+1 \right) }{4\,\beta+4}\,,\,\quad
\grave{{\cal{J}}}=\frac {3{\phi}^{6\,\beta} \left( -{\kappa}^{2} \xi
\right) ^{3\,\beta} \left( \beta+1 \right) }{3\,\beta+3}\,,
\end{eqnarray*}
\end{widetext}
\begin{widetext}
\begin{eqnarray*}
\grave{{\cal{Q}}}=\frac {3{\phi}^{6\,\beta} \left( -{\kappa}^{2} \xi
\right) ^{3\,\beta} \left( \beta+1 \right) }{3\,\beta+3}\,,\,\quad
\grave{{\cal{R}}}=\frac {3{\phi}^{6\,\beta} \left( -{\kappa}^{2} \xi
\right) ^{3\,\beta} \left( \beta+1 \right)
}{3\,\beta+3}\,,\,\quad\grave{{\cal{S}}}=\frac {{\phi}^{2\,\beta+1}
\left( -{\kappa}^{2}\xi \right) ^{\beta+5/2} \left( 5+2\,\beta
\right) }{ 2\left( \beta+5/2 \right) { \kappa}^{4}{\xi}^{2}}\,,
\end{eqnarray*}
\end{widetext}
\begin{widetext}
\begin{eqnarray*}
\grave{{\cal{T}}}=\frac {{\phi}^{6\,\beta+1} \left( -{\kappa}^
{2}\xi \right) ^{3\,\beta+5/2} \left( 6\,\beta+5 \right) }{4 \left(
3\,\beta+5/2 \right) {\kappa}^{4}{\xi}^{2}}\,,\quad
\grave{{\cal{U}}}=\frac {{\phi}^{6\,\beta+1} \left( -{\kappa}^
{2}\xi \right) ^{3\,\beta+7/2} \left( 6\,\beta+7 \right) }{ 8\left(
3\,\beta+7/2 \right) {\kappa}^{6}{\xi}^{3}}\,,\,\quad
\grave{{\cal{V}}}={\frac {{\phi}^{4\,\beta} \left( -{\kappa}^{2}\xi
\right) ^{2\,\beta} \left( \beta+1 \right) }{2\,\beta+2}}\,,
\end{eqnarray*}
\end{widetext}

\begin{widetext}
\begin{eqnarray*}
\grave{{\cal{W}}}=\frac{\phi^{4\beta} \left( -\kappa^{2}\xi \right)
^{2\beta} \left(\beta+1 \right) }{2\beta+2}\,,\quad
\grave{{\cal{X}}}=\frac {2{\phi}^{4\,\beta} \left( -{\kappa}^{2} \xi
\right) ^{2\,\beta} \left( \beta+1 \right)
}{2\,\beta+2}\,,\quad\grave{{\cal{Z}}}={\frac {{\phi}^{4\,\beta}
\left( -{\kappa}^{2}\xi \right) ^{2\,\beta} \left( \beta+1 \right)
}{2\,\beta+2}}\,.
\end{eqnarray*}
\end{widetext}

\section{Eigenvalues corresponding to critical line ${\cal{L}}_{1}$}
\label{B}
\begin{eqnarray*}
\lambda_{2{\cal{L}}_{1}}=\frac{3}{2\big(1-{x}^{2}-{z}^{2}+{z}^{2}{x}^{2}\big)}
\Bigg(-{y}^{2}\sqrt
{1-{x}^{2}}\hspace{0.5cm}\nonumber\\
+{y}^{2}{x}^{2}\sqrt{1-{x}^{2}}
+1-{x}^{2}-2\,{z}^{2}+2\,{z}^{2}{x}^{2}+\omega-\omega\,{x}^{2}
\hspace{0.5cm}\nonumber\\
-2
\omega{z}^{2}+2\,\omega\,{z}^{2}{x}^{2}+{y}^{2}{z}^{2}\sqrt{1-{x}^{2}}
-{y}^{2}\omega\sqrt{1-{x}^{2}}\hspace{0.5cm}\nonumber\\+{y}^{2}\omega\,{z}^{2}\sqrt{1-{x}^{2}}
+{z}^{4}-{z}^{4}{x}^{2}+{z}^{4}\omega-{z}^{4}\omega\,{x}^{2}\Bigg)\,,
\hspace{0.5cm}\nonumber\\
\end{eqnarray*}
\begin{eqnarray*}
\lambda_{3{\cal{L}}_{1},4{\cal{L}}_{1}}=\frac{-1}{4\big(1-{x}^{2}-{z}^{2}+{z}^{2}{x}^{2}\big)}
\Bigg(-21{x}^{2}-18{z}^{2}{x}^{4}\hspace{1.3cm}\nonumber\\-3{z}^{4}\omega+18\,{x}^{4}-3\omega-3
{z}^{4}-6\omega{z}^{2}{x}^{2}+3+12\,\sqrt {3}y{x}^{3}{z}^{2}
\hspace{1.2cm}\nonumber\\
+6\omega{z}^{2}-12\sqrt {3}yx{z}^{2}-9\sqrt
{1-x^{2}}\,{y}^{2}\omega{z}^{2}+18{z}^{2}{x}^{2}+3\omega{x}^{2}
\hspace{1cm}\nonumber\\+3{z}^{4}{x}^{2}+9\sqrt
{1-x^{2}}{y}^{2}-9\sqrt{1-x^{2}}{y}^{2}{z}^{2}-9\sqrt{1-x^{2}}{y}^{2}{x}^{2}\hspace{0.5cm}
\hspace{0.5cm}\nonumber\\
-12\sqrt {3}y{x}^ {3}+3{z}^{4}\omega{x}^{2}+12\sqrt
{3}yx+9\sqrt{1-x^{2}}{y}^{2}\omega\hspace{1cm}
\hspace{0.5cm}\nonumber\\
\pm\Bigg[81-81{y}^{4}{
\omega}^{2}{z}^{4}{x}^{2}+162{y}^{4}{\omega}^{2}{z}^{2}{x}^{2}+81{
y}^{4}-162{y}^{4}\omega{z}^{4}{x}^{2}\hspace{0.9cm}\nonumber\\-162{y}^{4}\omega{z}^{2}{
x}^{4}-672{x}^{2}{y}^{2}{z}^{2}+486{y}^{4}\omega{z}^{2}{x}^{2}-
486{x}^{2}-216\,{z}^{2}
\hspace{0.7cm}\nonumber\\
-96{y}^{2}-972{x}^{6}-72{z}^{6}+336{x
}^{2}{y}^{2}{z}^{4}+144{y}^{2}{x}^{6}{z}^{4}+9{\omega}^{2}\hspace{1.2cm}\nonumber\\
+324{x }^{8}-756{z}^{4}\omega{x}^{2} +864{z}^{4}\omega{x}^{4}+684
\omega{z}^{2}{x}^{2}+768{y}^{2}{x}^{4}{z}^{2}\hspace{0.8cm}\nonumber\\-828\omega{z}^{2}
{x}^{4}+324\omega{z}^{2}{x}^{6}-324{z}^{4}\omega{x}^{6}-288{
y}^{2}{x}^{6}{z}^{2}\hspace{1.5cm}\nonumber\\
-324{y}^{4}\omega{z}^{2}+162{y}^{4}\omega{
z}^{4}-162{y}^{4}{\omega}^{2}{z}^{2}+81{y}^{4}{\omega}^{2}{z}^{4}\hspace{1.3cm}\nonumber\\
+324{y}^{4}{z}^{2}{x}^{2}-324{y}^{4}\omega{x}^{2}+324{z}^{6}
\omega\,{x}^{2}-324{z}^{6}{x}^{4}\omega\hspace{1.5cm}\nonumber\\+108{z}^{6}{x}^{6}\omega
-162{y}^{4}{z}^{2}{x}^{4}+162{y}^{4}\omega{x}^{4}-81{y}^{4}{z}^
{4}{x}^{2}
\hspace{1cm}\nonumber\\
-36{
\omega}^{2}{z}^{2}{x}^{4}-81{y}^{4}{\omega}^{2}{x}^{2}+54{\omega}^{2}{z}^{4}{x}^{
4}-36{\omega}^{2}{z}^{6}{x}^{4}\hspace{1cm}
\end{eqnarray*}

\begin{eqnarray*}
+72{\omega}^{2}{z}^{2}{x}^{2}-108{\omega}^{2}{z}^{4}{x}^{2}+72{\omega}
^{2}{z}^{6}{x}^{2}+9{z}^{8}{\omega}^{2}{x}^{4}\hspace{1.5cm}\nonumber\\
-36{z}^{8}\omega{x
}^{2}+18{z}^{8}\omega{x}^{4}-18{z}^{8}{\omega}^{2}{x}^{2}-384{
z}^{4}{x}^{4}{y}^{2}\hspace{1.5cm}\nonumber\\+144\sqrt
{3}y{x}^{3}\omega{z}^{2}-144\sqrt{3}y{x}^{3}{z}^{4}\omega+72\sqrt
{3}yx{z}^{4}\omega\hspace{1.5cm}\nonumber\\
-72\sqrt{3}yx\omega{z}^{2}-1296\sqrt
{1-x^{2}}{y}^{2}{x}^{2}\omega{z}^{2}+1053x^{4}\hspace{1.5cm}\nonumber\\
+120\sqrt{3}{y}^{3}{x}^{3}{z}^{4}\sqrt{1-x^{2}}-120\sqrt {3}\sqrt
{1-x^{2}}{y}^{3}x\omega\hspace{1.5cm}\nonumber\\
+144\sqrt{3}\sqrt{1-x^{2}}{y}^{3}x{z}^{2}
-120\sqrt{3}{y}^{3}x{z}^{4}\sqrt{1-x^{2}}\hspace{1.5cm}\nonumber\\
-168\sqrt{3}{y}^{3}{x}^{3}{z}^{2}\sqrt{1-x^{2}}
+24\sqrt{3}{x}^{5}{y}^{3}{z}^{2}\sqrt{1-x^{2}}\hspace{1.5cm}\nonumber\\
-144{z}^{4}{x}^{7}\sqrt {3}y-72\omega{z}^{2}{x}^{5}\sqrt{3}y
-24\sqrt{3}{x}^{5}y{z}^{6}\omega\hspace{1.5cm}\nonumber\\
+48{z}^{6}\sqrt {3}y{x}^{3}\omega-162{\omega}^{2}{z}^{2}{x}^{2}\sqrt
{1-x^{2}}{y}^{2}\hspace{2.5cm}\nonumber\\
-378{z}^{4}{x}^{4}{y}^{2}\sqrt{1-x^{2}}\omega
+756{y}^{2}{x}^{4}{z}^{2}\sqrt{1-x^{2}}\omega\hspace{1.5cm}\nonumber\\
+162{\omega}^{2}{z}^{4}{x}^{2}\sqrt{1-x^{2}}{y}^{2}
+810{z}^{4}\omega{x}^{2}{y}^{2}\sqrt{1-x^{2}}\hspace{1.5cm}\nonumber\\
-108{z}^{6}\sqrt{1-x^{2}}{y}^{2}\omega{x}^{2}
-54\sqrt{1-x^{2}}{y}^{2}{\omega}^{2}{z}^{6}{x}^{2}\hspace{1.5cm}\nonumber\\
-24{z}^{6}\sqrt{3}yx\omega
+120\sqrt{3}\sqrt{1-x^{2}}{y}^{3}{x}^{3}\omega+72{z}^{4}\omega{x}^{5}\sqrt{3}y\hspace{0.5cm}\nonumber\\
+624\sqrt{3}y{x}^{3}{z}^{2}+240\sqrt{3}\sqrt{1-x^{2}}{y}^{3}x\omega{z}^{2}-168\sqrt{3}yx{z}^{2}\hspace{0.5cm}\nonumber\\
+120\sqrt{3}{y}^{3}{x}^{3}{z}^{4}\sqrt{1-x^{2}}\omega
-240\sqrt{3}{y}^{3}{x}^{3}{z}^{2}\sqrt{1-x^{2}}\omega\hspace{0.5cm}\nonumber\\
+24\sqrt{3}yx\omega -120\sqrt{3}{y}^{3}x{z}^{4}\sqrt{1-x^{2}}\omega+120\sqrt{3}yx{z}^{4}\hspace{0.5cm}\nonumber\\
+540\sqrt{1-x^{2}}{y}^{2}\omega{z}^{2}
-744\sqrt{3}{x}^{5}y{z}^{2}-384\sqrt{3}y{x}^{3}{z}^{4}\hspace{0.5cm}\nonumber\\
+594\sqrt{1-x^{2}}{y}^{2}{x}^{2}\omega-1242\sqrt{1-x^{2}}{y}^{2}{x}^{2}{z}^{2}\hspace{1.5cm}\nonumber\\
-48\sqrt{3}y{x}^{3}\omega-24\sqrt{3}\sqrt{1-x^{2}}{y}^{3}x
+48\sqrt{3}\sqrt{1-x^{2}}{y}^{3}{x}^{3}\hspace{0.4cm}\nonumber\\
+1188{y}^{2}{x}^{4}{z}^{2}\sqrt{1-x^{2}}
-378{z}^{4}{x}^{4}{y}^{2}\sqrt{1-x^{2}}+54\omega\hspace{0.5cm}\nonumber\\
+408{z}^{4}{x}^{5}\sqrt{3}y+288{z}^{2}{x}^{7}\sqrt {3}y
-324{z}^{2}{x}^{6}\sqrt{1-x^{2}}{y}^{2}\hspace{0.5cm}\nonumber\\
-378\sqrt{1-x^{2}}{y}^{2}{x}^{4}\omega
+162\sqrt{1-x^{2}}{y}^{2}{\omega}^{2}{z}^{2}+9z^{8}\hspace{0.8cm}\nonumber\\
+48z^{6}\sqrt{3}yx^{3}-24z^{6}\sqrt{3}yx
+108z^{6}\sqrt{1-x^{2}}y^{2}\omega\hspace{1.5cm}\nonumber\\
-432z^{4}\sqrt{1-x^{2}}y^{2}\omega+648z^{4}\sqrt{1-x^{2}}y^{2}x^{2}\hspace{2cm}\nonumber\\
-24\sqrt{3}x^{5}yz^{6}
-162\sqrt{1-x^{2}}y^{2}\omega^{2}z^{4}+24\omega x^{5}\sqrt{3}y\hspace{0.8cm}\nonumber\\
\end{eqnarray*}

\begin{eqnarray*}
+54\sqrt{1-x^{2}}y^{2}\omega^{2}z^{6}
+54\omega^{2}{x}^{2}\sqrt{1-x^{2}}y^{2}\hspace{1.5cm}\nonumber\\
-54z^{6}{x}^{2}{y}^{2}\sqrt{1-x^{2}}-24\sqrt{3}x^{5}y^{3}\sqrt{1-x^{2}}-36\omega^{2}z^{6}\hspace{0.5cm}\nonumber\\
-144x^{7}\sqrt{3}y+324x^{6}\sqrt{1-x^{2}}{y}^{2}
+378\sqrt{1-x^{2}}{y}^{2}{z}^{2}\hspace{0.5cm}\nonumber\\
+648\sqrt{1-x^{2}}y^{2}x^{2}
-216\sqrt{1-x^{2}}y^{2}\omega+9z^{8}\omega^{2}\hspace{1cm}\nonumber\\
+72\sqrt{3}yx -
288\sqrt{3}yx^{3}-54\sqrt{1-x^{2}}y^{2}\omega^{2}\hspace{1.5cm}\nonumber\\+54z^{6}\sqrt{1-x^{2}}{y}^{2}
-270z^{4}\sqrt{1-x^{2}}y^{2} +360\sqrt{3}x^{5}y\hspace{0.5cm}\nonumber\\
-810\sqrt{1-x^{2}}y^{2}x^{4}-162\sqrt{1-x^{2}}{y}^{2}
-2376z^{2}x^{4}\hspace{0.5cm}\nonumber\\-180\omega
z^{2}+1188z^{2}x^{2}-936z^{4}x^{2} +198z^{4}+216z^{4}\omega\hspace{0.5cm}\nonumber\\
+192y^{2}{z}^{2}+336x^{2}y^{2}+144y^{2}x^{6}
+2052z^{2}x^{6}+18z^{8}\omega\hspace{0.5cm}\nonumber\\+270\omega
x^{4}-108\omega
x^{6}+1602z^{4}x^{4}-1188z^{4}x^{6}+81y^{4}\omega^{2}\hspace{0.5cm}\nonumber\\
+162y^{4}\omega-162y^{4}z^{2}-108z^{6}\omega
-288z^{6}x^{4}+108z^{6}x^{6}\hspace{0.5cm}\nonumber\\-243y^{4}x^{2}+81y^{4}z^{4}
+252z^{6}x^{2}+243y^{4}x^{4}-81y^{4}x^{6}\hspace{0.5cm}\nonumber\\
-96y^{2}z^{4}-36\omega^{2}z^{2}+54\omega^{2}{z}^{4}
-648x^{8}z^{2}-18\omega^{2}x^{2}\hspace{0.5cm}\nonumber\\+9\omega^{2}x^{4}-18z^{8}x^{2}\hspace{0.5cm}\nonumber\\
+9z^{8}x^{4}+324z^{4}x^{8}-216\omega
x^{2}\hspace{0.5cm}\nonumber\\-384y^{2}x^{4}\Bigg]^{\frac{1}{2}}\Bigg)\hspace{1.5cm}\nonumber\\
\end{eqnarray*}

\end{document}